\documentclass[12pt,dvips]{article}
\usepackage{amssymb} 
\usepackage{epsfig}

\newcommand{\be}{\begin{equation}}
\newcommand{\ee}{\end{equation}}
\newcommand{\bea}{\begin{eqnarray}}
\newcommand{\eea}{\end{eqnarray}}
\newcommand{\RE}{\mathbb{R}}

\newcommand{\QA}{\;, \quad }
\newcommand{\condensate}{\ensuremath{\langle\bar{\psi}\psi\rangle}{}}
\newcommand{\subcond}{\ensuremath{\langle\bar{\psi}\psi\rangle_{\rm sub}}{}}
\newcommand{\contcond}{\ensuremath{\langle\bar{\psi}\psi\rangle_{\rm cont}}{}}

\newcommand\D{\ensuremath{{\cal D\,}}{}}
\newcommand\DWI{\ensuremath{{\cal D}_{\rm Wi}\,}{}}
\newcommand\DFP{\ensuremath{{\cal D}_{\rm Fp}\,}{}}
\newcommand\DOV{\ensuremath{{\cal D}_{\rm Ne}\,}{}}
\newcommand\unitmatrix{\ensuremath{\textrm{\boldmath{$\mathsf{1}$}}}}
\newcommand{\gaugeexp}[1]{\left\langle #1 \right\rangle_{\rm gauge}}

\begin{document}

\thispagestyle{empty}

\title{
\vspace{-5.0cm}
\vspace*{2cm}
Eigenvalue spectrum of massless Dirac operators on the 
lattice\thanks{Supported by Fonds 
zur F\"orderung der Wissenschaftlichen Forschung 
in \"Osterreich, Project P11502-PHY.} }
\author
{\bf F. Farchioni\footnote{E-mail: fmf@physik.kfunigraz.ac.at}, I. Hip,  C.~B.
Lang and M. Wohlgenannt~\\  \\ Institut f\"ur Theoretische Physik,\\
Universit\"at Graz, A-8010 Graz, AUSTRIA}
\maketitle
\begin{abstract}
We present a detailed study of the interplay between chiral symmetry
and spectral properties of the Dirac operator in lattice gauge
theories. We consider, in the framework of the Schwinger model, the
fixed point action and a fermion action recently proposed by Neuberger.
Both actions show the remnant of chiral symmetry on the lattice as
formulated in the Ginsparg-Wilson relation. We check this issue for
practical implementations, also evaluating the fermion condensate in a
finite volume by a subtraction procedure. Moreover, we investigate  the
distribution of the eigenvalues of a properly defined anti-hermitian
lattice Dirac operator, studying  the statistical properties at the low
lying edge of the  spectrum. The comparison with the predictions of
chiral Random Matrix Theory enables us to obtain an estimate of the
infinite volume fermion condensate.
\end{abstract}


\newpage

\section{Introduction}
\label{sec:intro}

A local lattice theory describing the physical degrees of freedom of 
the corresponding continuum theory necessarily breaks chiral symmetry
\cite{NiNi81a}.  In a usual discretization, like the Wilson  action,
this breaking is so violent, that no trace of the chiral properties of
the continuum is kept in the lattice theory. This involves many
inconveniences -- in particular for lattice QCD. At the classical
level, for fixed gauge field configurations, zero modes of  the
(Euclidean) Dirac operator have no definite chirality, and the
Atiyah-Singer theorem of the continuum theory \cite{AtSi71}, relating
the index of the Dirac operator to the topological charge of the
background gauge configuration, finds no strict correspondence  on the
lattice. At the quantum level, the chirality-breaking terms -- when
combined with  the ultraviolet divergences of the theory --  give raise
to non-universal  finite renormalizations. Examples are the additive
renormalization  of the bare quark mass $m_q$ and the finite
multiplicative renormalizations  of the chiral currents, which spoil
the usual current algebra. Since the chirality is explicitly broken,
the definition of an order parameter for the spontaneous  breaking of
the chiral symmetry is not straightforward.

In an early paper \cite{GiWi82} of lattice quantum field theory,
Ginsparg and Wilson made definite the concept of a chiral limit in the 
framework of a lattice theory breaking explicitly chiral symmetry, 
providing a general condition -- we will refer to it as the
Ginsparg-Wilson Condition (GWC) -- for the fermion matrix of the
lattice theory, i.e. the lattice Dirac operator. For a long time
Ginsparg and Wilson's ideas remained academic,  since no acceptable
solution of the GWC was found in the case of dynamical gauge fields,
locality of the lattice action being  the real bottle-neck in this
matter.

New attention on the GWC was raised by Hasenfratz \cite{Ha98c}, who
pointed out that the fixed point (FP) action, which is local by
construction, satisfies this relation. This result is natural, since
the FP action is a classically perfect action. 

Chiral symmetry on the lattice is made explicit in the overlap
formalism \cite{NaNe93}; recently Neuberger \cite{Ne98} found a form
of the corresponding Dirac operator, which also satisfies the GWC
\cite{Ne98a}.

In a series of papers, the GWC was theoretically analyzed, showing that
it is a sufficient condition for the restoration of  the main features
of the continuum (symmetric) theory. In the following we refer to
actions/fermions satisfying the GWC as to GW actions/fermions. At the
classical level, for FP actions, the Atiyah-Singer  theorem finds
correspondence on the lattice \cite{Ha98c,HaLaNi98};  at the quantum
level, no fine tuning, mixing and current renormalization  occur, and a
natural definition for an order parameter of the  spontaneous breaking
of the chiral symmetry is possible \cite{Ha98a}. The explicit form of
the lattice symmetry corresponding to the   chiral symmetry of the
continuum has been identified \cite{Lu98}.  On the basis of this
symmetry it has been shown \cite{Ch98a}  that the low-energy mechanisms
of the continuum QCD  (e.g. Goldstone's theorem for pions, solution of
the U(1) problem) emerge also in the theory with finite cut-off.

Monte Carlo calculations require actions with finite  number of
couplings (ultra-local actions),  while any solution of the GWC --
even when local -- is expected \cite{Ho98} to be extended all over 
the lattice. In the case of the FP action, the renormalization group
theory ensures exponential damping of the couplings with the
distance, and a parametrization procedure is viable in principle; 
the point to be verified is to what extent an approximation of the
FP  action  in terms of a restricted set of couplings is  able to
reproduce  the nice theoretical properties of the `ideal' FP action.

In this paper we check these issues in the case of the Schwinger model
($d=2$ QED for massless fermions),
testing a parametrization \cite{LaPa98}  of the FP action for the
non-overlapping block-spin transformation.  We focus mainly on the
interplay between chiral symmetry  and spectral properties of the Dirac
operator.  Preliminary results were presented in \cite{FaLaWo98} and 
\cite{FaHiLa98a}. We also test Neuberger's proposal \cite{Ne98}.
Previous investigations on this Dirac operator in the framework of the
Schwinger model  were accomplished in \cite{FaHiLa98} and \cite{Ch98g}
(for an application of the standard overlap formalism, see
\cite{NaNeVr95}). Neuberger's Dirac operator is determined through a
configuration-wise operatorial (numerical) projection \cite{Ne98a} of
the Wilson operator. In this case an explicit parametrization of the 
Dirac operator is not yet available; also the precise locality
properties of this operator have not been established (see, however
\cite{HeJaLu98}).

Based on universality argument, suggested by Leutwyler and  Smilga's  
\cite{LeSm92} sum rules for the eigenvalues of the chiral Dirac operator in a
finite volume, Shuryak and Verbaarschot \cite{ShVe93} hypothesized that
statistical properties of the spectrum of this operator at its lower edge are 
described by  Random Matrix Theory (RMT, for a recent review cf.
\cite{GuMuWe98}). This ansatz has been explicitly verified \cite{DaAkOsDoVe}.
When chiral symmetry holds  in the lattice theory,  the chiral version of RMT
(chRMT) should apply  \cite{ShVe93,VeZa93,Ve94}. So, another way to check the
effective restoration of the chiral symmetry by GW actions, is to compare
\cite{JaSp98} the statistical properties  of their spectrum with the prediction
of chRMT. In this paper we also address this point.

The sequel of the paper is organized as follows. First we summarize the
main properties of GW lattice fermions in relation to chiral symmetry;
in Section \ref{sec:num} we report our numerical  results about the
spectrum of the Dirac operator, in particular  the chirality of (quasi)
zero modes, their dispersion on the real axis, and agreement of the
index of the Dirac operator with the geometric definition for the
topological charge.  We also show the results of the calculation of the
chiral condensate  in a finite volume according to the prescription
given in~\cite{Ha98a}. In Section \ref{sec:res_edge} we concentrate  on
the study of the statistical properties of the eigenvalues  of the
Dirac operator, investigating the universality class at the lower edge
of the spectrum. Comparison with the predictions of chRMT allows us to 
extract the infinite volume fermion condensate from lattice data.  The
last section is devoted to discussion and  conclusions.

\section{Ginsparg-Wilson fermions}
\label{sec:prop}

The GWC for the fermion matrix $\D_{x,x^{\prime}}$ describing
massless fermions reads \cite{GiWi82,Ha98c}
\be
\frac{1}{2}\left\{\, \D_{x,x^{\prime}},\gamma^5\,\right\}\;=\;\left(\,
\D\,\gamma^5\,R\,\D\,\right)_{x,x^{\prime}}\;\;,
\label{eq:gwcg}
\ee
where $R_{x,x^{\prime}}$ is a local matrix in coordinate space, i.e.
whose matrix elements vanish exponentially with the distance. Usual
lattice actions obey the general $\gamma^5$-hermiticity property 
$\D^{\dagger}=\gamma^5\D\gamma^5$. In the particular case
$R_{x,x^{\prime}}=\frac{1}{2} \delta_{x,x^{\prime}}$ the GWC then
assumes the elegant form
\be
\D\,+\,\D^{\dagger} = \D^{\dagger}\,\D = \D\,\D^{\dagger}\;.
\label{eq:gwc}
\ee

\paragraph{Fixed point action.}

Any FP action of a block spin transformation (BST) satisfies the GWC
\cite{Ha98c}. Such an action is the FP of a recursion relation of the
form,
\bea
\DFP_{x^\prime y^\prime}(V)&=&\kappa_{\rm bs}\,\delta_{x^\prime
y^\prime}\nonumber\\ &&-\kappa_{\rm bs}^2\,\sum_{x y}\,\omega_{x^\prime
x}(U)\,\left(\frac{1}{\DFP(U)\,+\,\kappa_{\rm bs}\,
\omega^\dagger\,\omega}\right)_{x y}\,
\omega^\dagger_ {y y^\prime}(U)\;,
\label{prorec}
\eea
where $U(V)$ is fixed by the solution of the pure-gauge  part of the FP
problem, $\omega (U)$ is the matrix  in space-time and color indices
defining the (gauge-covariant)  block-spin average, and $\kappa_{\rm
bs}$ is a free parameter. The operator $R_{x,x^{\prime}}$ has an
expression \cite{Ha98c}  in terms of $\omega (U)$, and it can in
general be assumed  to be trivial in Dirac space. In the case of the
non-overlapping BST considered here and in \cite{LaPa98}, the
particular form (\ref{eq:gwc}) is fulfilled.

\paragraph{Neuberger's Dirac operator.}

In this approach \cite{Ne98} the starting point is the Wilson Dirac
operator at some value of $m \in (-1,0)$ corresponding to $\frac{1}{2D}
< \kappa < \frac{1}{2D-2}$ ($\kappa$ is the well-known  hopping
parameter, not to be confused with the parameter  $\kappa_{\rm bs}$ of
the block-spin transformation) and then one constructs 
\be\label{DOV}
\DOV = \unitmatrix  + \gamma_5\, \epsilon(\gamma_5\,\DWI)
\;,\;\textrm{where}\quad
\epsilon(M)\equiv \frac{M}{\sqrt{M^2}}\; .
\ee
From the definition (\ref{DOV}) it is
evident that  $\DOV$ satisfies (\ref{eq:gwc}) \cite{Ne98a}.
Here we consider the case  $\kappa=\frac{1}{2}$, $m=-1$ (for a
discussion in the context of this study cf. \cite{FaHiLa98}). 

\subsection{Chiral properties}
\label{subs:spe}

Dirac operators $\D$ obeying the GWC do have non-trivial spectral
properties.  For simplicity we consider here operators, like \DFP and
\DOV, satisfying the special version (\ref{eq:gwc}) of the GWC. 

\begin{itemize}
\item[i.] $[\D,\, \D^{\dagger}] = 0$, i.e. $\D$ is a normal operator; as
a consequence, its eigenvectors form a complete orthonormal set.
\item[ii.] The spectrum lies on a unit circle
in the complex eigenvalue-plane centered at $\lambda=1$.
\item[iii.] The property (i), together with
the hermiticity property of the fermion matrix, implies:
\begin{eqnarray}\label{definitechirality}
\gamma^5\,v_{\lambda}&\propto &v_{\overline{\lambda}}
\quad\;\;\;,\qquad\textrm{if}\quad
\lambda\not=\overline{\lambda}\;,\nonumber\\
\gamma^5\,v_{\lambda}&=&\pm\,v_{\lambda}
\quad,\qquad\textrm{if}\quad\lambda=\overline{\lambda}\in\RE \;.
\end{eqnarray}
Here $v_{\lambda}$ denotes an eigenvector of $\D$ with  eigenvalue
$\lambda$. Just as in the continuum, the eigenvectors of
complex-conjugate eigenvalues form doublets related through
$\gamma^5$;  moreover all real-modes have definite chirality. For the
general form (\ref{eq:gwcg}) of the  GWC this property holds only for
the zero modes.
\end{itemize}

\subsection{The Atiyah-Singer theorem on the lattice.}
\label{subs:lattasit}

The Atiyah-Singer index theorem (ASIT) in the continuum relates 
the topological charge $Q(A)$
of a differentiable gauge field configuration $A$ to 
the difference between the numbers of positive and negative 
chirality zero modes of the Dirac operator,
\be
Q(A)=\textrm{index}(A)\equiv n_{+} -n_{-}\;.
\ee
On the lattice, an index for $\D$ may be defined in a way analogous 
to the continuum, explicitly expressed by the relation \cite{HaLaNi98}
\be\label{index}
\textrm{index}(U)=-\textrm{tr}(\gamma^5 \,R\, \D(U))\;.
\ee
In the case of GW Dirac operators satisfying the special GW condition
(\ref{eq:gwc}), the above relation comes out trivially considering that
(due to their properties listed in the previous section)  only the
modes with real non-vanishing eigenvalues contribute to the trace in
the r.h.s, where $R=1/2$; since the overall chirality must be zero, it
reproduces up to a sign index$(U)$. 

This index$(U)$ can be used to define a {\em fermionic}  lattice
topological charge
\be\label{q_ferm}
Q_{\rm ferm}(U)\equiv\textrm{index}(U)\;
\ee
for which the ASIT is satisfied by definition.

In the case of the FP action the fermionic definition (\ref{q_ferm})
coincides \cite{HaLaNi98} with  the pure-gauge quantity $Q_{\rm
Fp}(U)$, the FP topological  charge \cite{BlBuHa96} of the
configuration $U$:
\be\label{index_fp}
Q_{\rm ferm}(U)\equiv\textrm{index}(U) = Q_{\rm Fp}(U)\;.
\ee
The non-obviousness of this relation relies on the fact that  $Q_{\rm
Fp}(U)$ can be defined {\em a priori} in the pure gauge theory, 
without any regard to the fermion part.  We stress that this result is
particular for a FP action, having no counterpart for a general 
(non-FP) GW action. Of course, in practical implementations one relies
on approximate parametrizations of the FP Dirac operator and the
strictness of the relation is consequently lost.

Not all fermionic formulations of the topological charge, associated to
different lattice Dirac operators, are equivalent, since they can be
more or less reminiscent of the continuum \cite{Ni98}.  The FP
definition is in this sense the most reliable, since it is essentially
based on a renormalization group guided procedure of  interpolation
(but maybe difficult to implement in realistic  situations like QCD). 
Here, as a general criterion, we compare the results for \DFP and \DOV
with the geometric definition
\be\label{geocha}
Q_{\rm geo}(U)=\frac{1}{2\pi}\,\sum_x\,{\rm Im\, ln}(U_{12}(x))\;,
\ee
which is in this context the most natural one and the closest to the
continuum definition. In $d=2$ the geometric definition is the FP for
a  particular BST \cite{FaLa98}.

In two dimensions, a theorem of the continuum -- the so-called
Vanishing Theorem \cite{VanTh} -- ensures that only either  positive or
negative chirality zero modes occur. It seems plausible that this
theorem applies  for FP actions \cite{FaLa98}; we know of no proof that
it should  apply to GW fermions in general.

\subsection{The fermion condensate}
\label{subs:cond}

In \cite{Ha98c} Hasenfratz proposed a subtraction procedure for  the
fermion condensate inspired by (\ref{eq:gwcg}) (an analogous
prescription for a physical lattice fermion condensate was proposed by
Neuberger in an earlier paper \cite{Ne98d} in the specific framework of
the overlap formalism and for the particular  case $R=1/2$). The
definition of the subtracted lattice condensate reads
\be
\subcond = 
-\frac{1}{V}\,\gaugeexp{\textrm{tr}\left(\frac{1}{\D}-R\right)}\;\;,
\label{eq:subfc}
\ee
where $V$ is the (finite) space-time volume. The expectation value on
the right-hand side denotes the gauge averaging including the
determinant weight, corresponding to the full consideration of
dynamical fermions. With (\ref{eq:gwcg}) we may  rewrite the right-hand
side as
\be
-\frac{1}{2\,V}\,\gaugeexp{\textrm{tr}
\left(\frac{1}{\D}-\frac{1}{\D^\dagger}\right)}\;.
\ee
Because of $\gamma_5$-hermiticity the trace in the above expression
vanishes, except when a zero mode of $\D$ occurs, in which case a
regulator-mass $\mu$ must be introduced, 
\be
\D\quad\rightarrow\quad \D(\mu)=\D+\mu \,\unitmatrix\;.
\ee

In $d=4$ for more than one flavor, $N_f>1$, $\subcond$ is an order
parameter for the spontaneous breaking of the chiral  symmetry: the
contribution of the zero modes to the gluon average vanishes when
$\mu\rightarrow 0$ because of the damping effect of the fermion
determinant. We conclude that for finite volume $\subcond(\mu)$
vanishes in the chiral limit \cite{Ha98c}.  

Here we consider $d=2$ and $N_f=1$; the chiral symmetry is broken by
the $U(1)$  anomaly. Still we want to use the above expressions to
study the condensate. In this situation $\subcond\not=0$ even in a 
finite volume. The configurations responsible
for the non-zero fermion condensate in  a finite volume are those from
the $|Q|=1$ sector; indeed, if $\D$ has just one zero mode -- which is
possible due to the ASIT and Vanishing Theorem only for $|Q|=1$ -- the
quantity $({\rm det}\,\D\,\textrm{tr}\,\D^{-1})$ has a non-zero limit
when $\mu\rightarrow 0$.  The effect of the subtraction is in general
(for any $N_f$) just to  remove the contribution of the $Q=0$ sector.

The situation in the infinite volume, when the fermion condensate is
obtained through the sequence of limits $\lim_{\mu\rightarrow 0}\lim_{V
\rightarrow \infty}$, is different. In this case the role of
the exact zero modes is irrelevant, the quasi zero modes (the modes
with small but non-vanishing eigenvalues) are responsible for the
non-zero fermion condensate as the Banks-Casher formula \cite{BaCa80}
shows.

We observe (cf. \cite{Ne98,FaLaWo98} as well as the discussion in
\cite{ChZe98}) that (\ref{eq:subfc}) may be rewritten in the form
\be\label{directdef}
\subcond=
-\frac{1}{V}\,\gaugeexp{\textrm{tr}\left(\frac{1}{\tilde{\D}}\right)}
\quad\textrm{with}\quad
\tilde{\D} =  \D\, (1-R\,\D)^{-1}\;.
\ee
The redefined fermion matrix  $\tilde{\D}$ is anti-hermitian and 
anti-commutes with $\gamma^5$:
\be
\tilde{\D}^{\dagger}\,=\,-\tilde{\D}\;,\quad\{\tilde{\D},\gamma^5\}\,
=\,0\;,
\ee
from which it follows that $\tilde{\D}$ has a purely imaginary
spectrum,
\be
\tilde{\D}\,v=-i\,\lambda\,v\;,\quad\lambda\in\RE\;.
\ee
The zero modes of $\D$ coincide of course with those of $\tilde{\D}$.
The replacement $\D\rightarrow\tilde{\D}$ makes manifest the chiral 
invariance implicit in the original operator (observe however that
according to Hasenfratz's subtraction prescription the dynamics of
fermions is still given by the chirality-breaking matrix  $\D$).

The spectral density of $\tilde{\D}$, 
\be\label{RefSpecDensity}
\rho(\lambda)=\frac{1}{V}\frac{d N}{d\lambda}
\ee 
is obtained from the eigenvalue spectrum sampled with
the weight of $(\textrm{det}\D)^{N_f}$.
It complies with the Banks-Casher formula for the (infinite-volume) 
subtracted fermion condensate \cite{FaLaWo98},
\be\label{BanksCasher}
\lim_{\mu\to 0}\,\lim_{V\to\infty} \subcond(\mu) =  
-\pi\,\lim_{\lambda\to 0}\,\lim_{V\to\infty}\rho(\lambda)\;;
\ee
the sequence of the limits is essential.

In the case of the simpler GWC (\ref{eq:gwc}),  the spectrum of
$\tilde{\D}$ is obtained by mapping the spectrum of $\D$ (which in this
case, we recall, lies on a unit circle in the complex plane) onto the
imaginary axis  by the stereographic projection:
\be\label{projection}
\lambda\quad\rightarrow\quad\lambda\,\left(1-\frac{\lambda}{2}
\right)^{-1}\;.
\ee
The spectral density agrees with that of $\D$ in ${\cal O}(\lambda)$.

\subsection{Spectral microscopic fluctuations}
\label{subs:edge}

In the framework of QCD with arbitrary $N_f$, Leutwyler and Smilga 
\cite{LeSm92} derived, through the calculation of low--energy effective 
lagrangians, sum rules which are obtained  in a limit where QCD reduces to a
simple matrix model. These arguments suggest \cite{ShVe93}, that the symmetry
of the theory  (in the simplest case just the chiral one) rather than its
detailed dynamical structure dictates the microscopic fluctuations of the
spectrum of the massless Dirac operator; they should have a universal behavior
in a scaling variable  $z=V\Sigma\lambda$ in the limit $V\to\infty$.  Here
$\Sigma$ denotes a free parameter related to the dynamics of the system. For
fixed $z$, $\lambda\to 0$ in the thermodynamic  limit and the part of the
spectrum near zero -- which is also  the most relevant for the continuum limit
-- is under study.

The simplest model encompassing chiral symmetry is \cite{ShVe93} a chiral 
Random Matrix Theory. Since GW actions are expected to describe 
massless quarks and  implicitly realize chiral symmetry  on the lattice
(explicitly for the matrix $\tilde{\D}$ which anti-commutes with  $\gamma_5$),
it is natural to investigate  whether the predictions of chRMT for the
microscopic fluctuations of the spectrum apply; for GW fermions cf.
\cite{JaSp98,Sp98}. Previous studies for staggered lattice fermions, which
partially restore chirality, where already accomplished  for the SU(2)
\cite{BeMeSc98,MaGuWe98,BeMeWe98,BeGoMe98}  and the SU(3)
\cite{DaHeKrGoHeRa98} lattice gauge theory in four dimensions.

Three classes of universality are predicted within chRMT \cite{Ve94},
corresponding to orthogonal,  unitary or symplectic ensembles (chOE, chUE or
chSE respectively);  universality of related correlation functions has been
proved  in \cite{AkDaMaNiSeVe}. Starting from Leutwyler and Smilga's effective
lagrangians, it is possible to show \cite{DaAkOsDoVe} that the microscopic 
fluctuations of QCD (with three colors) are described by chUE.  In our context,
QED$_2$, no additional symmetry for the Dirac operator beyond the chiral one is
present; in this situation one also expects chUE universality \cite{Ve94}.

Spectral fluctuations may be studied through a variety  of probes (for an
incomplete list of such studies within lattice gauge theory, cf.
\cite{BeMeSc98,MaGuWe98,BeMeWe98,BeGoMe98,DaHeKrGoHeRa98}). Here we consider 
just three of them:
\begin{itemize}
\item the probability distribution of the 
smallest eigenvalue $P(\lambda_{\rm min})$;
\item
the microscopic spectral density $\rho_s(z)$
\be\label{eq:sp_dens}
\rho_s(z) = \lim_{V\to\infty}\frac{1}{\Sigma}\,
\rho\left (\frac{z}{V\Sigma}\right)\;,
\ee
where $\rho(\lambda)$ is the spectral density (\ref{RefSpecDensity}); \item the
number variance\footnote{We thank J.-Z. Ma for drawing our attention on this
particular test.},
\be
\Sigma^2(S_0,S) =
\gaugeexp{(N(S_0,S)-\bar{N}(S_0,S))^2}\;,
\ee
i.e. the variance of the number of eigenvalues in the interval
$[S_0,S]$.
\end{itemize}

Different numbers of flavors $N_f$  correspond to  different
predictions from chRMT. In the terminology of lattice calculations, 
$N_f=0$ corresponds to the quenched situation, while $N_f=1$ (or more) 
to the dynamical one. Therefore comparison with chRMT allows us to
check whether the dynamics of the fermions has been effectively
included in the simulations \cite{BeMeWe98}.  In the dynamical setup,
the Banks-Casher relation (\ref{BanksCasher})  relates the parameter
$\Sigma$ to the (subtracted) fermion condensate in the infinite volume:
\be
\Sigma=-\lim_{\mu\rightarrow 0}\lim_{V\rightarrow\infty} \subcond\;.
\ee

In chRMT there are specific predictions concerning the spectral
distribution in different topological sectors. This theory knows
nothing about the dynamical content of gauge fields; the parameter
$\nu$, which defines the topological sector, is the difference between
the numbers of left- and right-handed modes in the matrix
representation. In the framework of gauge theories  $\nu$ corresponds
to the index  of the Dirac operator: for GW fermions it can be defined
as in the continuum  and produces a fermionic definition of the
topological charge on the lattice (see Section \ref{subs:spe}, eq.
(\ref{index})  and (\ref{q_ferm})). This is not the case for other
(non-GW) lattice actions, where the  index -- and as a consequence the
topological charge -- is not well defined  causing ambiguities in the
comparison with chRMT \cite{BeGoMe98}.

In our discussion of the numerical results we  rely on dimensionless
quantities. The lattice spacing $a$ and the (physical) gauge coupling
constant $e$ are dimensionful and we assume the usual asymptotic
relation to the dimensionless  coupling $\beta$,
\be\label{LattScalePar}
\sqrt{\beta}=\frac{1}{e\,a}\;.
\ee
In order to compare with the (theoretical) continuum values we
therefore have to include corresponding factors of $e$.
The physical lattice size $(L\,a)$ is then in dimensionless units
\be
L\,a\;e=\frac{L}{\sqrt{\beta}}\;.
\ee
For $N_f=1$ the Schwinger mass is
\be\label{SchwingerMass}
a\, m_{Schwinger} = \frac{1}{\sqrt{\beta\,\pi}}\;.
\ee
The continuum, infinite volume value for the condensate is
\be\label{SigmaPhys}
-\contcond
=c\,e \quad\textrm{where}\quad 
c=\frac{\exp{(\gamma)}}{2\,\pi\,\sqrt{\pi}}\approx 0.15989\;,
\ee
($\gamma$ denotes the Euler constant)
and its dimensionless lattice partner (corresponding to $\contcond/e$)
is
\be
\frac{\condensate_{\rm lat}}{e\,a}=
\condensate_{\rm lat}\,\sqrt{\beta}\;.
\ee

\section{The numerical analysis}
\label{sec:num}

The lattice gauge fields $U_{x,\mu}$ are in the compact representation
with the usual Wilson plaquette action.

In \cite{LaPa98} the FP Dirac operator was parametrized as
\be\label{DFP}
\DFP(x,y) =
\sum_{i=0}^3\sum_{x\, , f}\,  \rho_i(f)\, 
\sigma_i\, U(x,f)\;,\quad\textrm{with} \; y\equiv x+\delta f\;.
\ee
Here $f$ denotes a closed loop through $x$ or a path from  the lattice
site $x$ to $y=x+\delta f$ (distance vector $\delta f$) and $U(x,f)$ is
the parallel transporter along this path. The $\sigma_i$-matrices
denote the Pauli matrices for $i= 1,2,3$ and the unit matrix for
$i=0$.  Geometrically the couplings have been restricted to lie within
a  $7\times7$  lattice. The action obeys the usual symmetries; 
altogether it has 429 terms per site. The action was determined as the
(numerically approximate) FP of the Dirac operator for gauge fields
distributed according to the non-compact formulation with the Gaussian
measure. Excellent  scaling properties, rotational invariance and
continuum-like dispersion relations were observed at various values of
the gauge coupling $\beta$ \cite{LaPa98}. Here we study the action
(\ref{DFP}) only for the compact gauge field distributions (see,
however, the results in \cite{FaLaWo98}). In this case the action is
not expected to exactly reproduce  the FP of the corresponding BST, but
nevertheless it is still  a solution of the GWC; violations of the GWC
are instead introduced  by the parametrization procedure, which cuts
off the less local couplings. 

The situation is different for Neuberger's operator, which is
determined for each configuration on the basis of (\ref{DOV}): in our
simple context computer time was not really an obstacle and therefore
we computed  $\epsilon(\gamma_5\,\DWI)$ through diagonalization with
machine accuracy (cf. \cite{FaHiLa98} for more details, as well as
\cite{Ne98e,EdHeNa98,HeJaLu98,Bo98} for more efficient approaches in
$d=4$). 

Essentially uncorrelated gauge configurations have been generated in
the  quenched setup. As a measure we have used the autocorrelation
length for  the geometric topological charge, which -- for Metropolis
updating -- behaves like ${\rm exp}(1.67\beta-3)$  for the range of
gauge couplings studied. Unquenching is obtained by including the
fermionic determinant in the observables; better ways to include  the
dynamics of fermions with the present actions -- in particular
Neuberger's operator -- are still being developed \cite{Li98}.  From
earlier experience \cite{LaPa98,FaLaWo98} we learned that our
procedure  is justifiable at least for the discussed model and the
presented  statistics.   We perform our investigation on sets of  5000
configurations (for $\beta=$ 2, 4) and 10000 configurations (for
$\beta=$ 6) for lattice size $L=16$, and 5000 configurations (for
$\beta=$4, 6) and size $L=24$.

\subsection{The spectrum}
\label{subsec:res_spe}

\begin{figure}[t]
\begin{center}
\epsfig{file=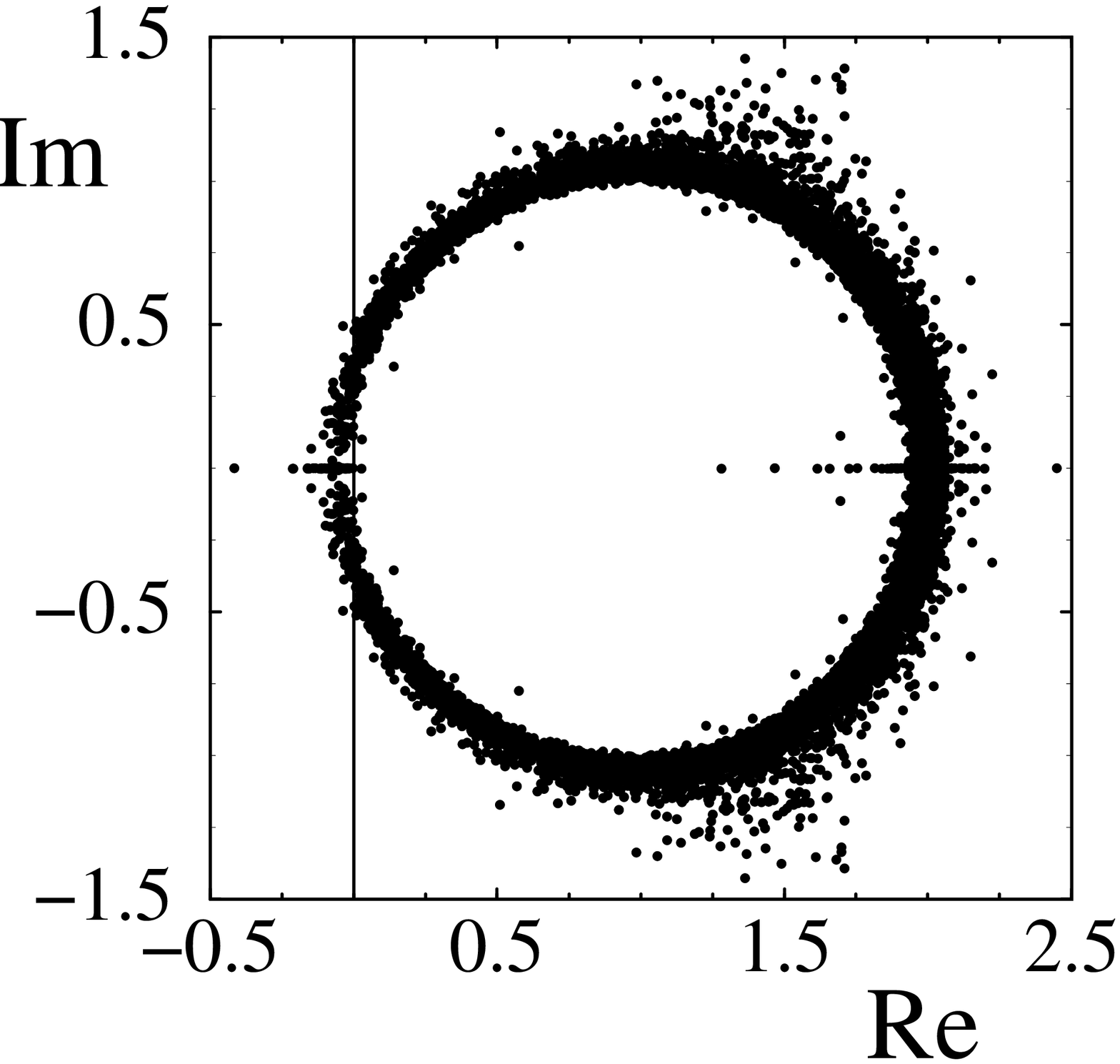,width=4.2cm,angle=-90}
\epsfig{file=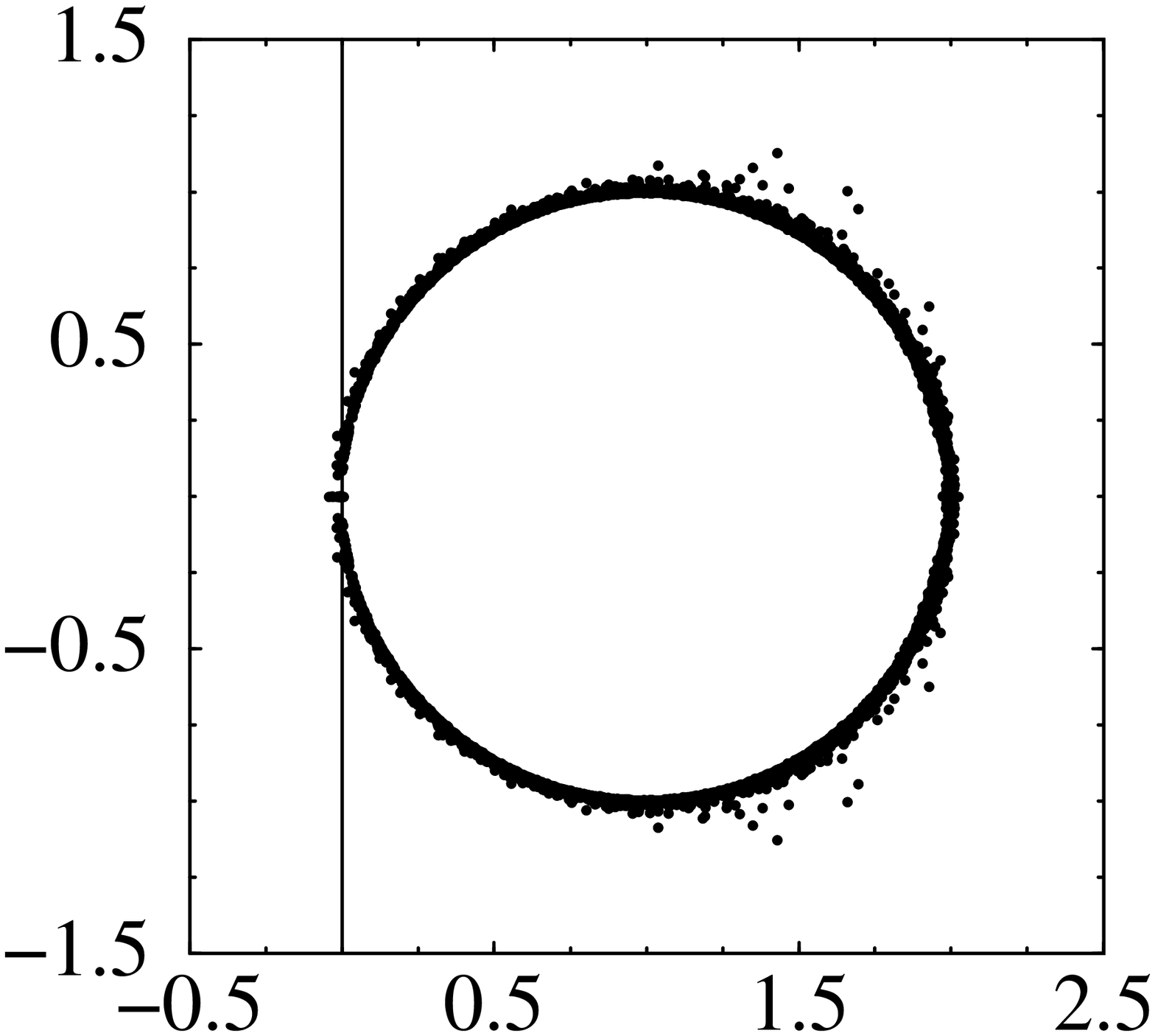,width=4.2cm,angle=-90}
\epsfig{file=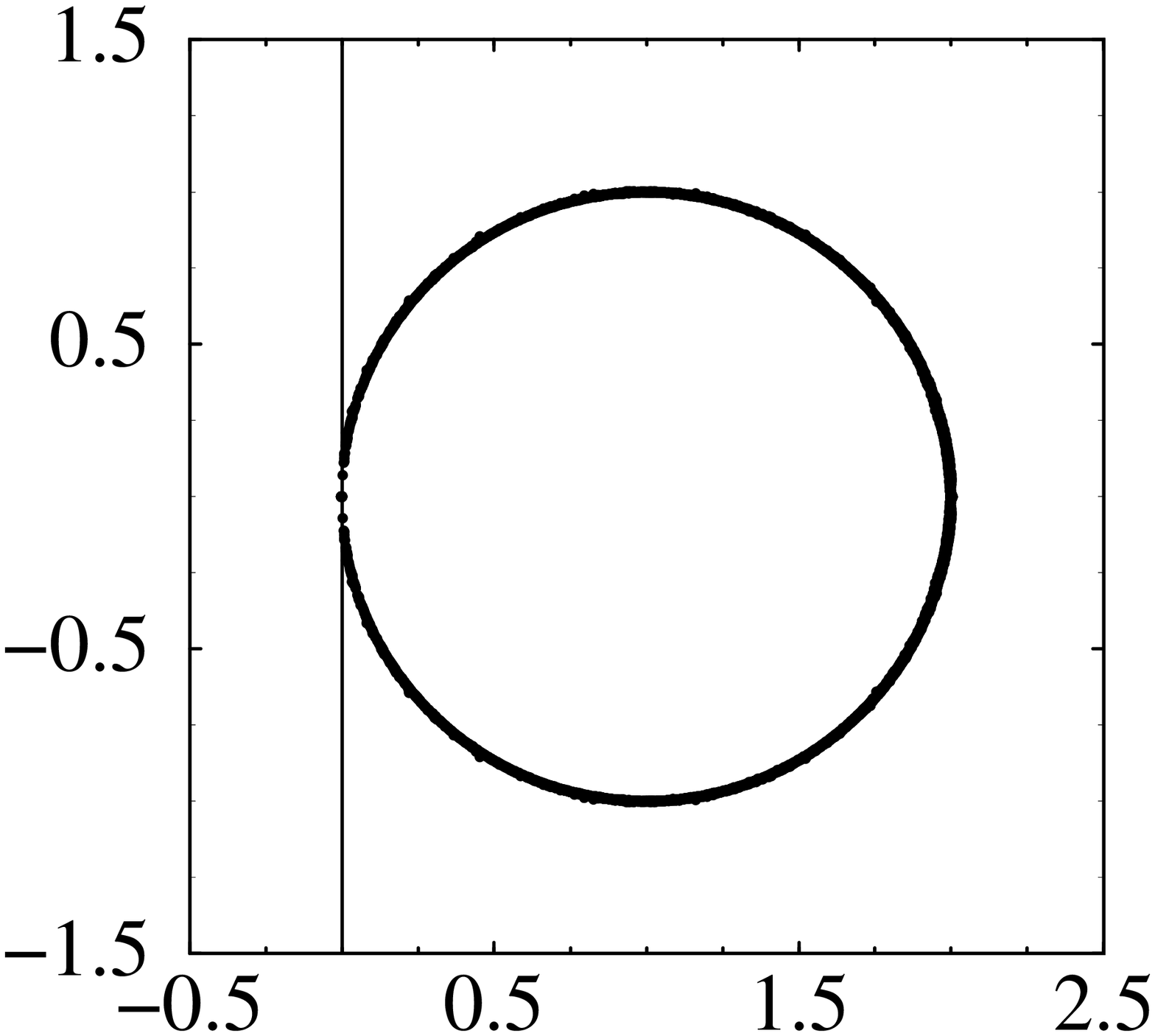,width=4.2cm,angle=-90}\\
\end{center}
\caption{\label{fig:spe}
Eigenvalues of the parametrized FP Dirac operator 
at the values of $\beta=2, 4, 6$ (from left to right), 
on $16^2$-lattices and sampled over 25 gauge configurations each.}
\end{figure}

Fig. \ref{fig:spe} shows the spectrum of the studied FP Dirac operator
collectively for a subset of 25 uncorrelated gauge configurations in
thermal equilibrium. While Neuberger's Dirac operator has (within
computational precision,  in our case 14 digits) exactly circular
spectrum, in the case of the FP action the effect of the truncation in 
coupling space due to the ultra-local parametrization is displayed  by
an increased fuzziness for lower $\beta$. 

The spectrum of the Dirac operator is gauge invariant and so it can be
expanded in series of loop operators. At lowest order in the lattice
spacing only few of them are independent, but when considering higher
order corrections operators of larger (geometric) extent have to be
introduced in order to have a complete set. So it may  be argued that
the truncation of the couplings to a $7\times 7$ lattice  for the
present parametrization of the FP action implies  an error in the form
of an operator of some (high) dimension:
\be
\lambda_{\rm Fp}(U)-\lambda_{\rm par}(U) = O^{(k)}(U)\QA 
k={\rm dim}[O^{(k)}]\;\;.
\ee
In order to estimate quantitatively the scaling behavior  of the
deviations of the spectrum from the ideal circular shape,  we defined a
mean deviation $|\lambda-1|$ from the unit circle  in an angular
window  of $|\mbox{arg}(1-\lambda)|<\pi/4$.  The displayed behavior
with $\beta$ of the average width (standard deviation)  $\sigma$ of
that distribution is $\sigma$ $\propto 1/\beta^{2.41}\simeq a^5$, thus
suggesting  a dimension-5 operator \cite{FaLaWo98}.

\begin{figure}[t]
\begin{center}
\epsfig{file=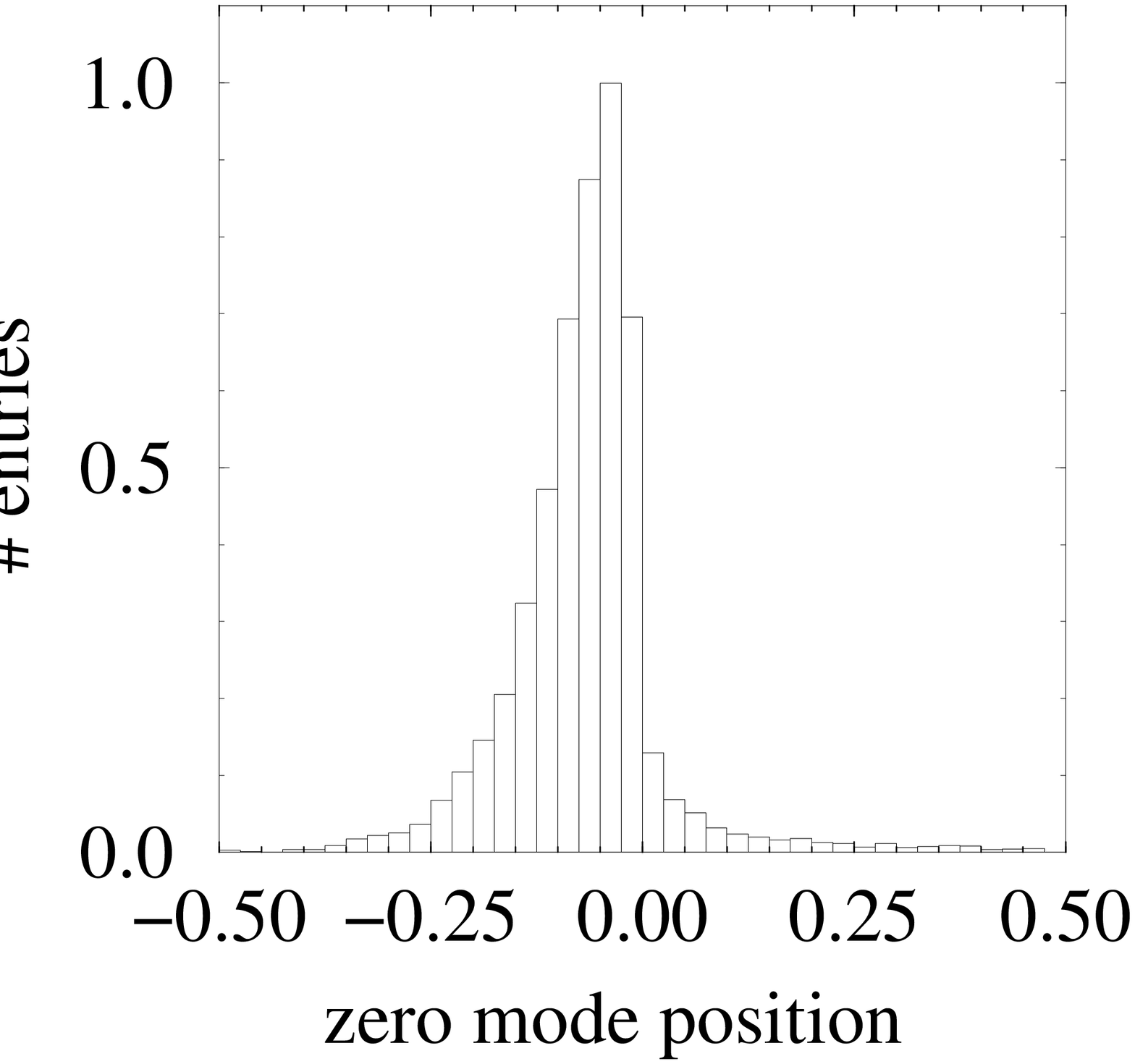,width=4.2 truecm, angle=-90}
\epsfig{file=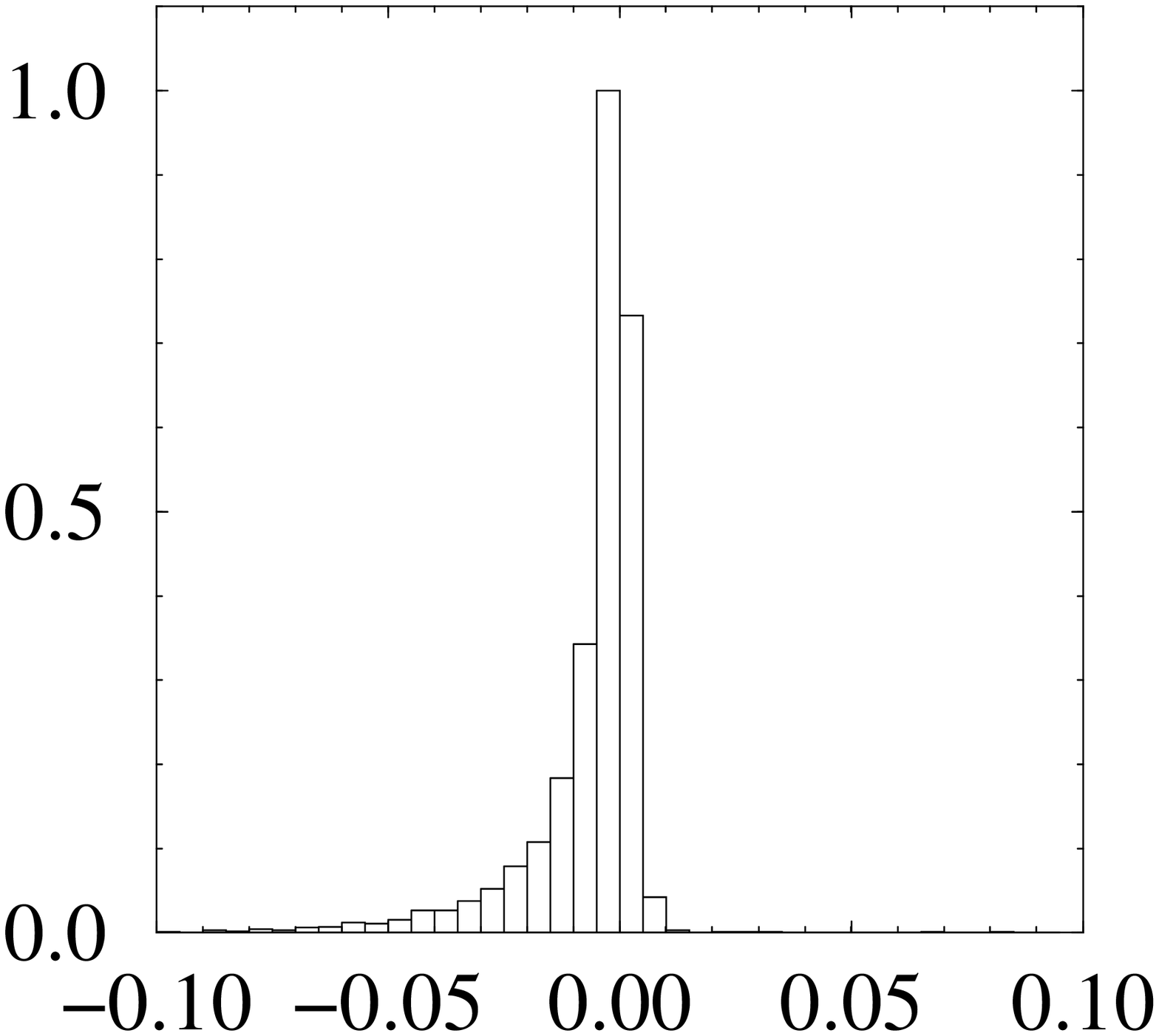,width=4.2 truecm, angle=-90}
\epsfig{file=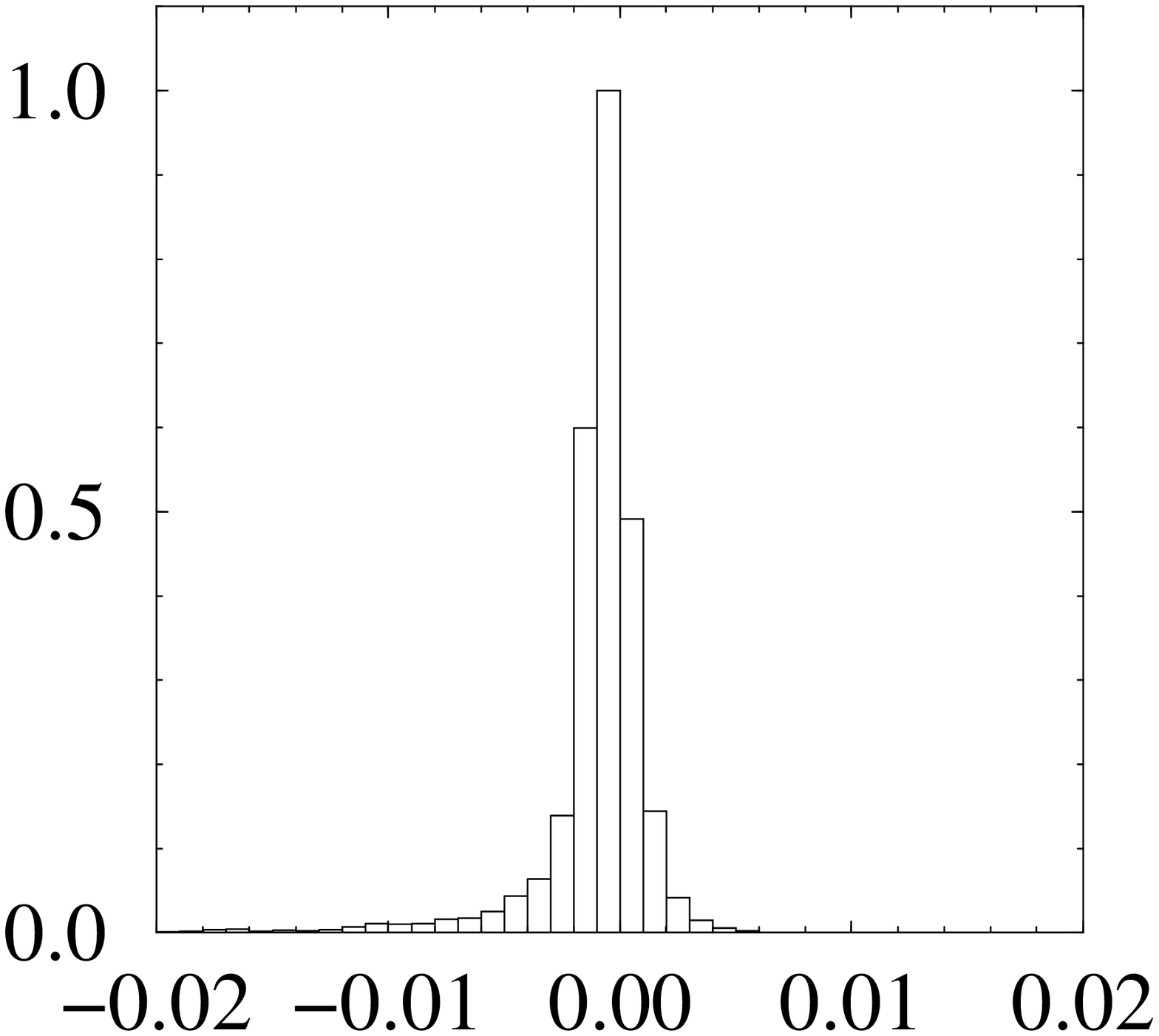,width=4.2 truecm, angle=-90}
\end{center}
\caption{\label{fig:poszm}
Distribution of the (real) eigenvalues for the quasi zero modes for the
FP action on a $16^2$ lattice; from left to right: $\beta=$2, 4, 6
(observe the different scales  of the abscissa). For convenience the
entries have been normalized  to the maximum entry.} 
\end{figure}

Another effect observed for the FP action and  due to the truncation,
is the scattering of the zero modes on the real axis. The relative
spread is tiny at large values of $\beta$ but becomes relevant when
decreasing $\beta$. It damages the nice theoretical properties  of the
operator causing difficulties in the numerical approach (see the
following). From Fig. \ref{fig:poszm} we see that the spread is $\sim
0.002$ for $\beta=6$, $\sim 0.01$ for $\beta=4$,  while for $\beta=2$
it becomes substantial.

\begin{figure}[t]
\begin{center}
\epsfig{file=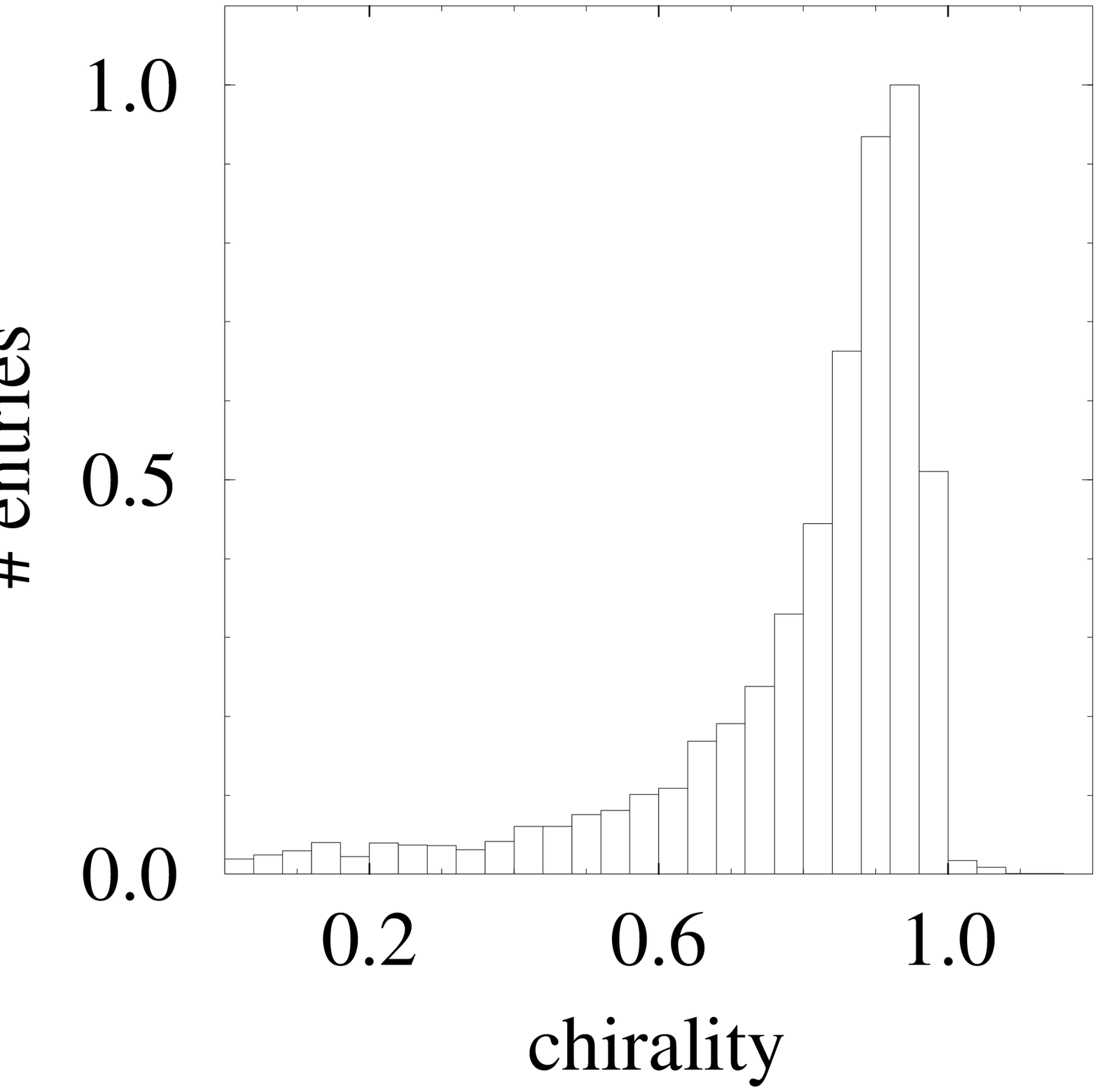,width=4.2 truecm, angle=-90}
\epsfig{file=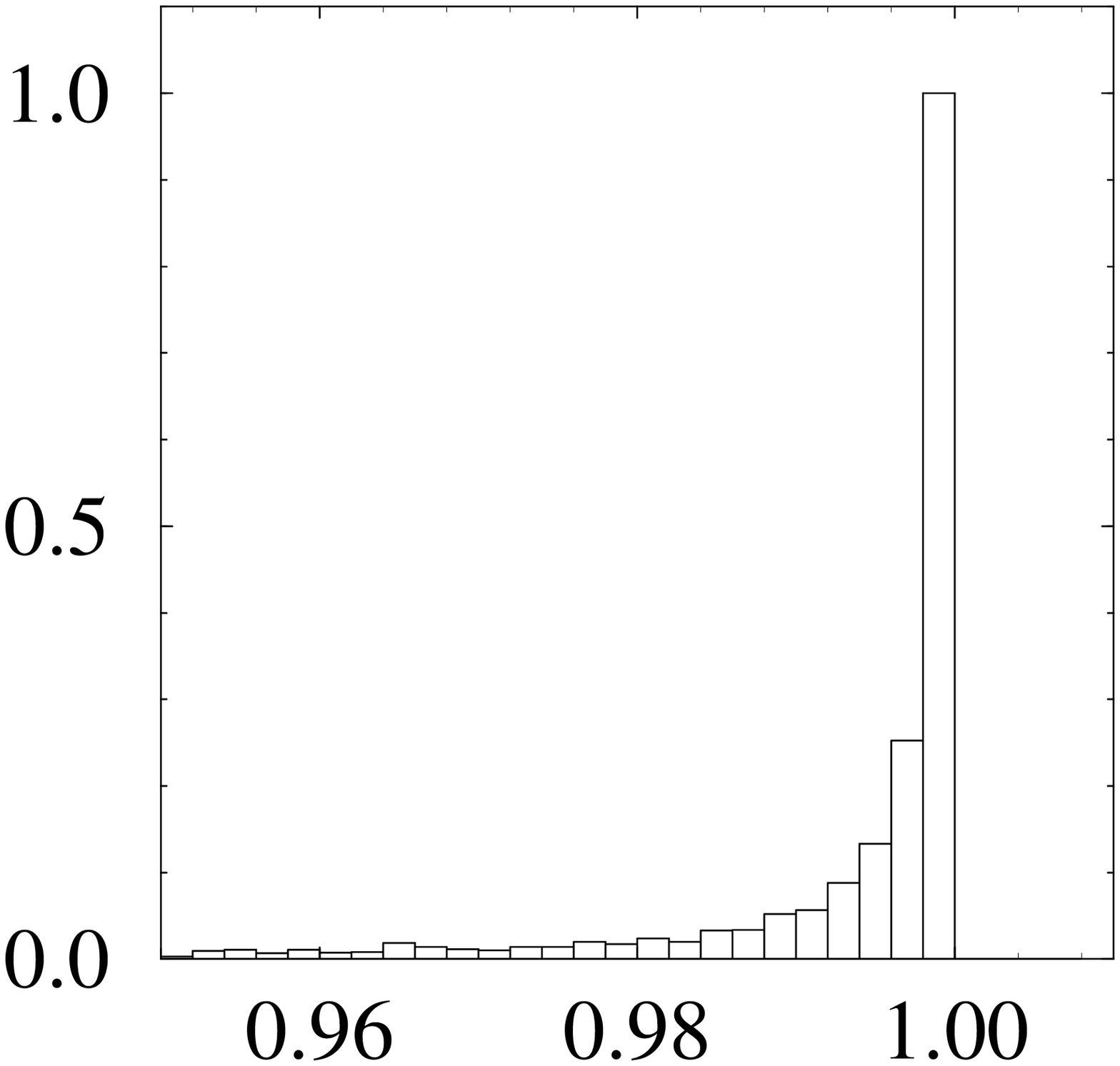,width=4.2 truecm, angle=-90}
\epsfig{file=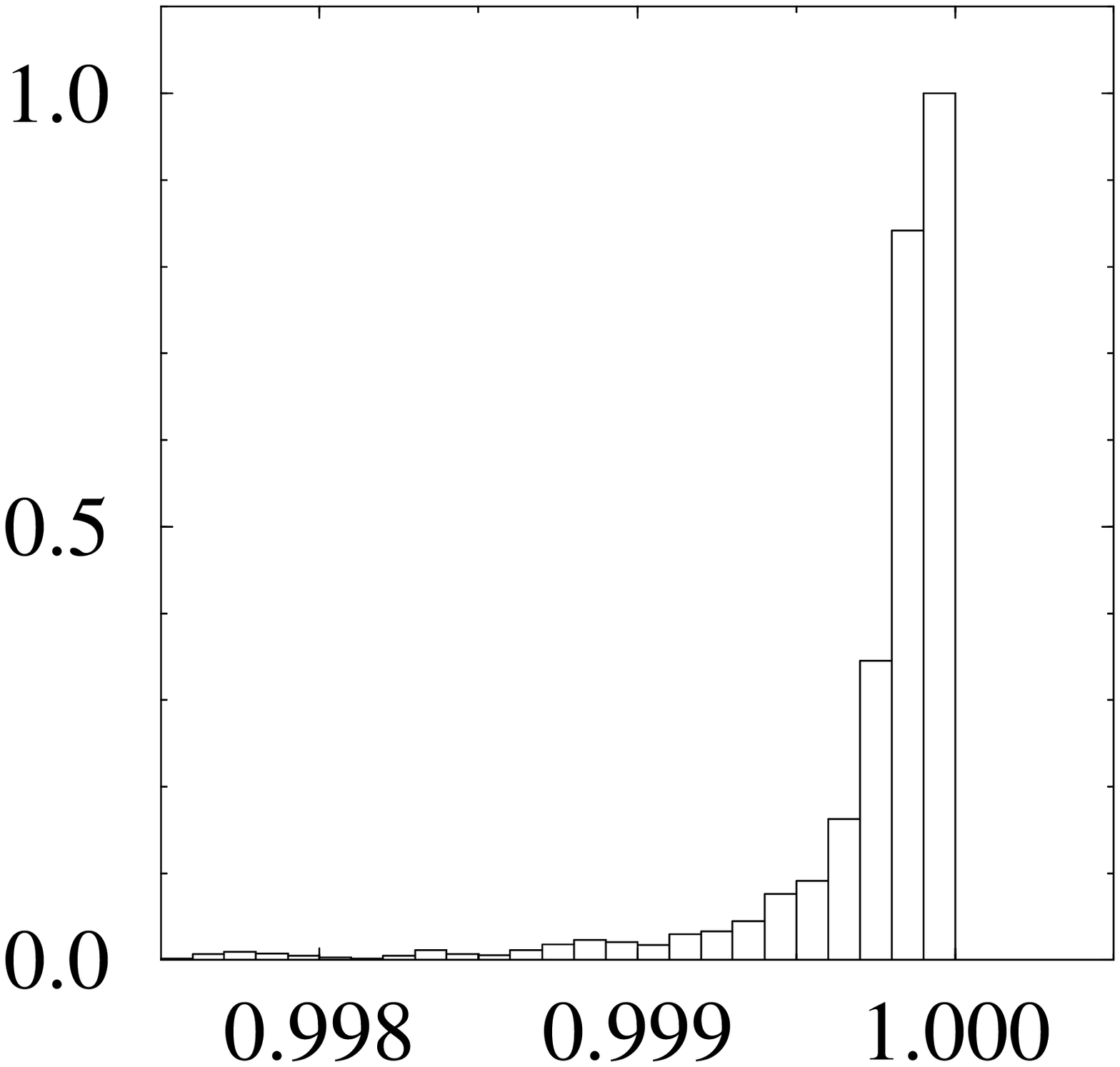,width=4.2 truecm, angle=-90}
\end{center}
\caption{\label{fig:chi}
Distribution of the chirality $(\overline{v}_\lambda,
\gamma_5\,v_\lambda)$ of the quasi zero modes of the  Dirac operator
for the FP  action on a $16^2$ lattice; from left to right: $\beta=2$,
4, 6 (observe the different scales of the abscissa). We show only the 
positive chirality contribution.  For convenience the entries have been
normalized  to the maximum entry.} 
\end{figure}

A similar effect can be observed in the distribution of the chirality
of the (almost) zero modes, displayed in Fig. \ref{fig:chi}. In this
case, the accuracy of realization of chiral symmetry  is directly
probed. For Neuberger's operator the chirality of the zero modes is
$\pm 1$ within numerical precision.

\begin{table}
\begin{center}
\begin{tabular}{crrr}
\hline
Op. & $p(\beta=2)$ & $p(\beta=4)$ & $p(\beta=6)$\\
\hline
\DOV &  74.22   &  99.70   & 100.00\\
\DFP &  96.58   & 100.00   & 100.00\\
\hline
\DOV & 100.00   & 100.00   & 100.00\\
\DFP &  91.00   &  99.84   &  99.96\\
\hline
\end{tabular}
\end{center}
\caption{\label{tabratioASIT} Upper part: percentage $p(\beta)$
of configurations where the index of the Dirac operator index$(U)$
agrees with the geometric charge.
Lower part: percentage $p(\beta)$
of configurations where the Vanishing Theorem is fulfilled.}
\end{table}

Table \ref{tabratioASIT} gives the results of the comparison between
the  index of the Dirac operator index$(U)$ (the modes are counted
according to the sign of their chiralities) and the geometric charge of
the gauge  configuration. Observe that in the case of the FP action the
agreement is excellent already at $\beta=2$, thus indicating a close
correspondence between the geometric definition and the FP topological
charge.  For this value of $\beta$ the deviation is comparatively large
for  Neuberger's operator, reflecting \cite{FaHiLa98} in this respect 
features of the Wilson operator entering (\ref{DOV}).

The lower part of the table demonstrates that for \DOV zero modes have
just one chirality here. For \DFP this situation is also rapidly
approached towards the continuum limit, as predicted by the Vanishing
Theorem.

\subsection{The fermion condensate in a finite volume}
\label{subs:res_cond}

The finite-volume fermion condensate is determined following the 
prescriptions of Section \ref{subs:cond}.  In order to check the
correctness of our determination we verify the Ward 
identity (see also \cite{Ch98a}), valid in the limit $\mu\to 0$
\begin{eqnarray}
\label{eq:ward}
-\lim_{\mu\to 0}\,\condensate_{\rm sub}(\mu)&= &
\lim_{\mu\to 0}\,\chi'(\mu)\\
\textrm{with}\quad
\chi'(\mu)&\equiv &\frac{1}{V\,\mu}\gaugeexp{(\textrm{index}(U))^2}(\mu)
\;.
\end{eqnarray}

\begin{figure}[t]
\begin{center}
\epsfig{file=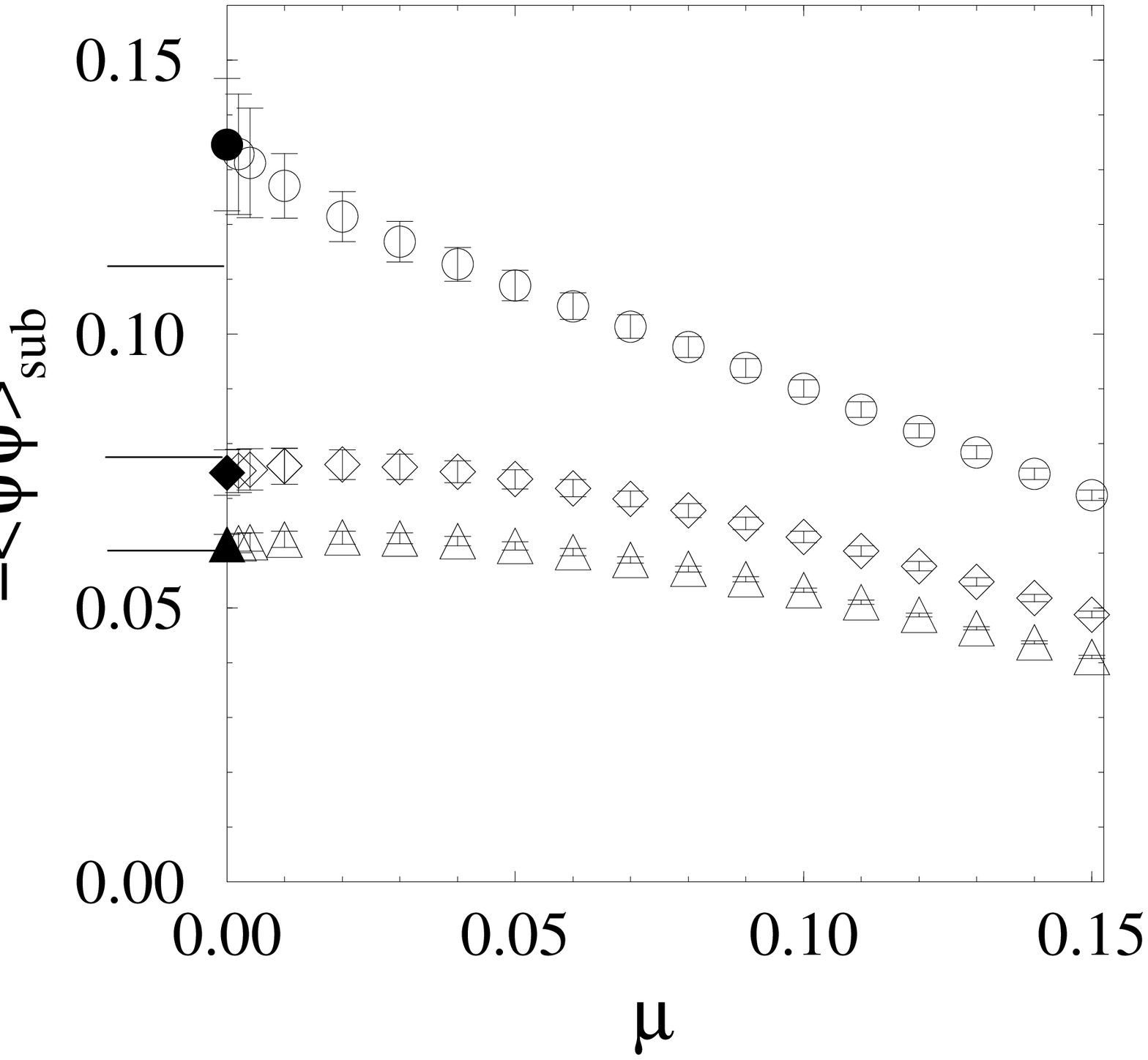,width=6. truecm, angle=-90}
\epsfig{file=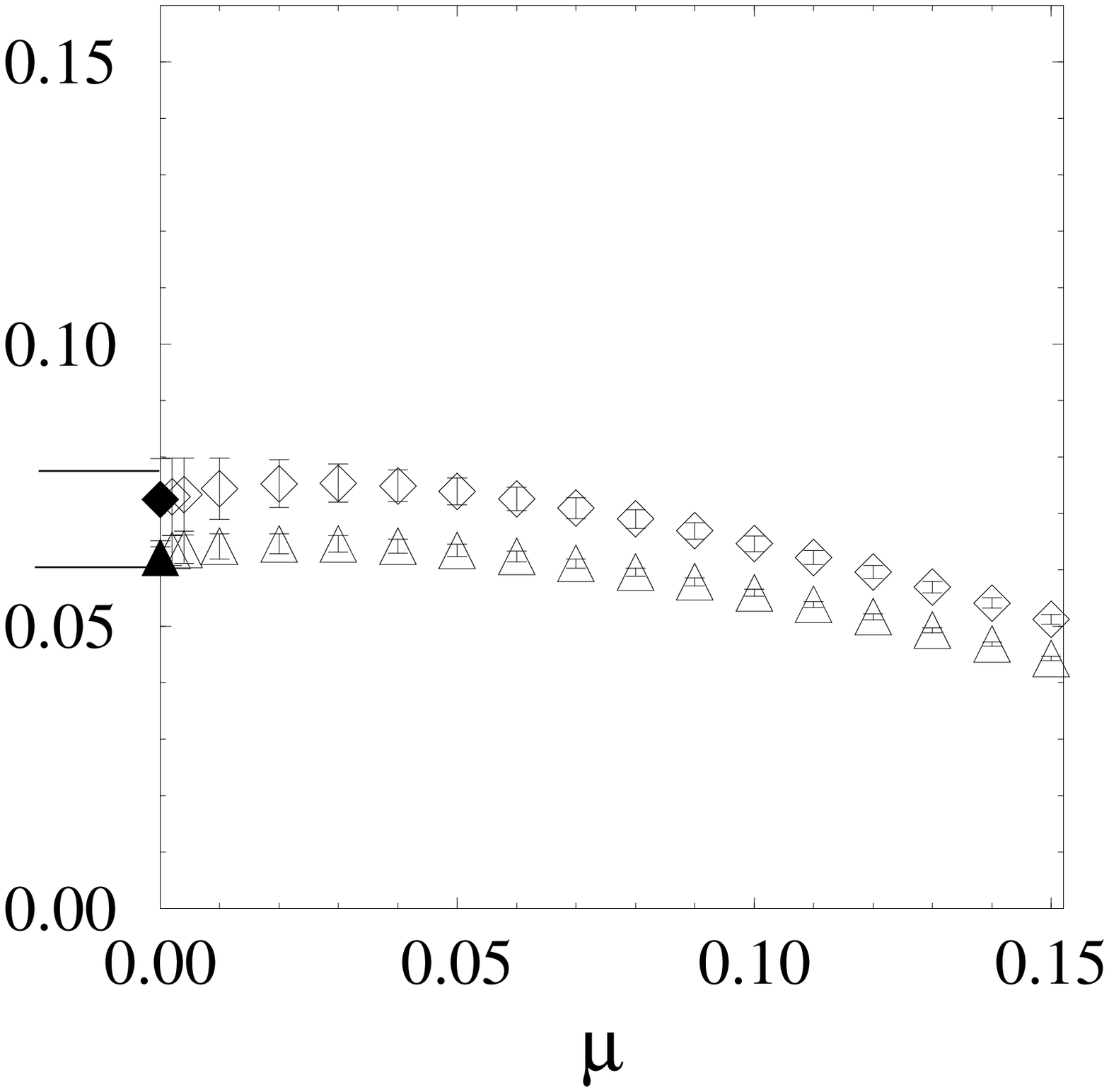,width=6. truecm, angle=-90}
\end{center}
\caption{\label{fig:cond_fv}
The subtracted chiral condensate as a function of $\mu$  for
Neuberger's (left) and FP (right) operator on  a $16^2$ lattice:
$\beta=2$ (circles),  $\beta=4$ (diamonds), $\beta=6$ (triangles); the
full symbols denote the zero-mass estimate obtained through  a linear
extrapolation. For $\beta \ge 4$ they agree within the error bars with
the finite volume numbers (indicated by the pointers on the left-hand
side) from theory \cite{SaWi92}.}
\end{figure}

\begin{table}
\begin{center}
\begin{tabular}{ccclll}
\hline
Op. &$L$& $\beta$ & $-\subcond$ & $\chi'$ & Cont.\\
\hline
\DOV & 16 & 2 & 0.135(12)     &  0.135(18)   &  0.1127 \\
\DFP & 16 & 4 & 0.0725(73)    &              &  0.0777 \\
\DOV & 16 & 4 & 0.0747(41)    &  0.0747(41)  &  0.0777 \\
\DFP & 24 & 4 & 0.104(38)     &              &  0.0798 \\
\DFP & 16 & 6 & 0.0622(19)    &              &  0.0604 \\
\DOV & 16 & 6 & 0.0617(18)    &  0.0617(18)  &  0.0604 \\
\DFP & 24 & 6 & 0.0621(36)    &              &  0.0650 \\
\hline
\end{tabular}
\end{center}
\caption{\label{tab:cond}
Subtracted condensate $\subcond$  and $\chi'$ for the FP and
Neuberger's action  (all obtained in the limit $\mu\to 0$). The last
column  indicates the continuum value in the corresponding 
physical volume \cite{SaWi92}.}
\end{table}

In Table \ref{tab:cond} we compare the values for both sides  of
(\ref{eq:ward}) (extrapolated at $\mu=0$): excellent agreement between
the two quantities is displayed by data for Neuberger's action.  In the
case of the FP action,  $\chi'(\mu)$ is unstable for $\mu\to 0$: this
is simply explained by the spread of the zero modes which prevents the
cancellation between the denominator factor $\mu$ with the lowest
eigenvalue in the fermion determinant (which is $\mu$ for an exact zero
mode).

The subtracted condensate for the (approximate) FP action is plagued by
similar problems. For $\beta=4$ and $L=24$ the determination of the
condensate has large error bars and for $\beta=2$ and $L=16$ it is not
possible at all due to the instability of the $\mu\to 0$ limit. Except
for these two cases the relative accuracy of our results is between 3
and 6 \%.  In Table \ref{tab:cond} we also compare our data with the
continuum value \cite{SaWi92} in the corresponding physical volume,
assuming the asymptotic relation (\ref{LattScalePar}); the discrepancy
is always smaller than one standard deviation.

In Fig. \ref{fig:cond_fv} we report $\subcond(\mu)$ for a range  of
$\mu$ values in the case of Neuberger (left) and FP (right) action; the
expected linear asymptotic behavior allows for the extrapolation at
$\mu=0$.

\section{Statistical properties of the spectrum}
\label{sec:res_edge}

GW actions realize chiral symmetry  on the lattice. This realization
is implicit, in the sense that the Dirac operator breaks the chiral
symmetry, as usually defined, by a local term ${\cal O}(a)$.  We 
make the symmetry manifest and construct a fermion matrix 
$\tilde{\D}$ which anti-commutes with $\gamma^5$. This is achieved
through the transformation proposed  in (\ref{directdef}). In this
section we study the microscopic fluctuations  of the spectrum of
this operator.

\subsection{The trivial sector: $N_f=0$ and $\nu =0$}
\label{subs:trivial}

Since chRMT gives specific predictions for given topological number
$\nu$ and number of flavors $N_f$, it allows to disentangle the
statistical properties of the spectrum of the lattice Dirac operators
from the more subtle questions of the correct reproduction of the
dynamics of fermions and of the identification of the topological
charge with the index (see the discussion on this latter  point in
Section \ref{subs:edge}).  Consequently, we start our investigation in
the simplest case $N_f=0$ and $\nu =0$.

We expect that a gauge theory of the type studied here -- without
further symmetries -- should lead to a spectral distribution in the
universality class of chUE \cite{Ve94}. This is to be checked with the
data.

Comparison of the statistics of the Dirac spectrum distribution with
chRMT requires the knowledge of the inherent scale parameter $\Sigma$
that is related to the dynamical  properties of the system. 
Technically $\Sigma$ may be determined by comparing spectral data  with
chRMT predictions for a particular statistics.  The consistency of the
determination can be then checked by using the obtained value of
$\Sigma$  as an input parameter for further comparisons with chRMT.

\paragraph{Smallest eigenvalue.}

A simple way to determine $\Sigma$ is to look at the probability
distribution of the smallest eigenvalue $P(\lambda_{\rm min})$. This
statistics contains spectral information on a region  of small
enough values of $\lambda$ not to be  affected by the macroscopic,
non-universal (i.e. {\em not} described by chRMT) component of the
spectrum. Of course the macroscopic component can be easily eliminated
from lattice data through an unfolding procedure,  which -- at this
level -- would introduce unnecessary arbitrariness, however.

\begin{figure}[t]
\begin{center}
\epsfig{file=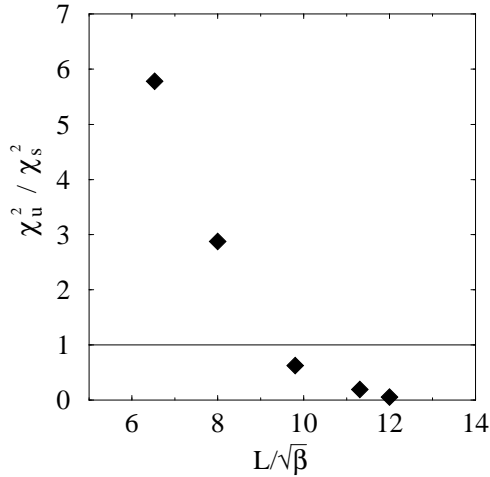,width=6.5 truecm, angle=-90}
\end{center}
\caption{\label{fig:chi_fit}
The ratio between the $\chi^2_u$ ($\chi^2$ with chUE) and  
$\chi^2_s$  ($\chi^2$ with chSE) in the case of the FP action as a
function of the physical lattice size $L/\sqrt{\beta}$.}
\end{figure}

As discussed, we expect results lying in the universality class of chUE.
However, in order to decide on the applicability of chRMT in an unbiased 
approach, we tested the predictions \cite{Fo93} for the distribution of the
smallest  eigenvalue from all three variants of chRMT through a standard
best-fit procedure on our lattice data. This test clearly ruled out the chOE, 
which has a completely different distribution shape. The two
remaining ensembles chUE and chSE seemed to fit data in disconnected --
complementary -- regions of the lattice parameter space $(L,\beta)$. 

\begin{figure}[tp]
\begin{center}
\epsfig{file=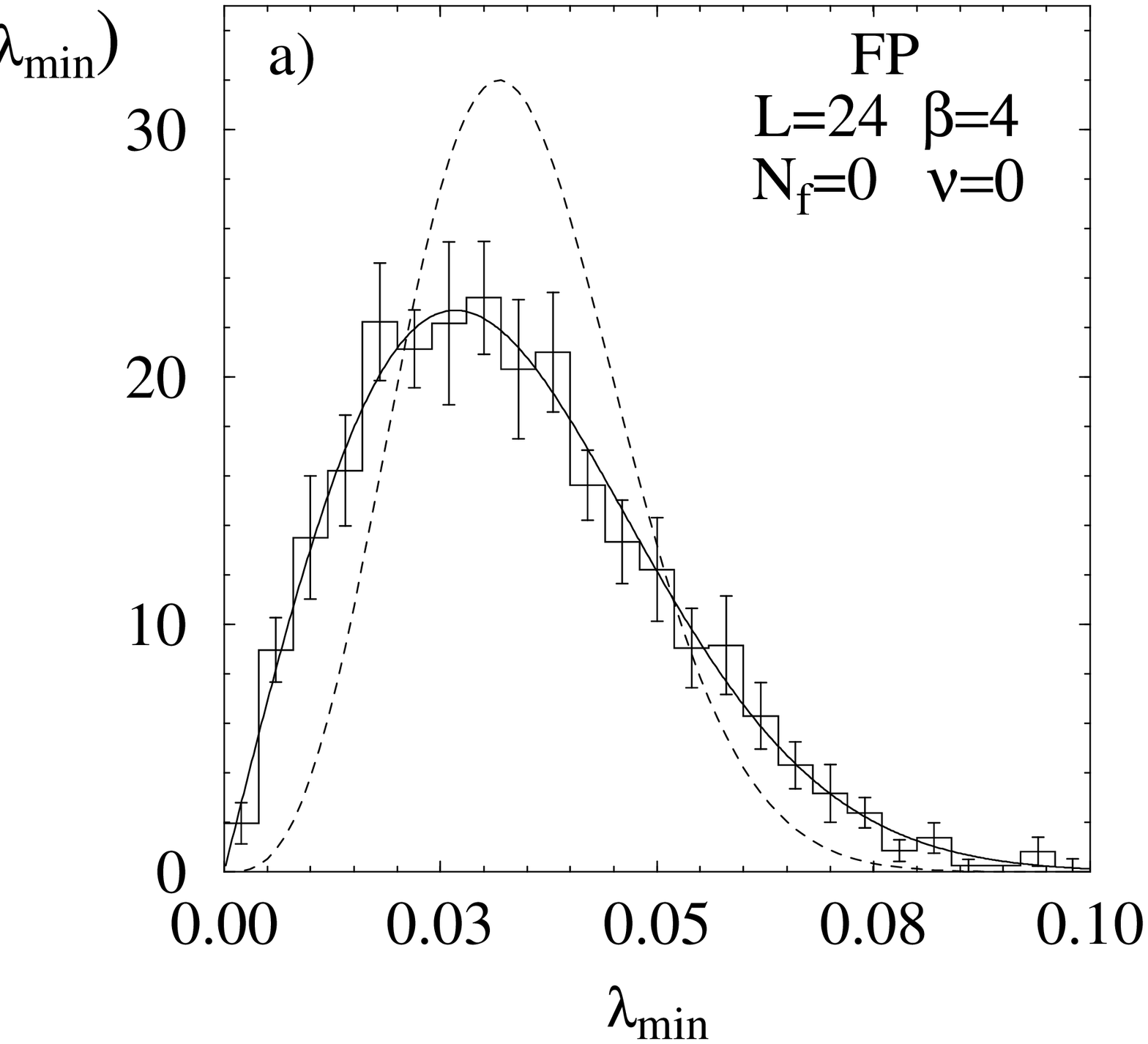,width=6 truecm, angle=-90}
\epsfig{file=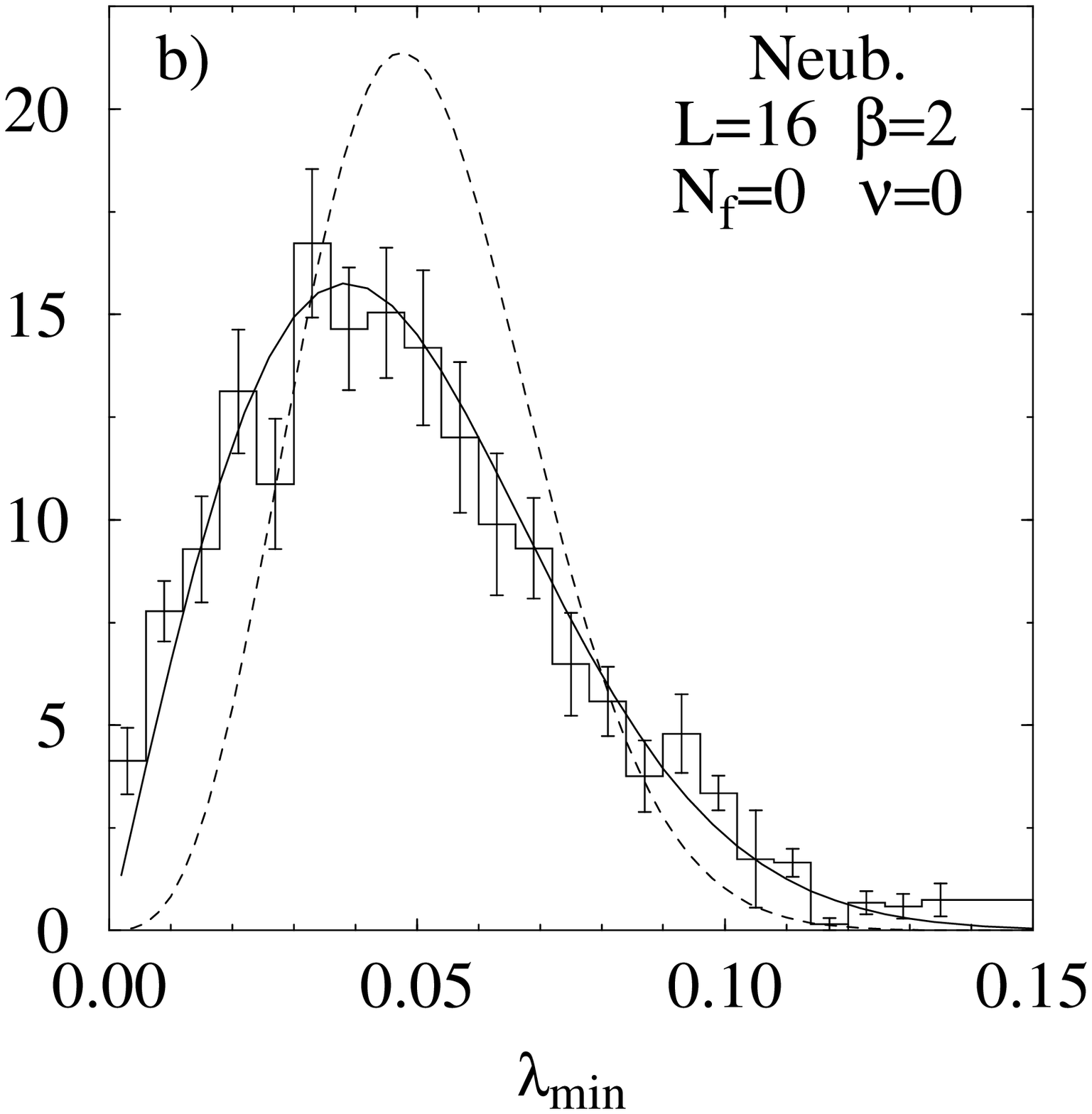,width=6 truecm, angle=-90}
\epsfig{file=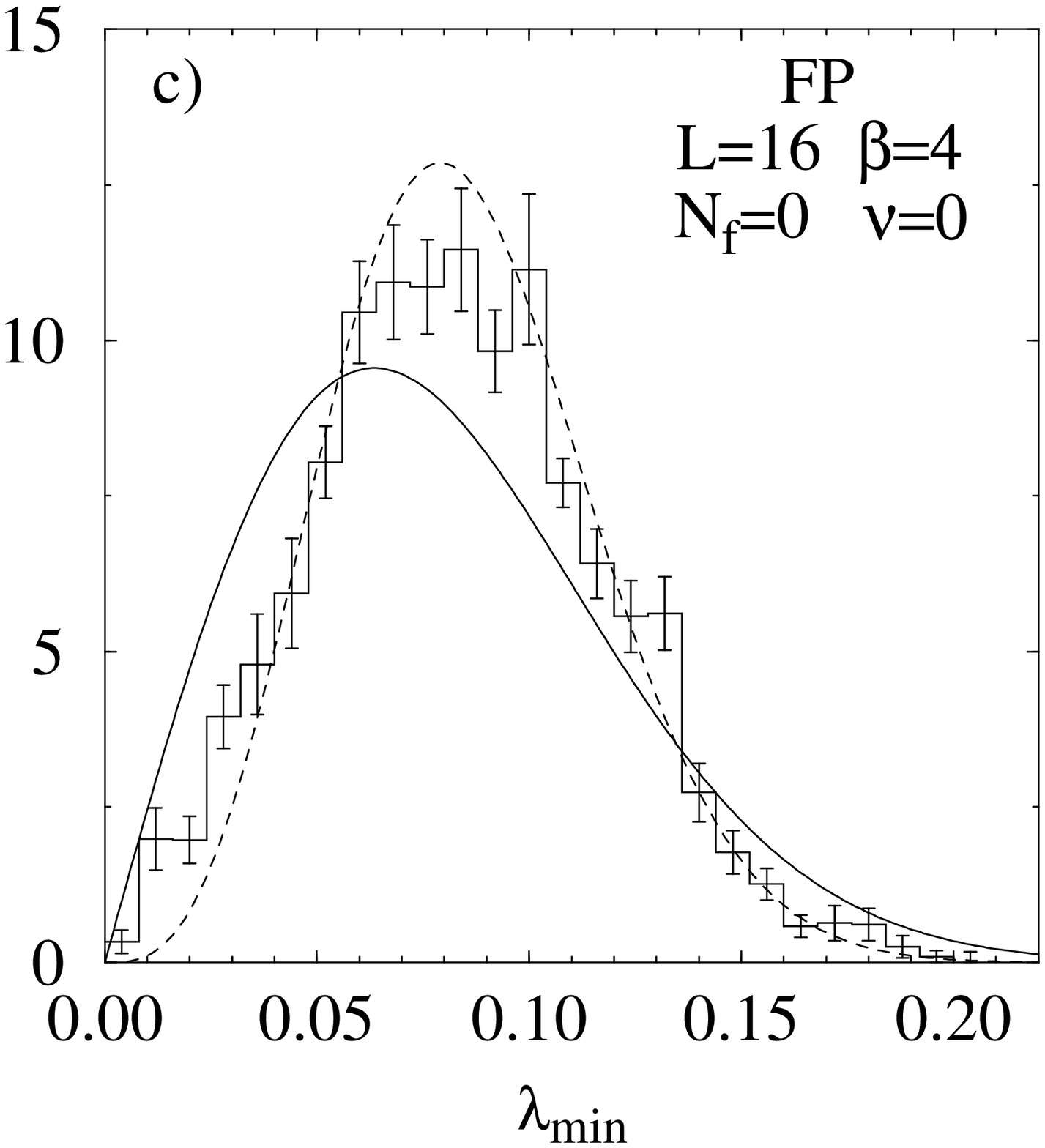,width=6 truecm, angle=-90}
\epsfig{file=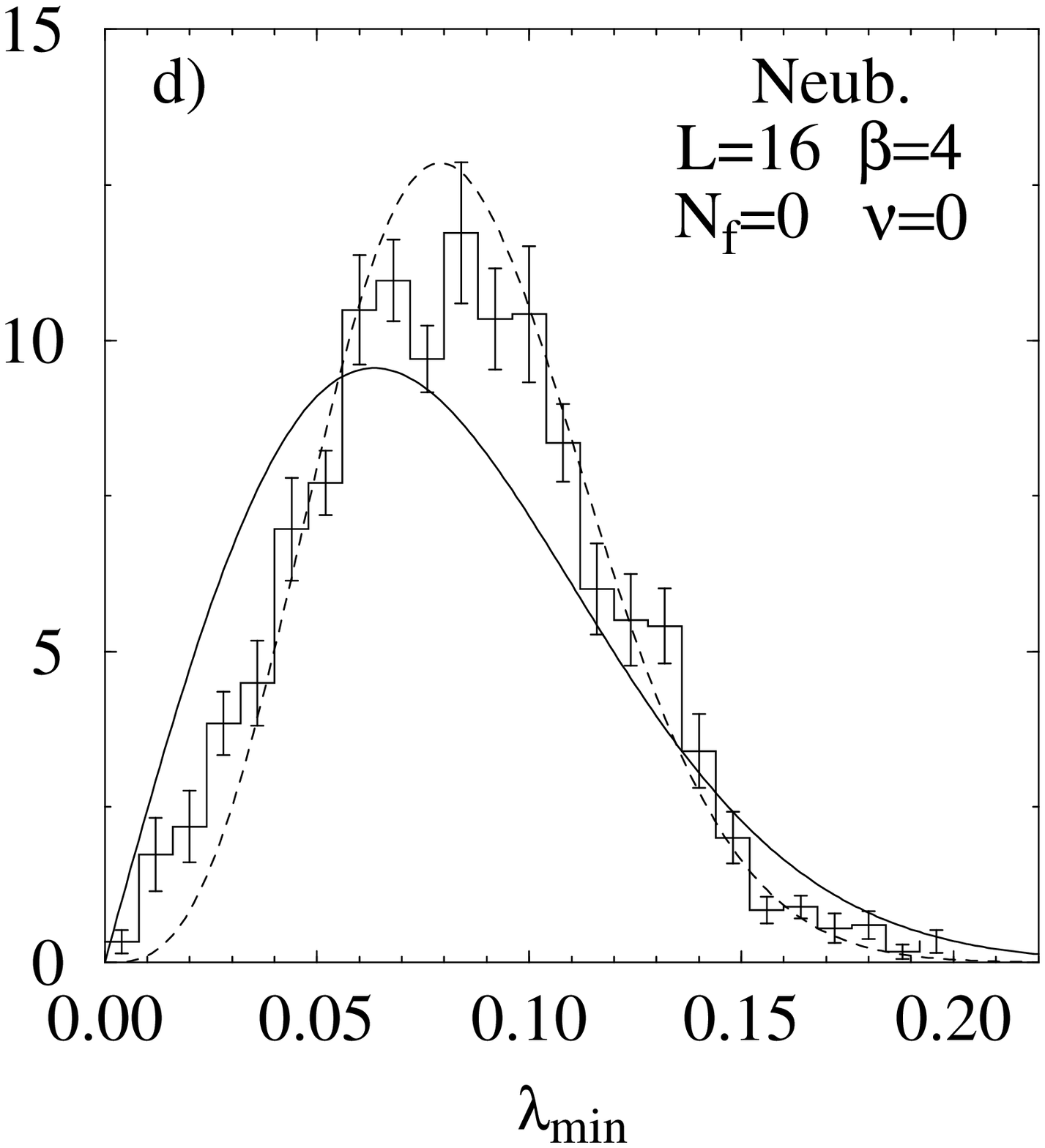,width=6 truecm, angle=-90}
\end{center}
\caption{\label{fig:min}
Distribution of the smallest eigenvalue $P(\lambda_{\rm min})$  in the
$\nu=0$ topological sector for different lattices and values of
$\beta$, for the FP  and Neuberger's Dirac operator. The full and
dashed curves represent the theoretical  predictions of chUE and
chSE, respectively.} \end{figure}

In Fig.  \ref{fig:chi_fit} we plot the ratio of $\chi^2$-values
obtained for the fits to the two distributions as a function of the
{\em physical} size. We find that in the region of large  physical
volumes chUE is preferable (the ratio is smaller than unity),  while
in the region  of small physical volumes chSE provides better fits to
the data.  The `transition' seems to take place at
$L/\sqrt{\beta}\simeq 9$. In Fig. \ref{fig:min} we compare our lattice
results for $P(\lambda_{\rm min})$ with the best fit prediction from
chUE and chSE for two cases in the large-volume region (upper part)
and in the small-volume one (lower part). In Table \ref{tab:rec_cond}
we give the resulting values of $\Sigma$. Within chRMT this value is
intrinsically volume-independent. The obtained  value of $\Sigma$
is taken as an input parameter for the subsequent investigation.

For small physical volumes we find chSE distribution shapes; however,
this result is in conflict with other properties of  such ensembles.
For chSE the eigenvalues should come in degenerate pairs, which is not
true in our case\footnote{We thank T. Wettig for pointing this
out to us.}. 

Such a volume dependence of our results does not come unexpected
and is related to the limitations of applicability of RMT. Actually, in
QCD one expects RMT to be applicable \cite{LeSm92,ShVe93} for
\be
\frac{1}{m_{QCD}} \ll a\,L \ll \frac{1}{m_\pi}
\ee
which, in our case with $N_f=1$  reduces to 
\be\label{InvSM}
\frac{1}{a\,m_{Schwinger}} \ll L\;. 
\ee
From the continuum results we expect on the lattice (\ref{SchwingerMass}), and
then (\ref{InvSM}) reads $\sqrt{\pi}\ll L/\sqrt{\beta}$, to be compared with the
discussed observations. We suspect that the observed behavior at small volume
is outside the scope of RMT and that one may be misled by the superficial
agreement with chSE. We return to a discussion in the last section.

\paragraph{Microscopic spectral density.}

\begin{figure}[tp]
\begin{center}
\epsfig{file=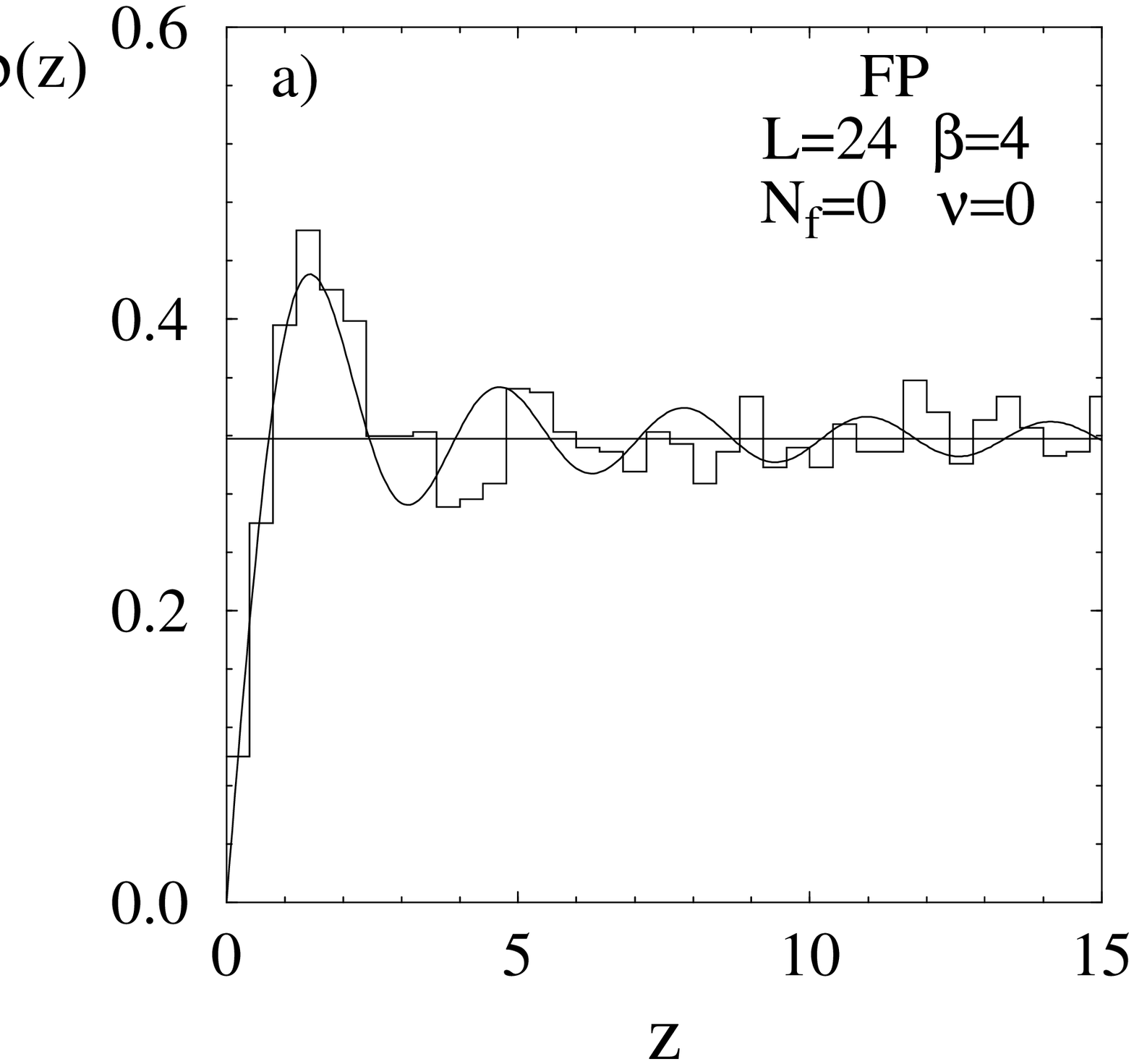 ,width=6 truecm, angle=-90}
\epsfig{file=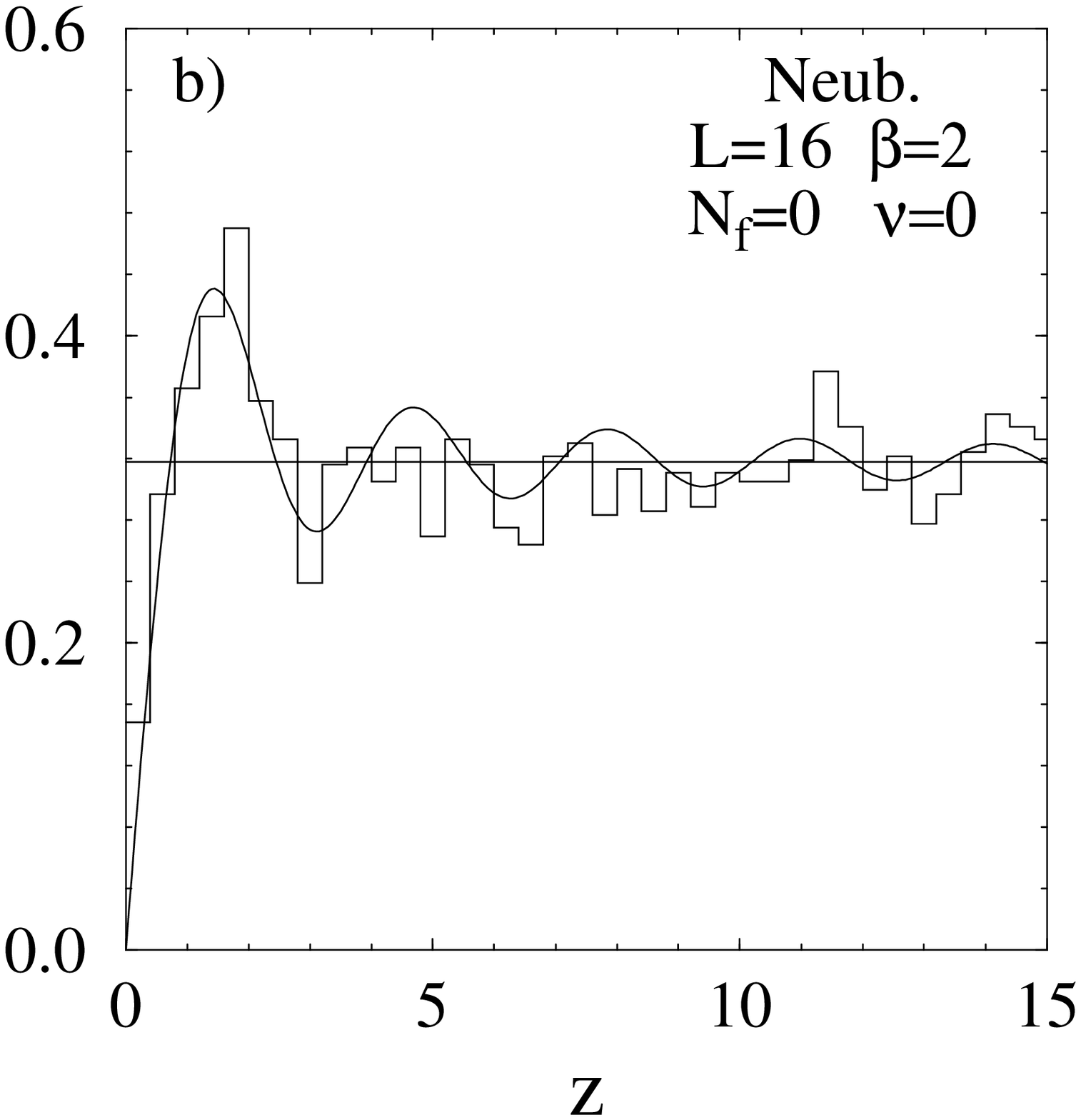,width=6 truecm, angle=-90}
\epsfig{file=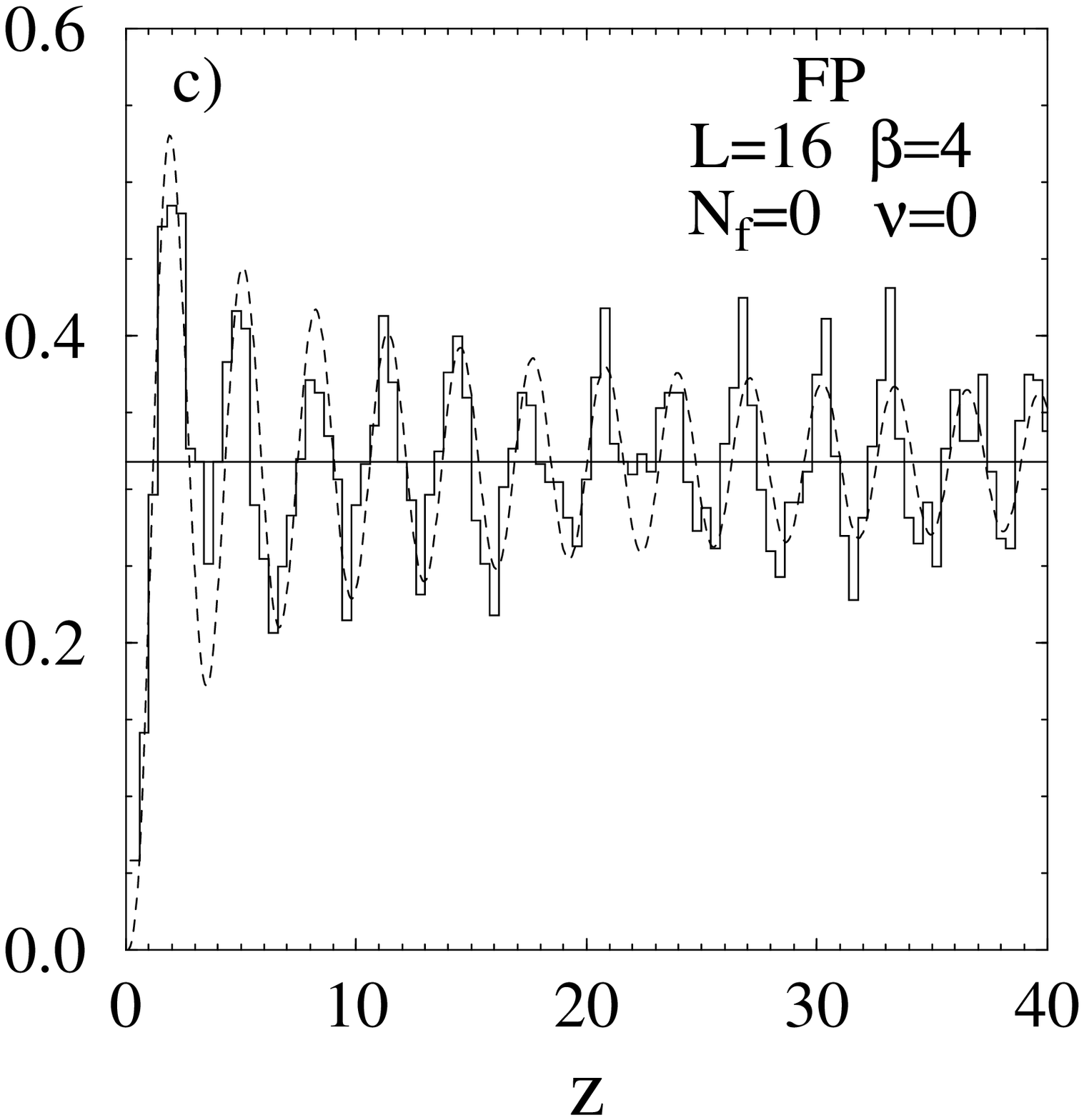,width=6 truecm, angle=-90}
\epsfig{file=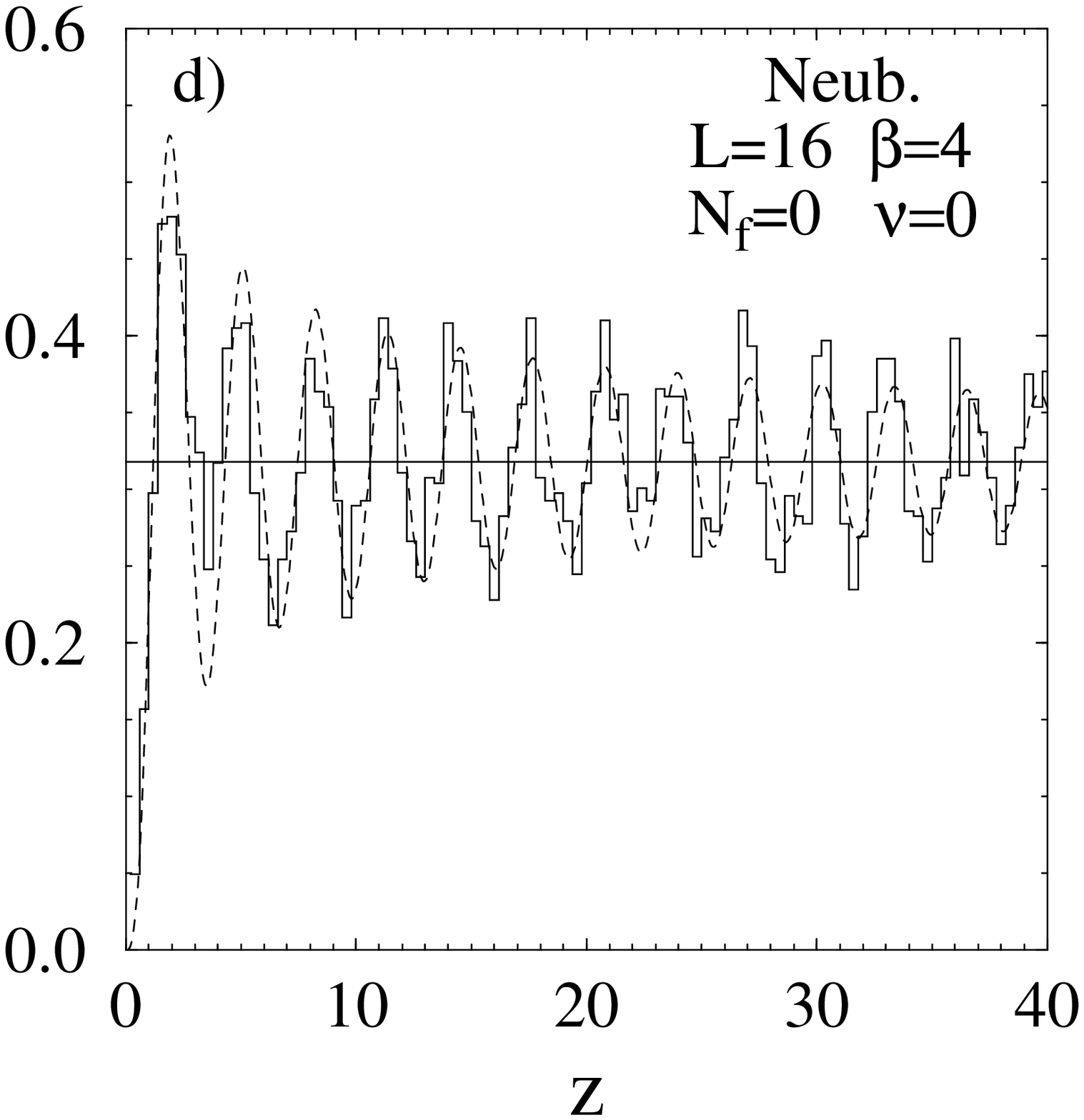,width=6 truecm, angle=-90}
\end{center}
\caption{\label{fig:dist}
Microscopic spectral density $\rho_{\rm s}(z)$
in the $\nu=0$ topological sector for
different lattices and values of $\beta$, for the FP  and Neuberger's
Dirac operator. The full and dashed curves represent the theoretical 
predictions of chUE and chSE, respectively.}
\end{figure}

Using $\Sigma$ as input parameter, as discussed,  we checked the
universal behavior of the microscopic spectral density $\rho_s(z)$. In
this case the macroscopic unfolding of the lattice  data was necessary
(in this we followed essentially the suggestions  in \cite{MaGuWe98}).
We report the results in Fig. \ref{fig:dist}, again comparing with the chUE
\cite{VeZa93} and chSE \cite{NaFo95,MaGuWe98} predictions.  We see a
(universal) behavior compatible -- within the accuracy of our data --
with that found in the case of the distribution of the smallest
eigenvalue, namely chUE in the  large-volume region and chSE in the
small-volume region.

\begin{figure}[tp]
\begin{center}
\epsfig{file=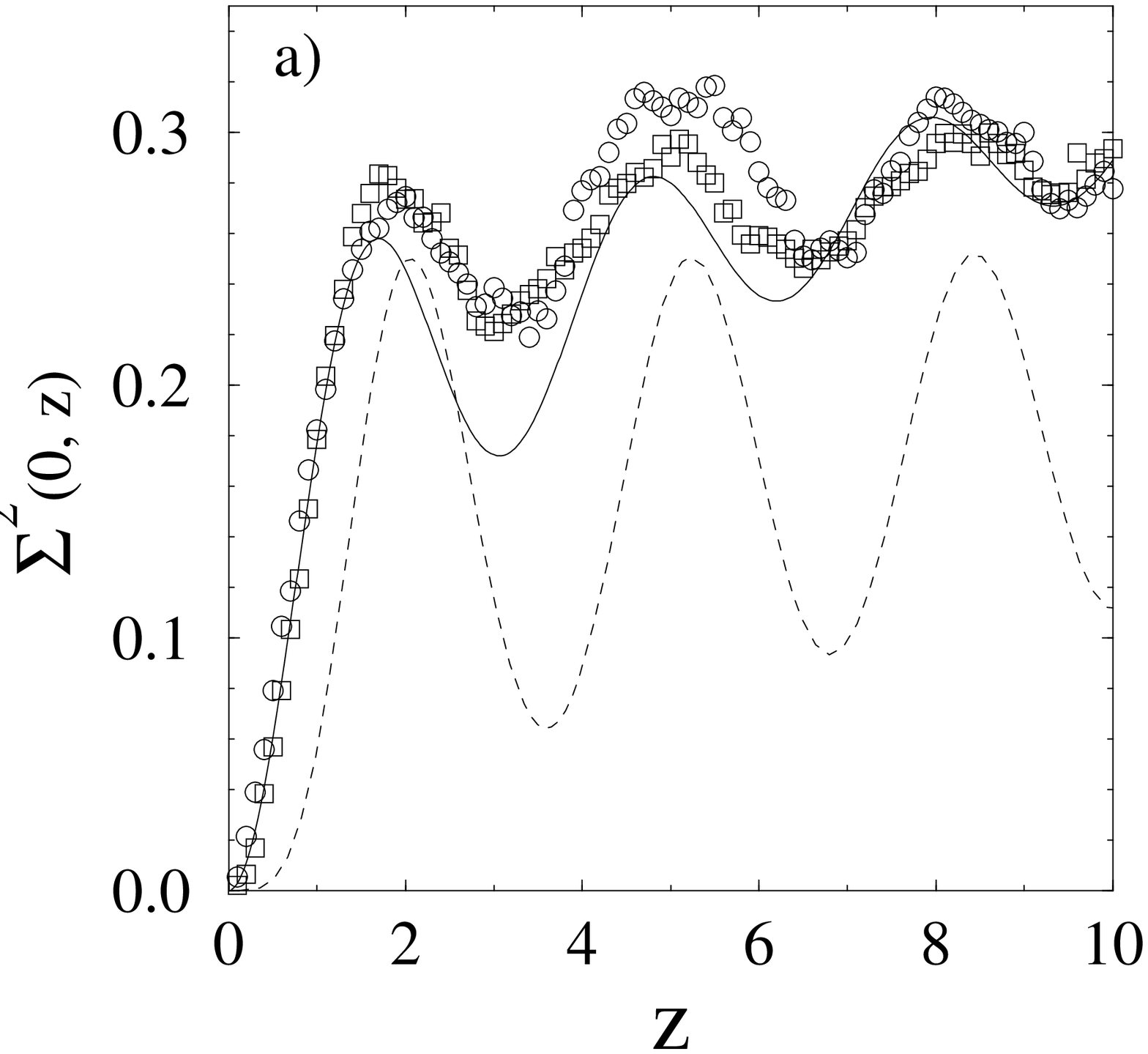,width=6.5 truecm, angle=-90}
\epsfig{file=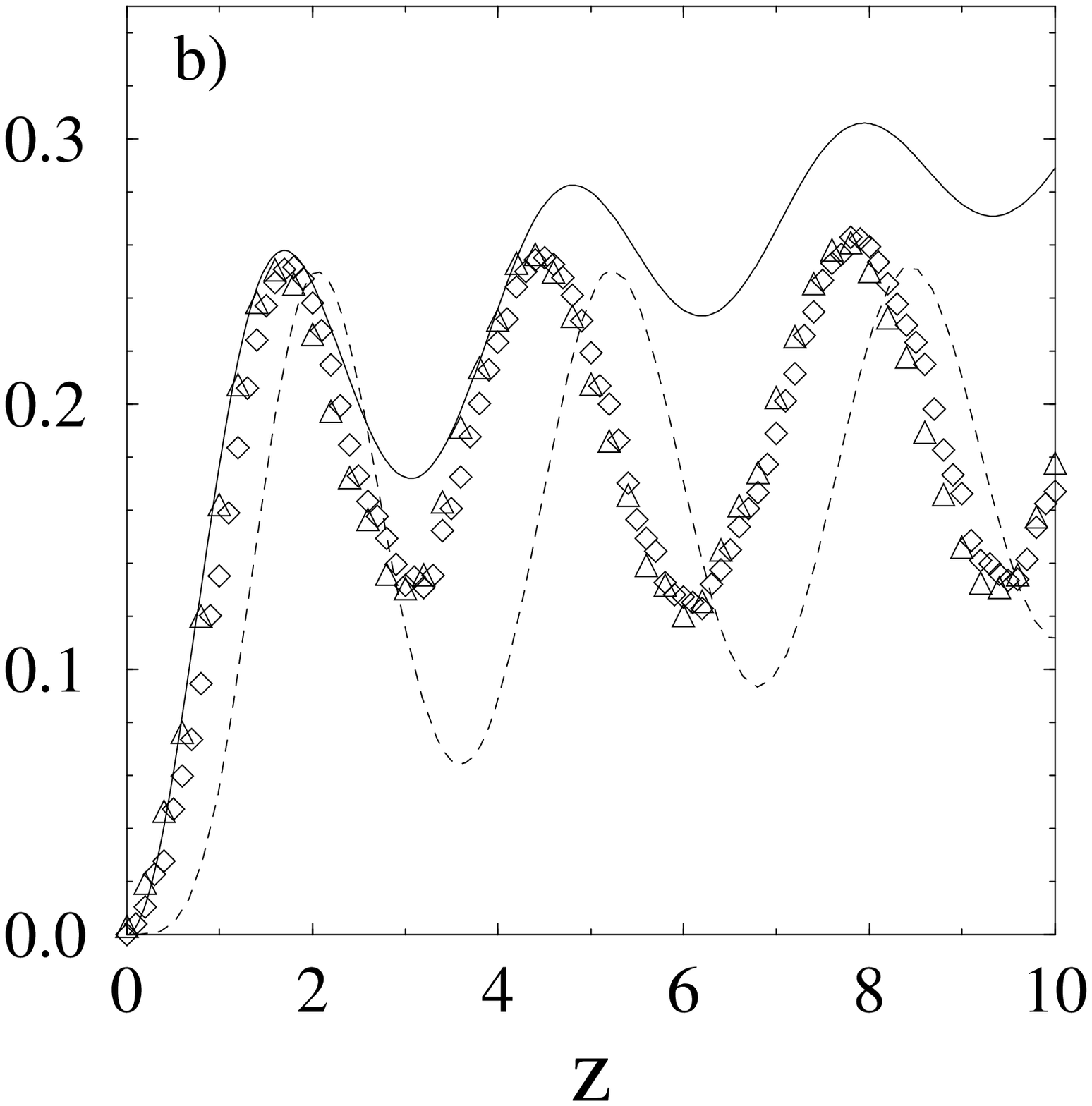,width=6.5 truecm, angle=-90}
\end{center}
\caption{\label{fig:nvar}
Number variance as a function of the unfolded variable $z$:  (a) large
volume region, $L=16$, $\beta=2$ and Neuberger's action (circles),
$L=24$, $\beta=4$ and FP action (boxes);   (b) small volume region,
$L=16$, $\beta=4$ for Neuberger's (triangles) and the FP action
(diamonds). The full and dashed curves represent the theoretical
predictions of chUE and chSE, respectively.}
\end{figure}

\paragraph{Number variance.}

A third test was accomplished by checking the number variance 
$\Sigma^2(S_0,S)$. We report the results (for the unfolded variable
$z$) in Fig. \ref{fig:nvar} for the same collection of cases 
considered in Fig. \ref{fig:min} and \ref{fig:dist}. In the large volume
case we see agreement with the theoretical  prediction \cite{MaGuWe98}
of chUE up to a value of $z\simeq2$. This outcome is also consistent
with the results for $\rho_s(z)$, where the lattice data agree with
chUE up to the same value of $z$, corresponding  to the support of the
distribution of the smallest eigenvalue  (see Fig. \ref{fig:min} and
\ref{fig:dist}, upper part). The existence of an upper bound of $z$,
$z_{max}$, for  the applicability of chRMT is well-known in the
literature  (see for example \cite{BeGoGu98}) and an estimate for
$z_{\rm max}$  is available \cite{StJaNoPaOsVe98} in $d=4$: 
$z_{max}\sim f_{\pi}^2 \,(L\,a)^2$; such an estimate (related to the
massless modes in $d=4$) is absent in $d=2$.

In the small volume case we observe instead {\em dis}agreement  with the
analytical prediction of chSE \cite{MaGuWe98} already for very small
values of $z$, even if the overall  shape seems to resemble the
symplectic behavior. 

\subsection{Topological sector: $N_f=0$ and $\nu=1$}

As discussed earlier (Section \ref{subs:edge}), in the case of GW
actions a definition of the topological charge  corresponding to the
parameter $\nu$ of chRMT is available: this is the  index of the
lattice Dirac operator. 

\begin{figure}[tp]
\begin{center}
\epsfig{file=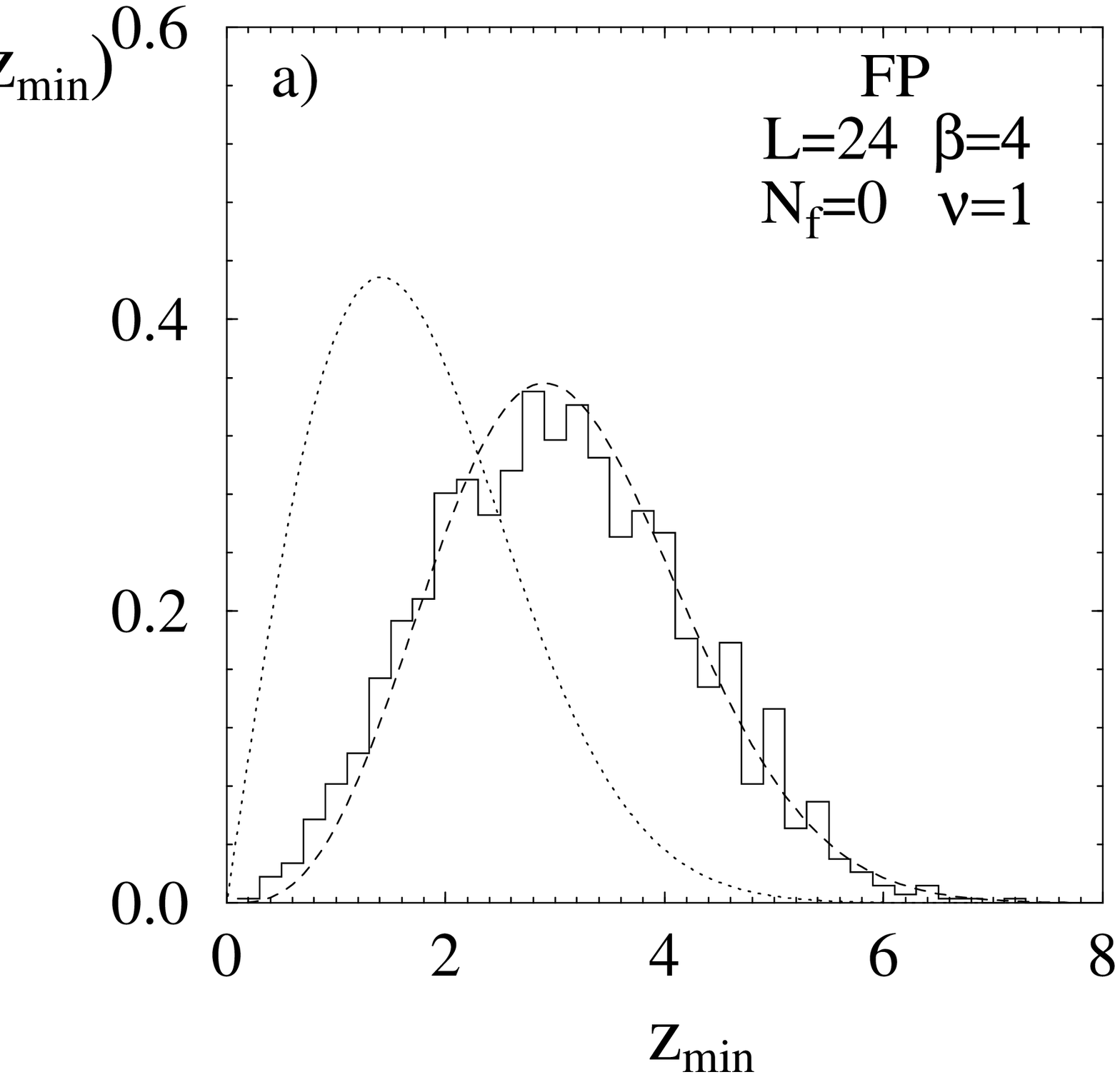,width=6 truecm, angle=-90}
\epsfig{file=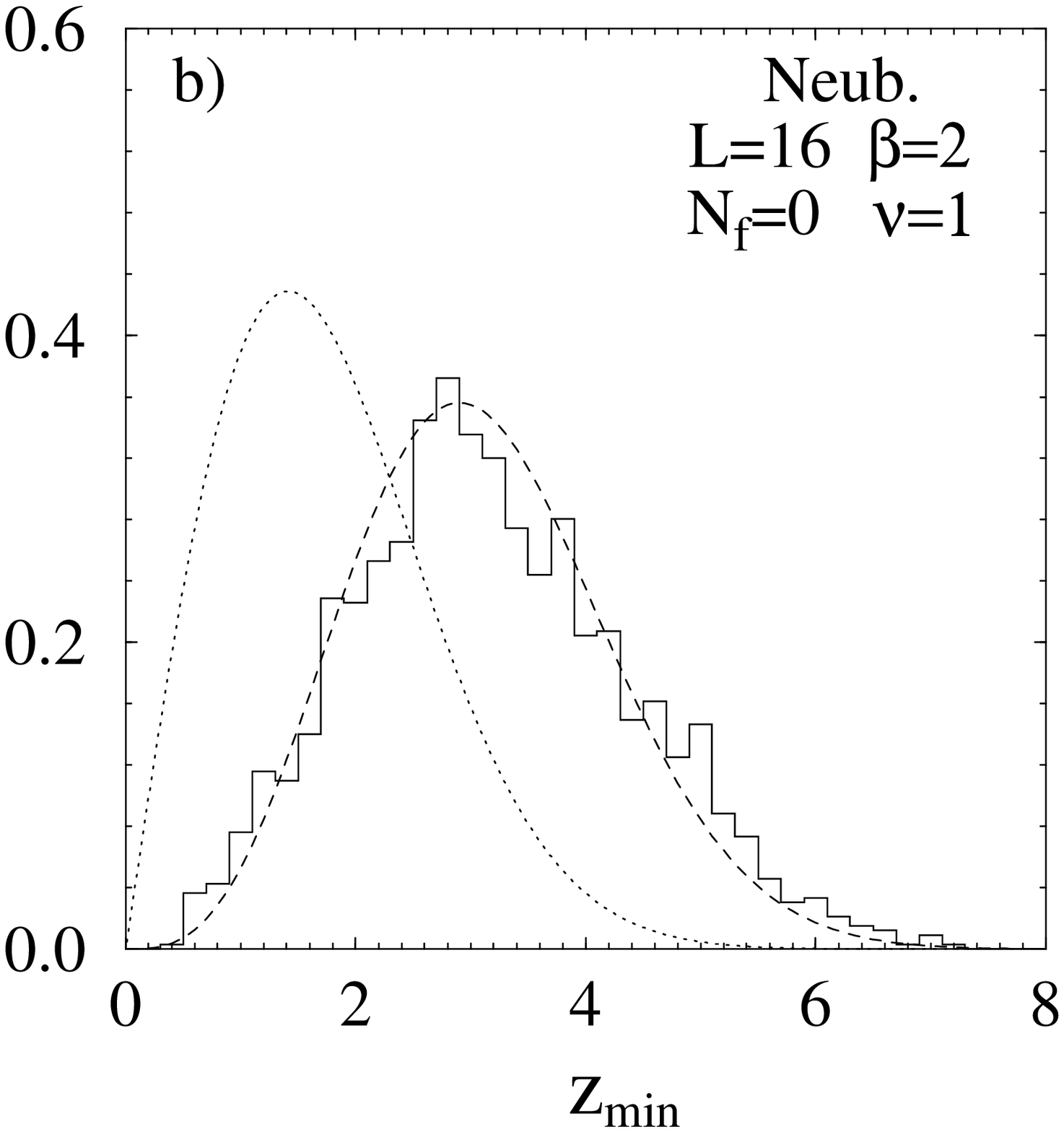,width=6 truecm, angle=-90}
\epsfig{file=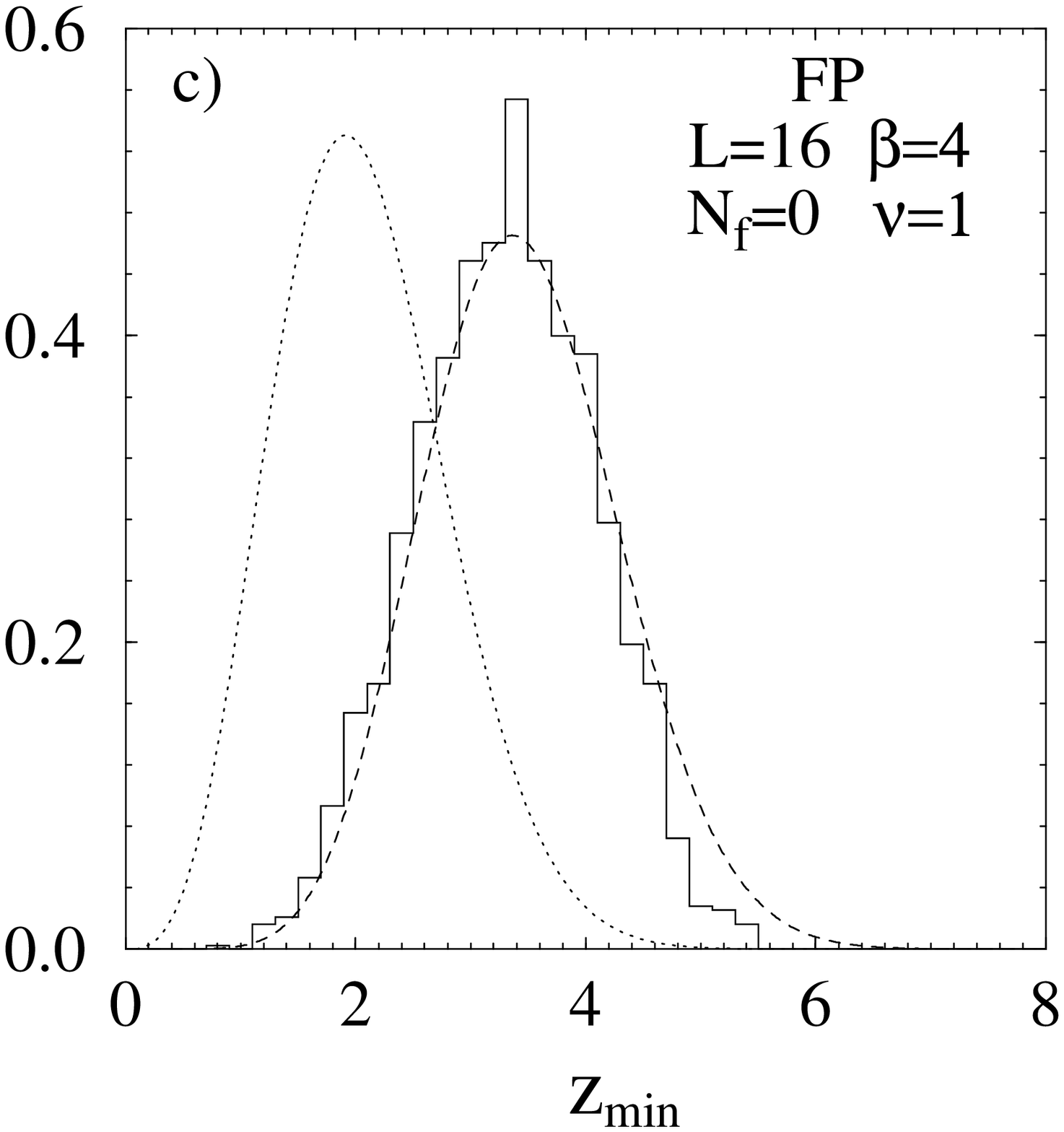,width=6 truecm, angle=-90}
\epsfig{file=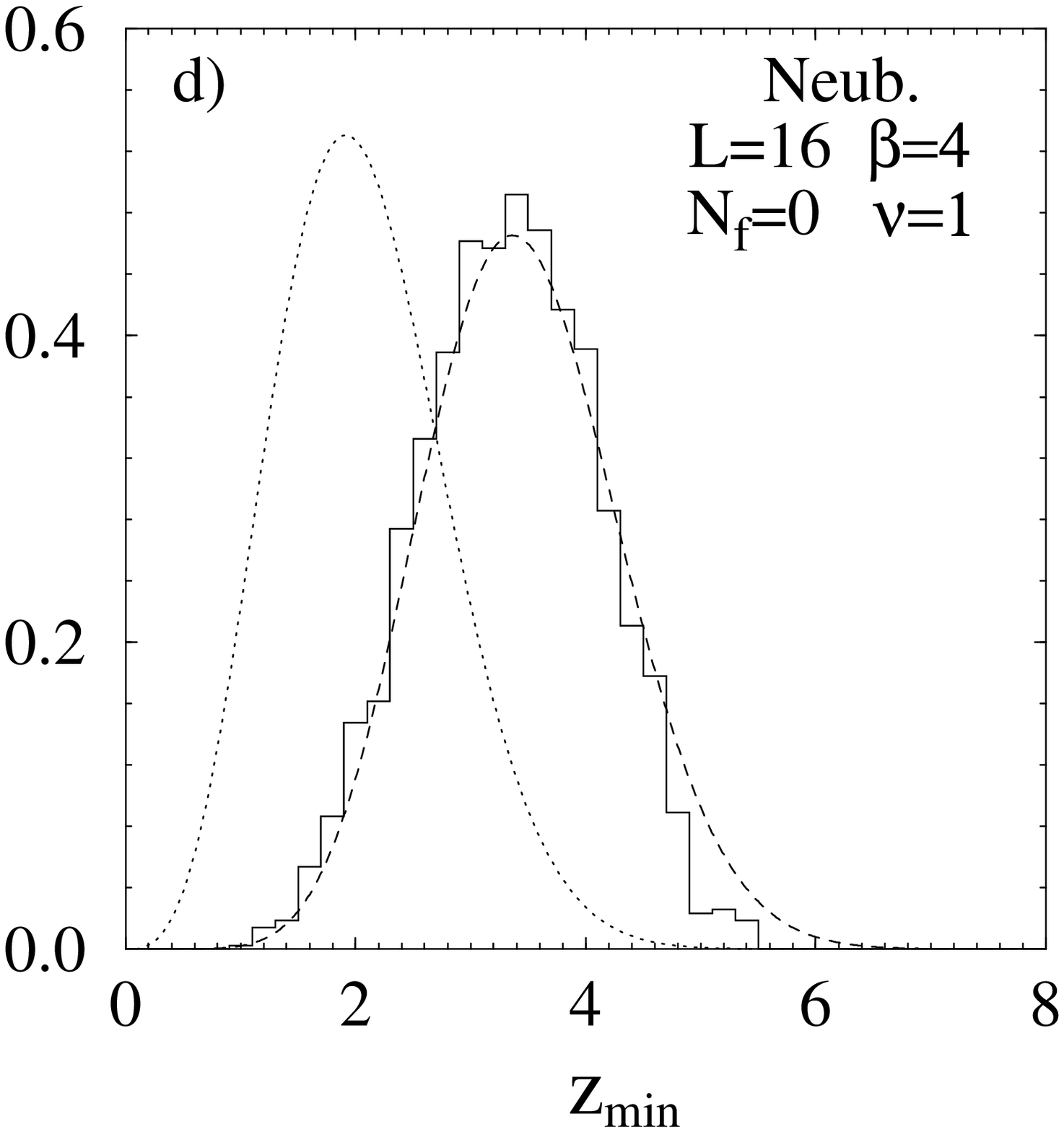,width=6 truecm, angle=-90}
\end{center}
\caption{\label{fig:topmin}
Distribution $P(z_{\rm min})$ of the smallest eigenvalue 
in the unfolded variable $z$ in the $\nu=1$ topological
sector for different lattices and values of $\beta$, for the FP  and
Neuberger's Dirac operator. The dotted and dashed curves represent the
theoretical  predictions of the chRMT ensemble 
for the values $\nu=0$ and $\nu=1$, respectively,
with chUE for (a, b) and chSE for (c, d), 
cf. the discussion in the text.}
\end{figure}

\begin{figure}[tp]
\begin{center}
\epsfig{file=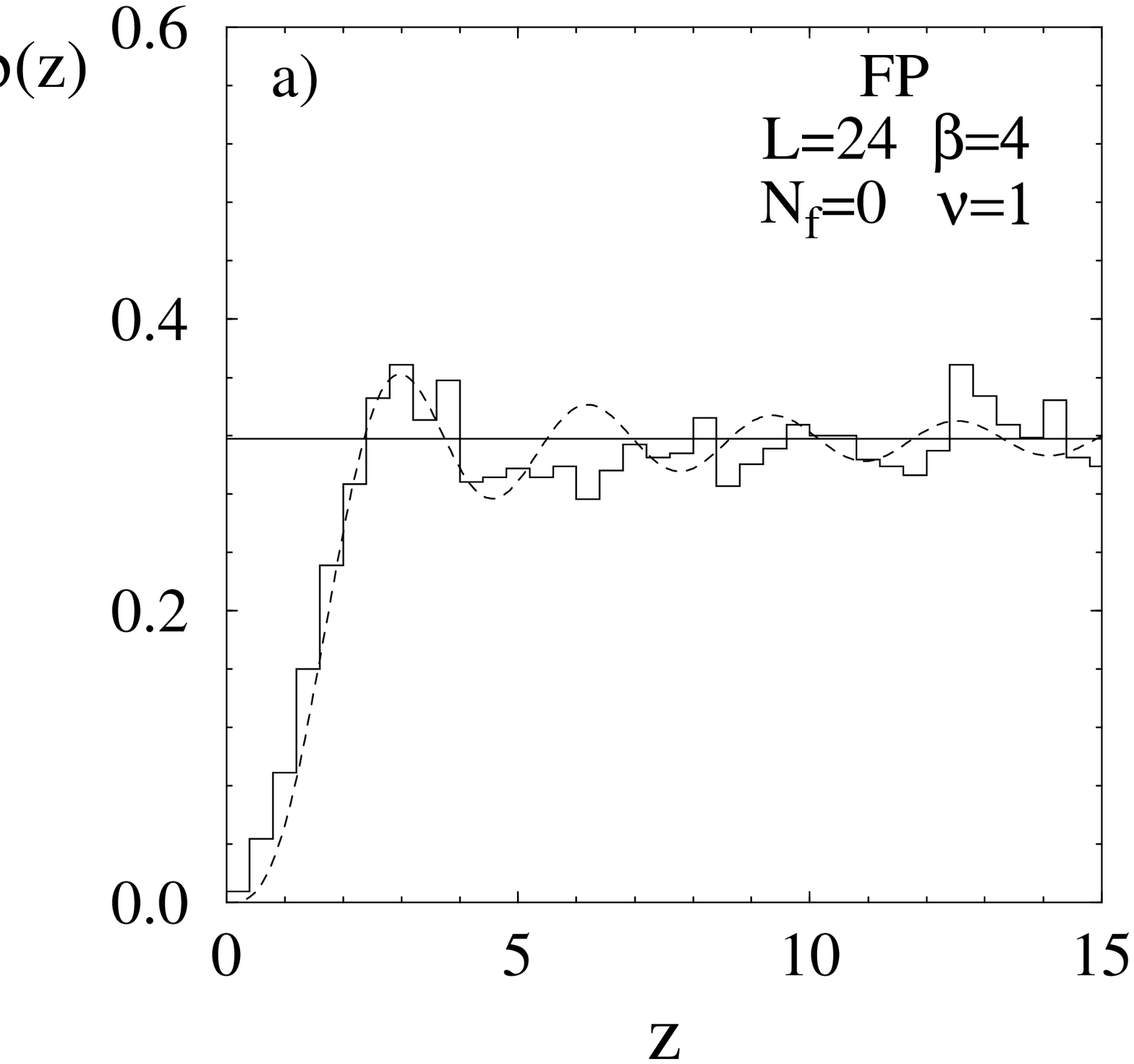,width=6 truecm, angle=-90}
\epsfig{file=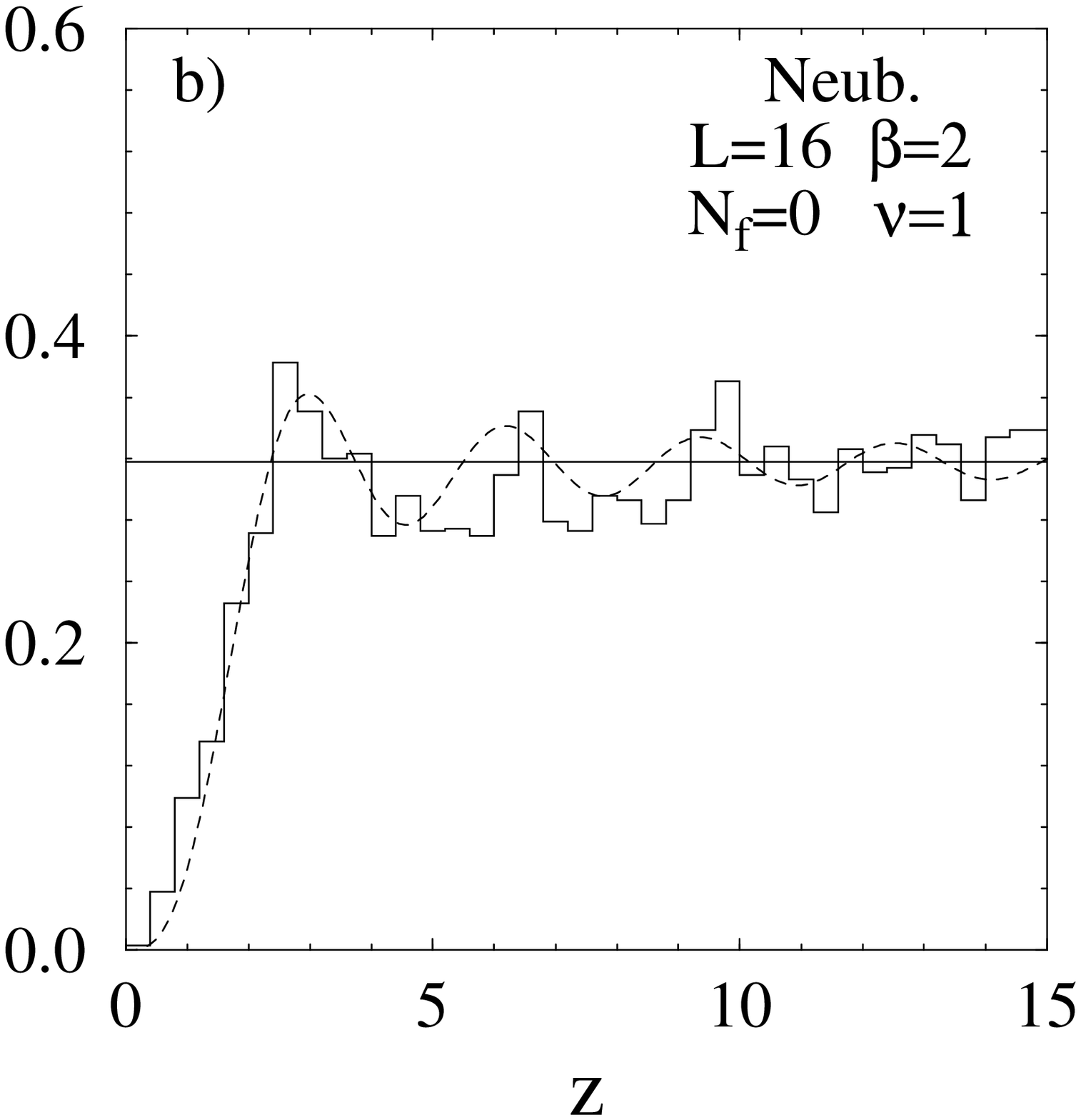,width=6 truecm, angle=-90}
\epsfig{file=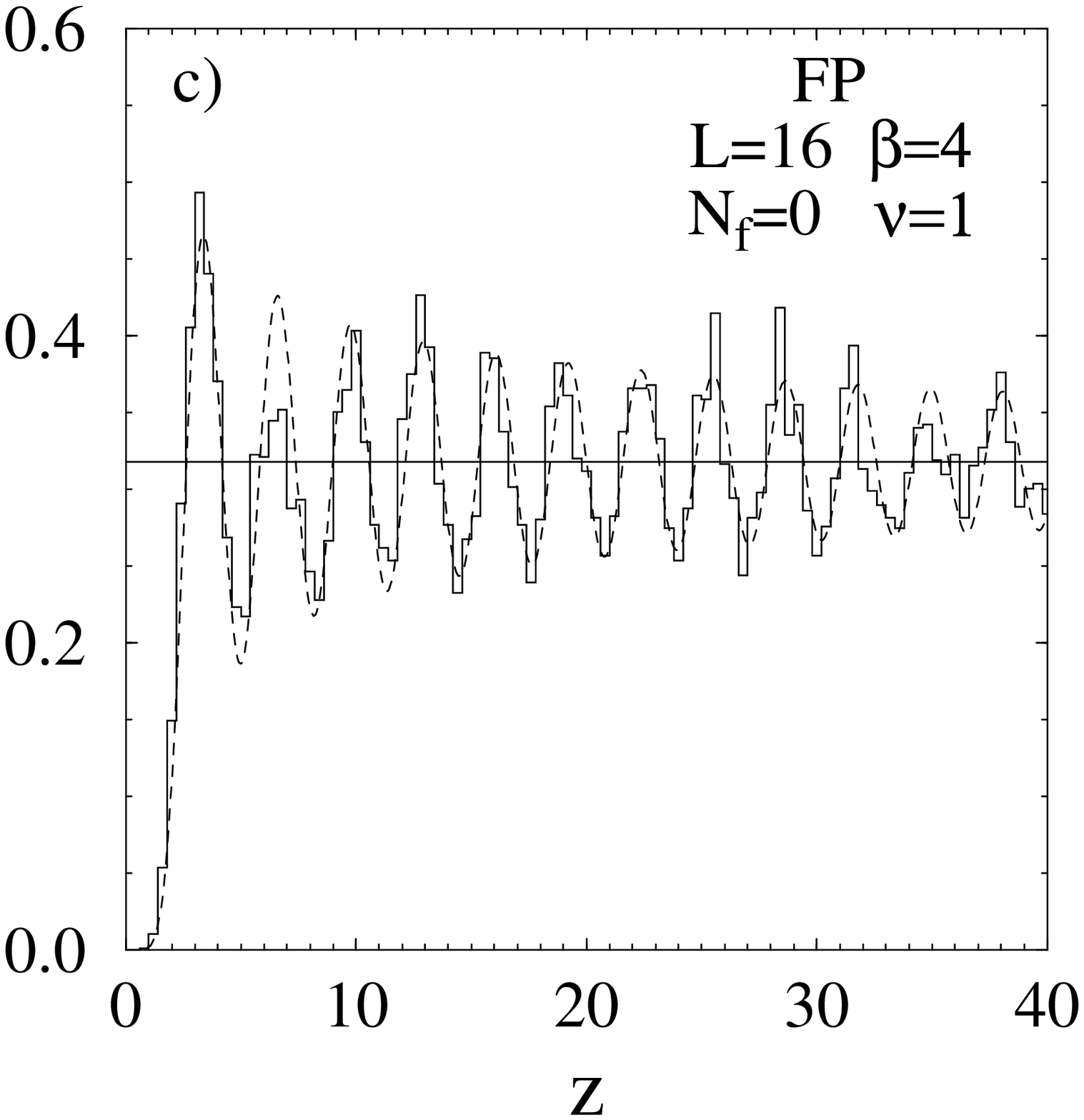,width=6 truecm, angle=-90}
\epsfig{file=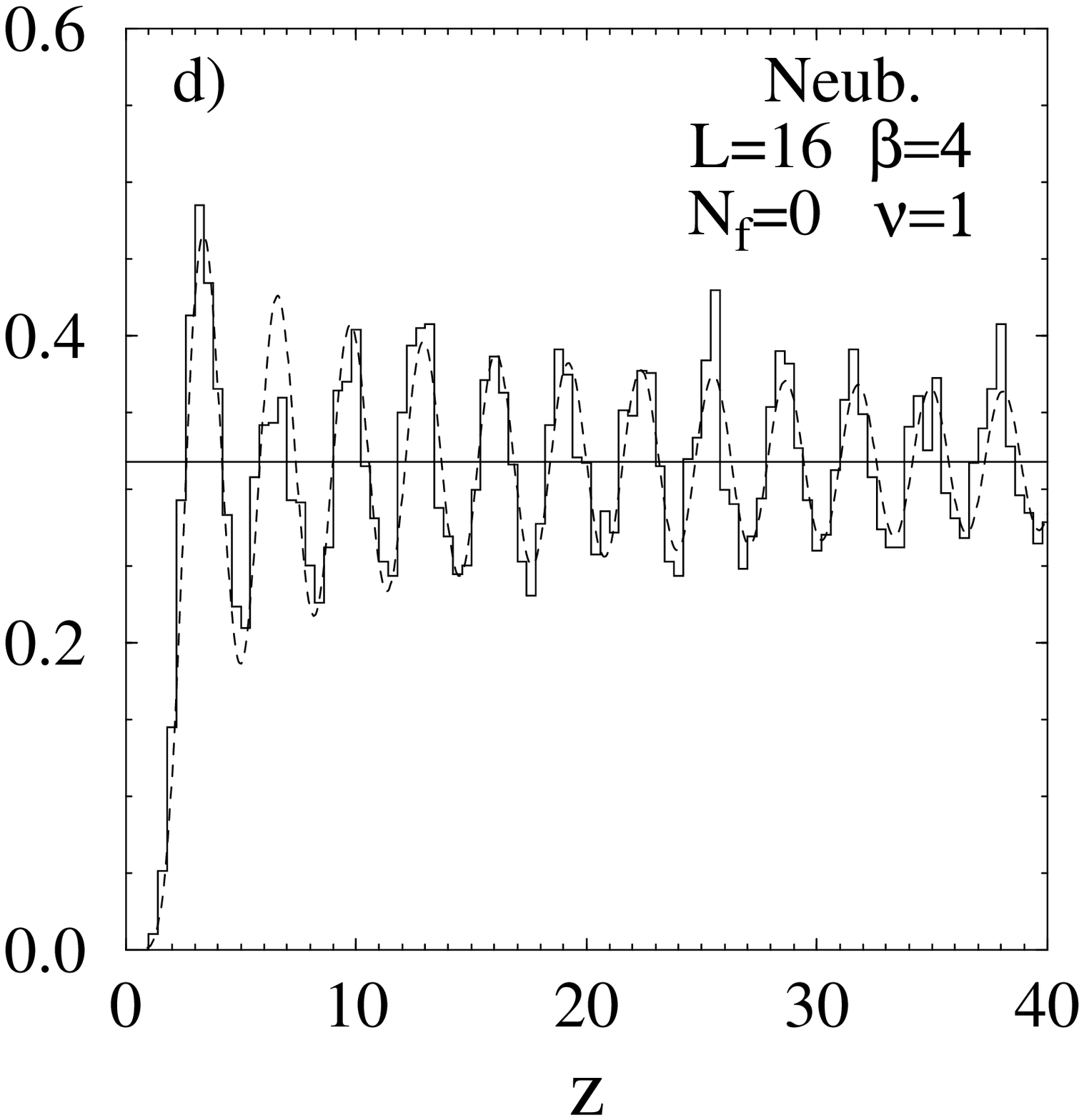,width=6 truecm, angle=-90}
\end{center}
\caption{\label{fig:topdist}
Microscopic spectral density $\rho_{\rm s}(z)$ in the $\nu=1$
topological sector for different lattices and values of $\beta$, for
the FP  and Neuberger's Dirac operator. The dashed curves represent the
theoretical  predictions of the chRMT ensemble for $\nu=1$, with chUE
for (a, b) and chSE for (c, d).}
\end{figure}

In Fig. \ref{fig:topmin} and \ref{fig:topdist} we present our results for the
distribution of the smallest eigenvalue and microscopic spectral density
respectively, for configurations with index$(U)$=1 and compare with the
corresponding predictions of chUE \cite{NaFo98,VeZa93} and chSE 
\cite{BeMeWe98,NaFo95,MaGuWe98}. For these configurations we use the identical
unfolding as in the previous section for  $\nu=0$ configurations (for
corresponding $L$ and $\beta$) and, in particular, the same value of $\Sigma$.
This procedure is consistent with the expectations from chRMT  that $\Sigma$
should not depend on the topological sector, since all such dependence is
incorporated in the functional behavior of the distributions. We observe for
$P(\lambda_{\rm min})$ and $\rho_s(z)$ a universal  behavior, again consistent
with the observations of the previous section.

\subsection{Dynamical fermions}

Consistent with our unquenching procedure, we  define the
distributions for the setup with dynamical fermions by weighting the
entries of the quenched distributions with  the fermion determinant. For
example, in the case of the microscopic spectral density we take:
\be
\rho_s(z)=\frac{\sum_C \rho_s(C,z)\,{\rm det}\D(C)}{\sum_C {\rm
det}\D(C)}\;\;,
\ee
where $\rho_s(C,z)$ is the spectral distribution for an individual 
configuration $C$; the quenched setup is recovered replacing ${\rm
det}\D(C)$ with unity. 
This ``unquenching'' is formally correct; it is not optimal from the point of
view of statistical reliability and one may expect large fluctuations of
intermediate results. However, in our $d=2$ studies (cf. \cite{LaPa98,FaHiLa98})
we found stable results for the given statistics.

\begin{figure}[tp]
\begin{center}
\epsfig{file=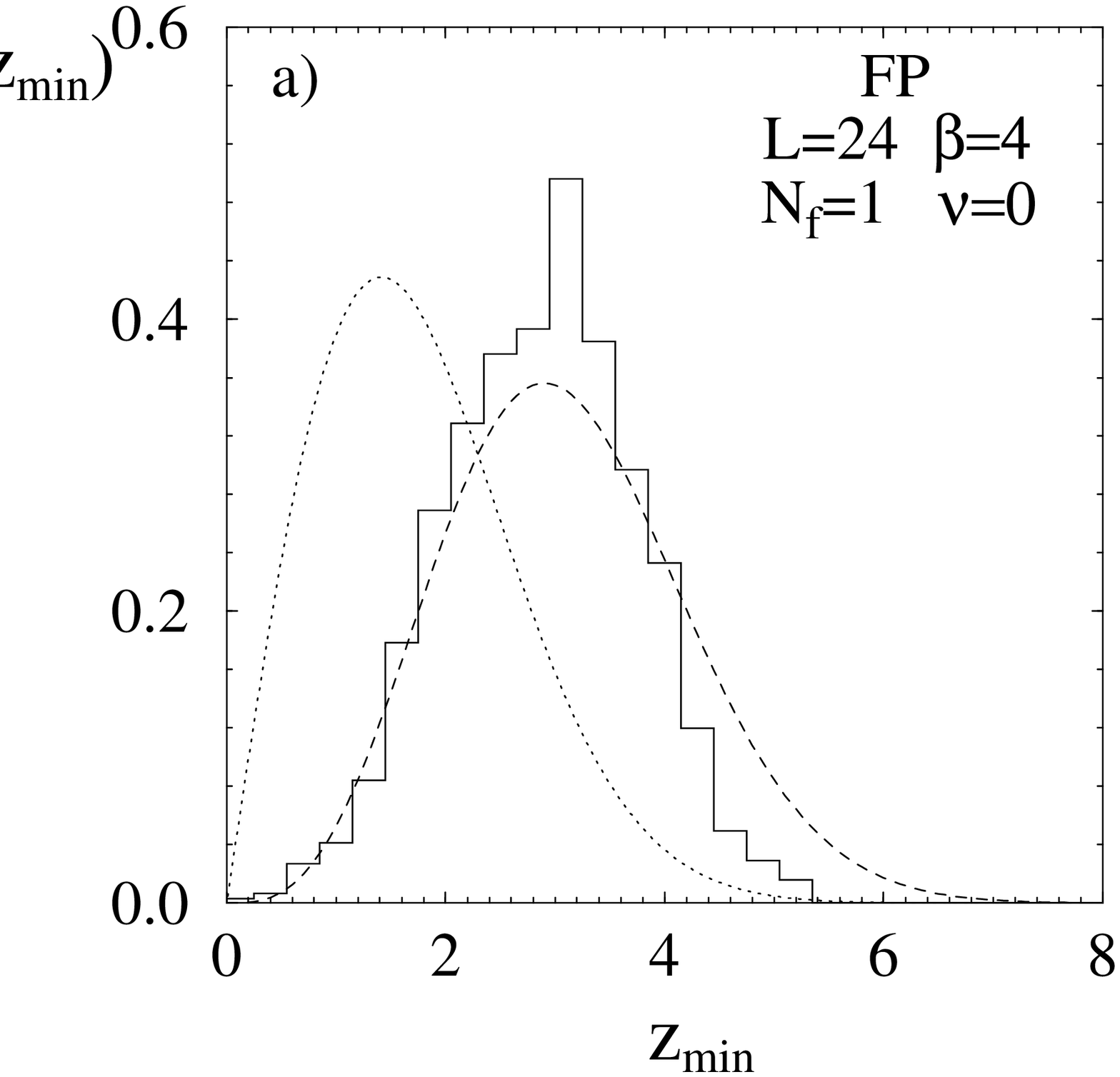,width=6 truecm, angle=-90}
\epsfig{file=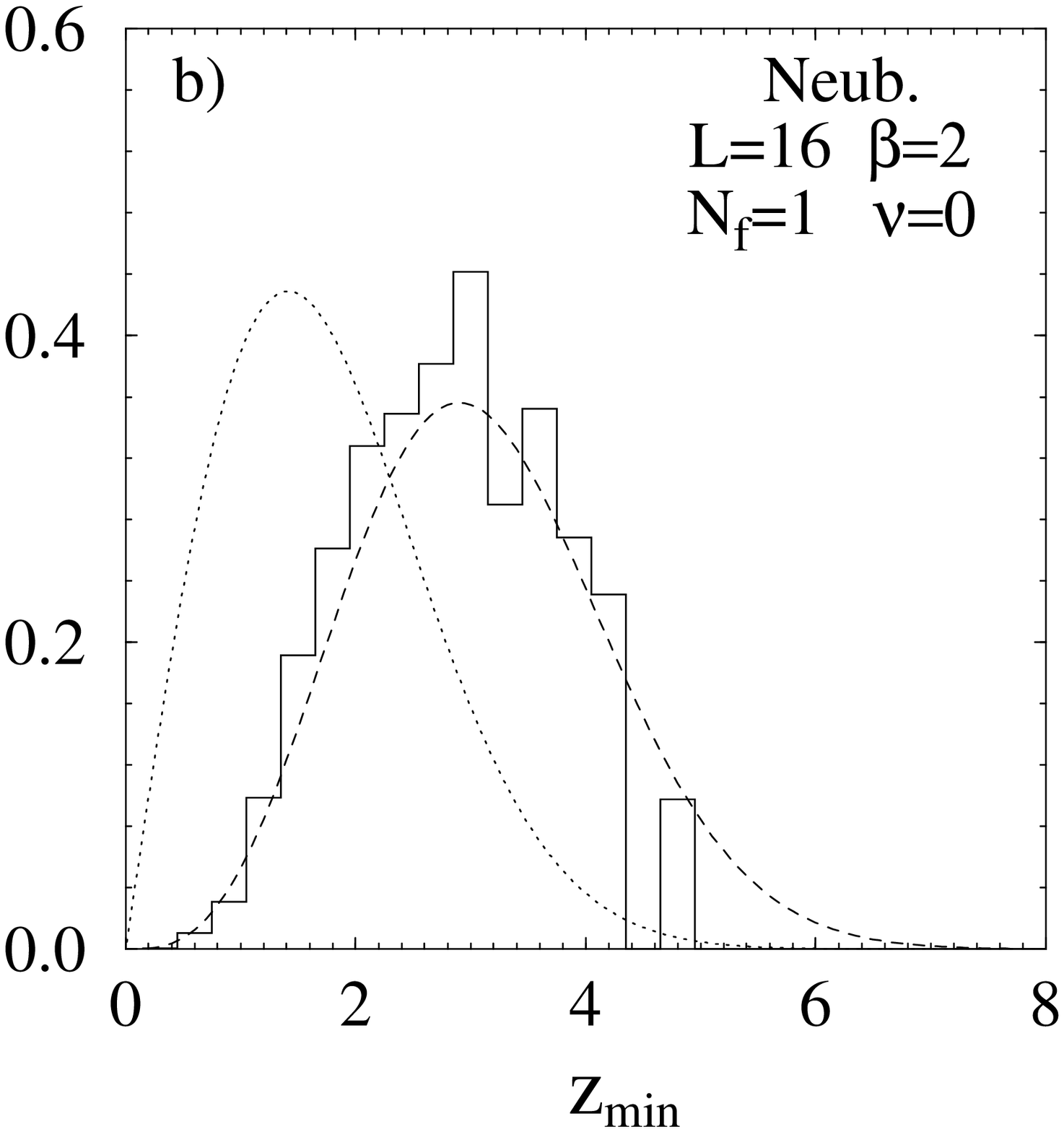,width=6 truecm, angle=-90}
\epsfig{file=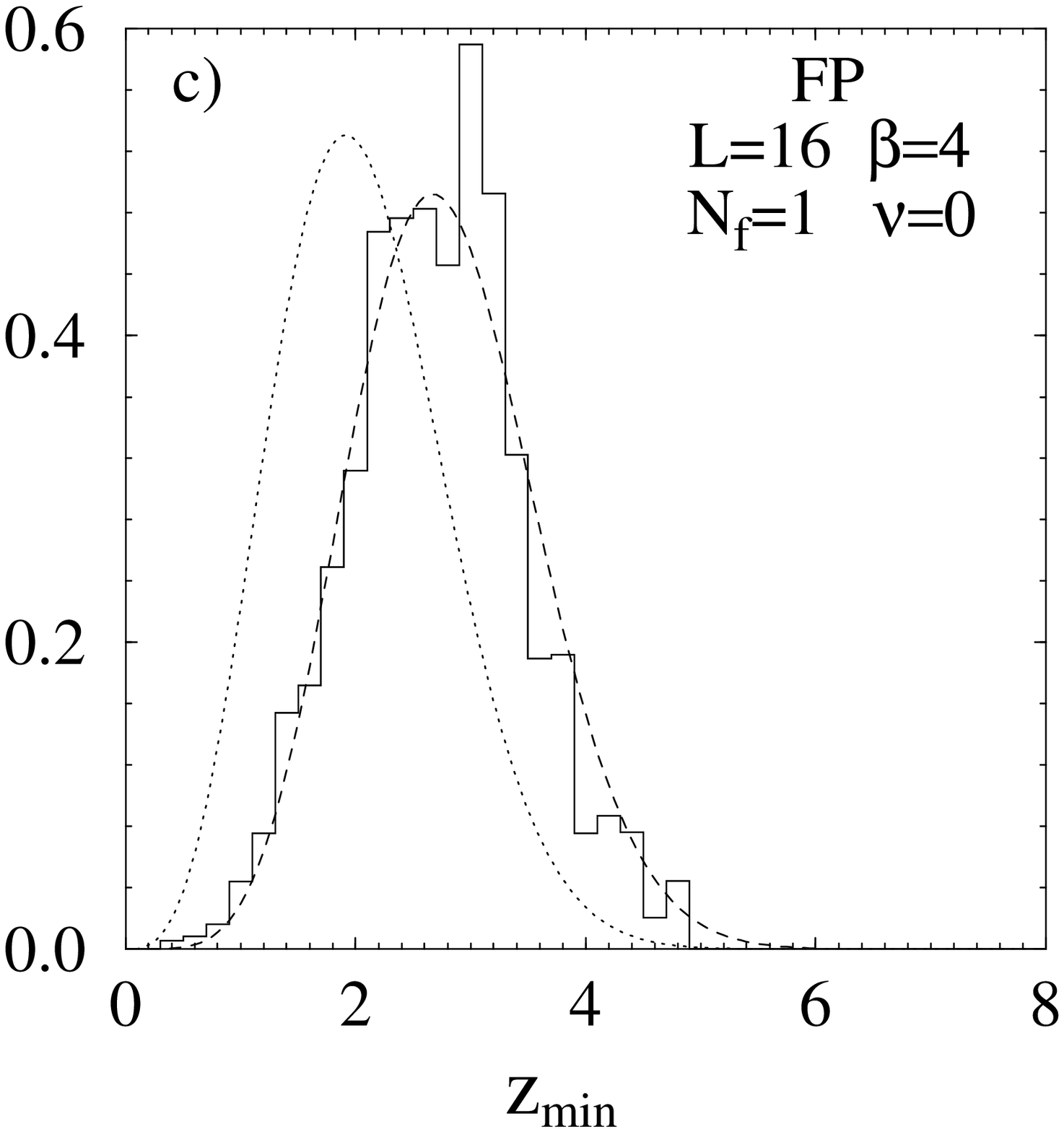,width=6 truecm, angle=-90}
\epsfig{file=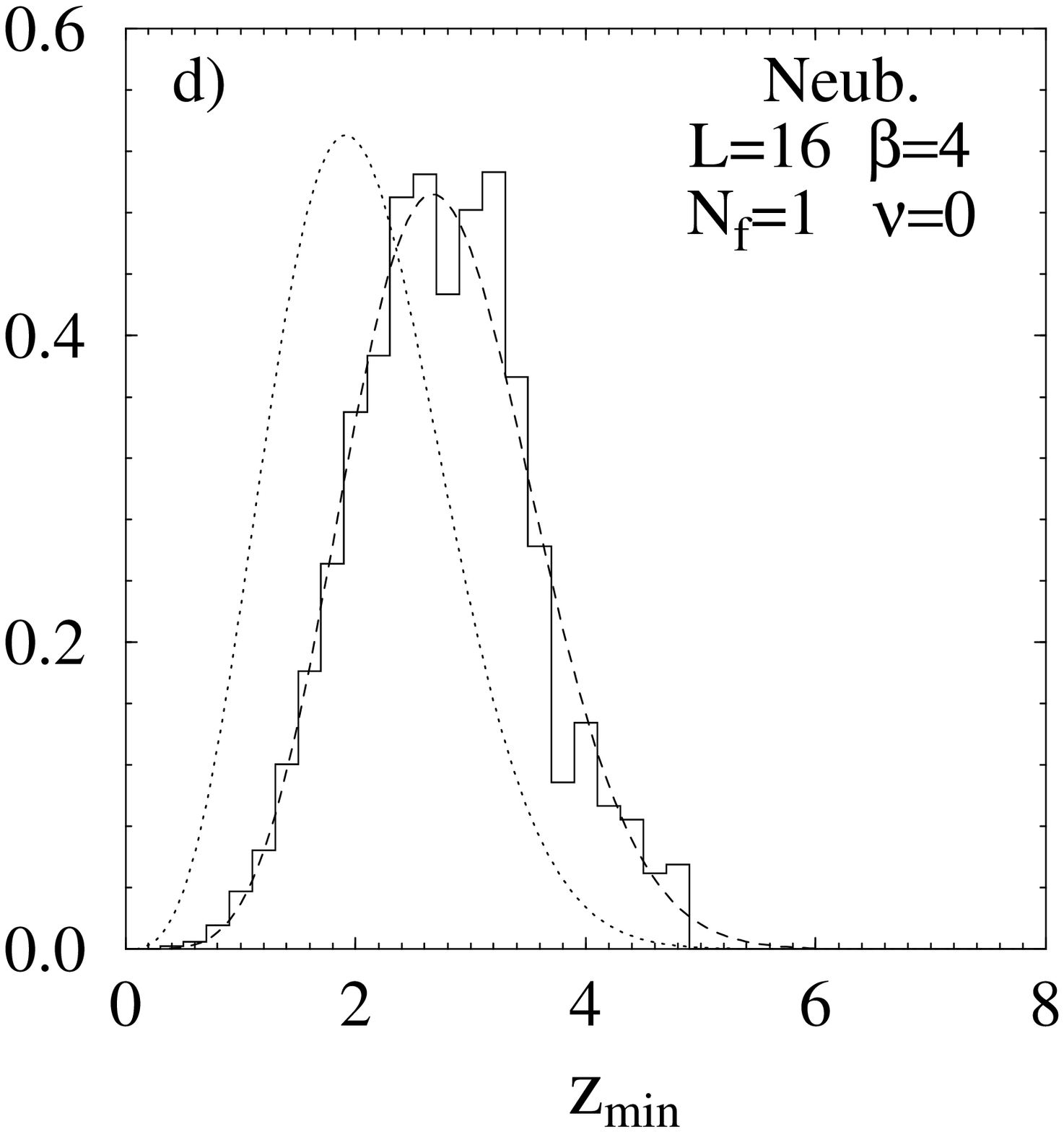,width=6 truecm, angle=-90}
\end{center}
\caption{\label{fig:unqmin}
Distribution  $P(z_{\rm min})$ of the smallest eigenvalue in the
unfolded variable $z$ for the unquenched ($N_f=1$) data in the $\nu=0$
topological sector, for the FP  and Neuberger's Dirac operator. The
dotted and dashed curves represent the theoretical  prediction of the
best fitting chRMT ensemble (cf. Fig. \ref{fig:topmin}) in the trivial
sector  for $N_f=0$ and $N_f=1$, respectively.}
\end{figure}

\begin{figure}[tp]
\begin{center}
\epsfig{file=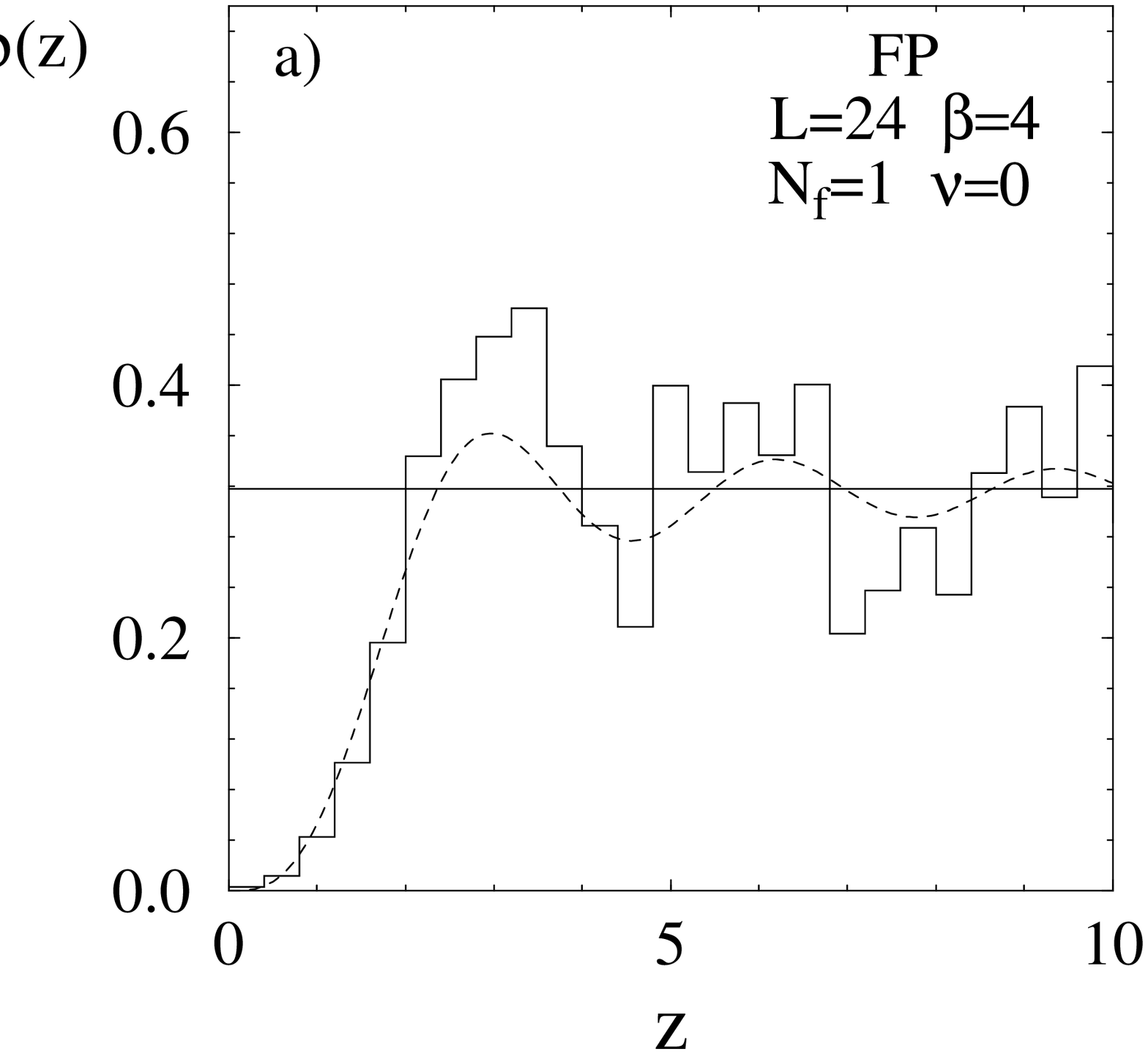,width=6 truecm, angle=-90}
\epsfig{file=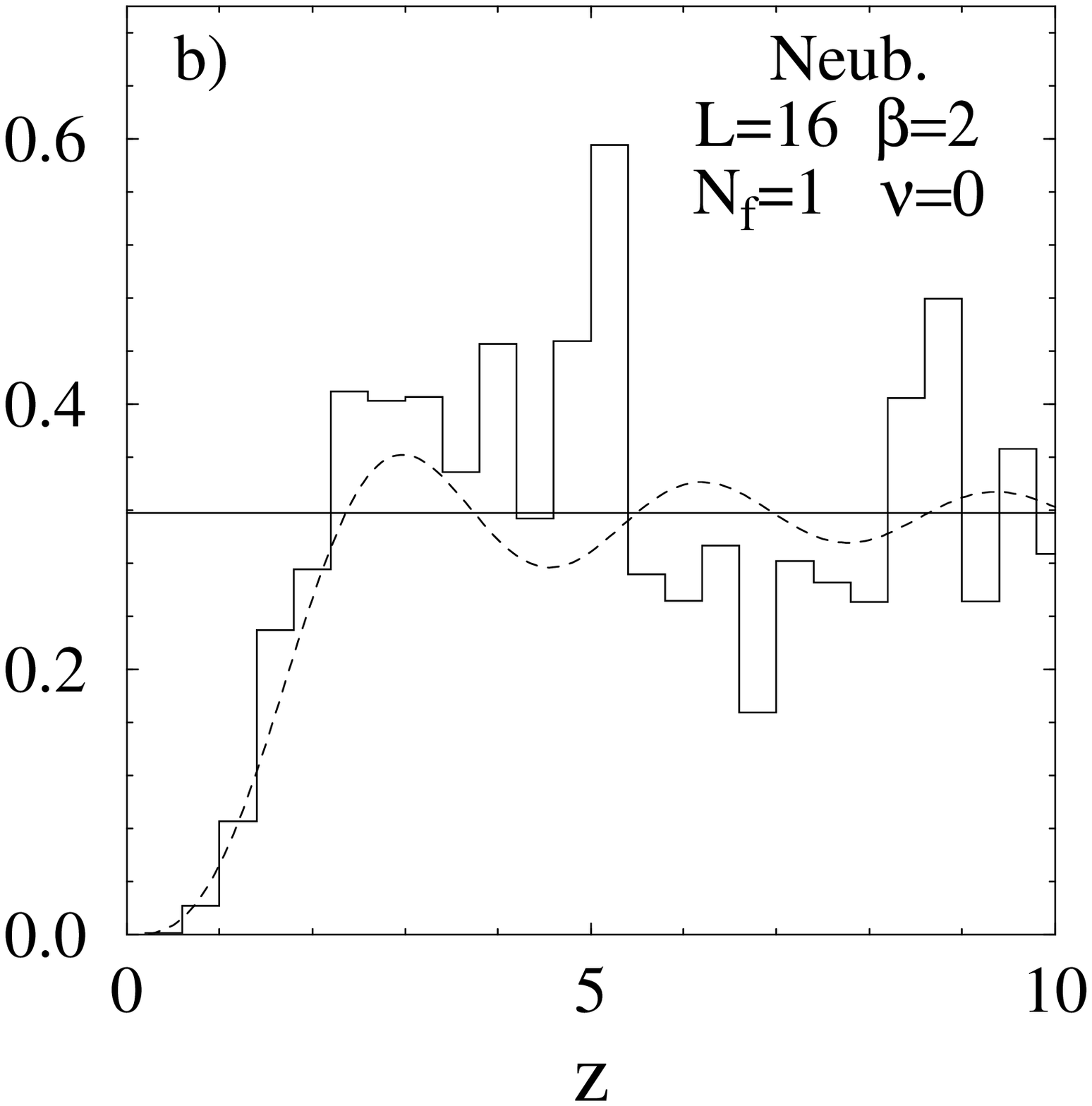,width=6 truecm, angle=-90}
\epsfig{file=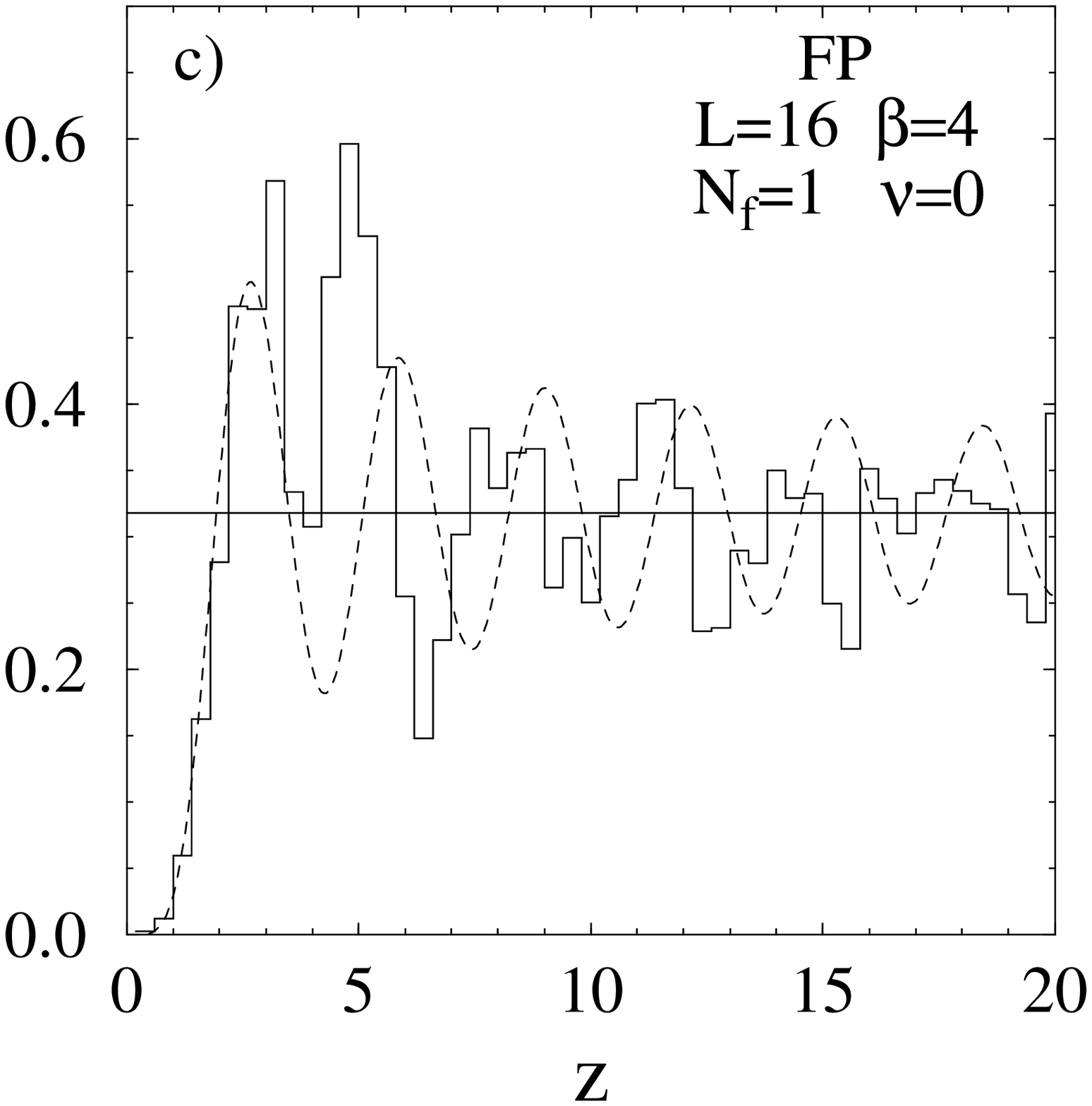,width=6 truecm, angle=-90}
\epsfig{file=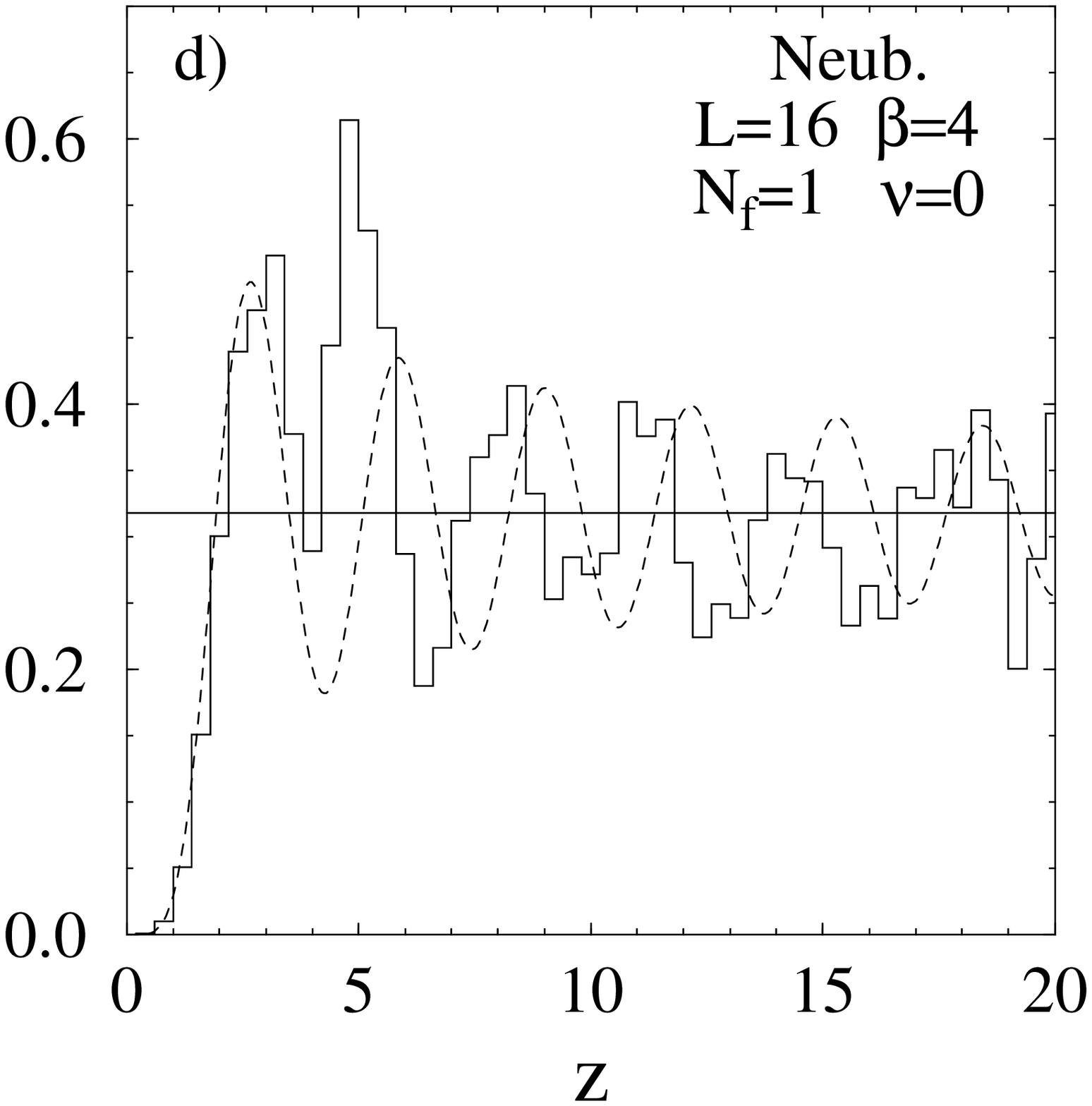,width=6 truecm, angle=-90}
\end{center}
\caption{\label{fig:unqdist}
Microscopic spectral density $\rho_{\rm s}(z)$
for the unquenched ($N_f=1$) data  in the
trivial topological sector for the FP  and Neuberger's Dirac operator.
The dashed curves represent the theoretical  predictions of the best
fitting chRMT ensemble  (cf. Fig. \ref{fig:topmin})  for $N_f=0$ and
$N_f=1$, respectively.}
\end{figure}

In Fig. \ref{fig:unqmin} and  \ref{fig:unqdist} we show the results for
$P(\lambda_{\rm min})$ and $\rho_s(z)$ respectively in the case of 
configurations with index$(U)$=0. We compare these with the
predictions  of chUE \cite{NaFo98,VeZa93} and chSE
\cite{NaFo98,NaFo95,MaGuWe98} in the case $N_f=1$, $\nu=0$. In
Fig. \ref{fig:unqdist} we see that higher statistics 
for dynamical fermions would be desirable in particular for $\rho_s(z)$.

The value of $\Sigma$ used for the definition of the unfolded  variable
$z$ was found by a best-fit procedure in the distribution of the
smallest eigenvalues like in Section \ref{subs:trivial}. In the
large-volume case  we find, consistent with the results of the quenched
setup, agreement  with chUE up to $z\simeq 2$.

\begin{figure}[tp]
\begin{center}
\epsfig{file=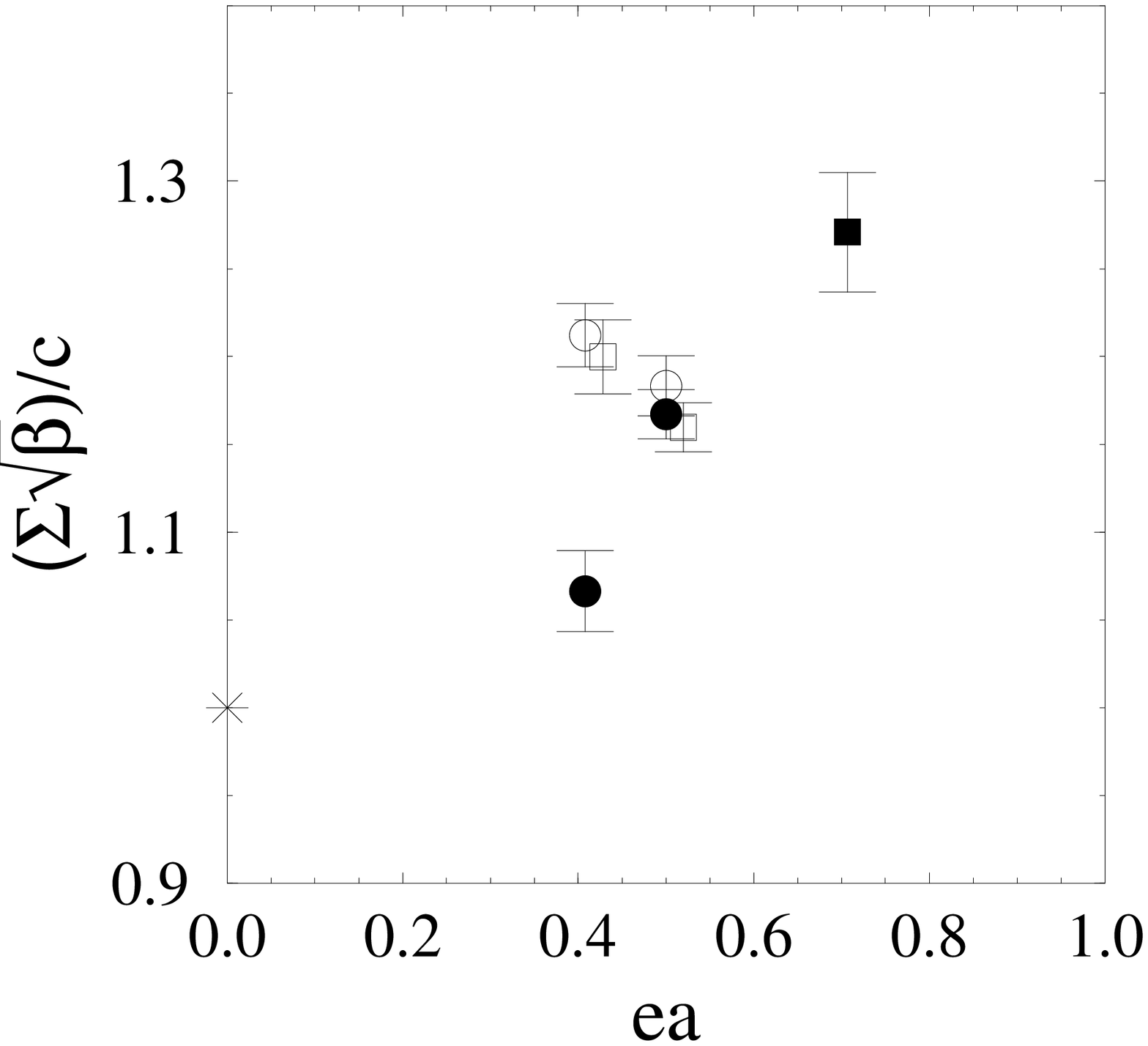,width=6 truecm, angle=-90}
\epsfig{file=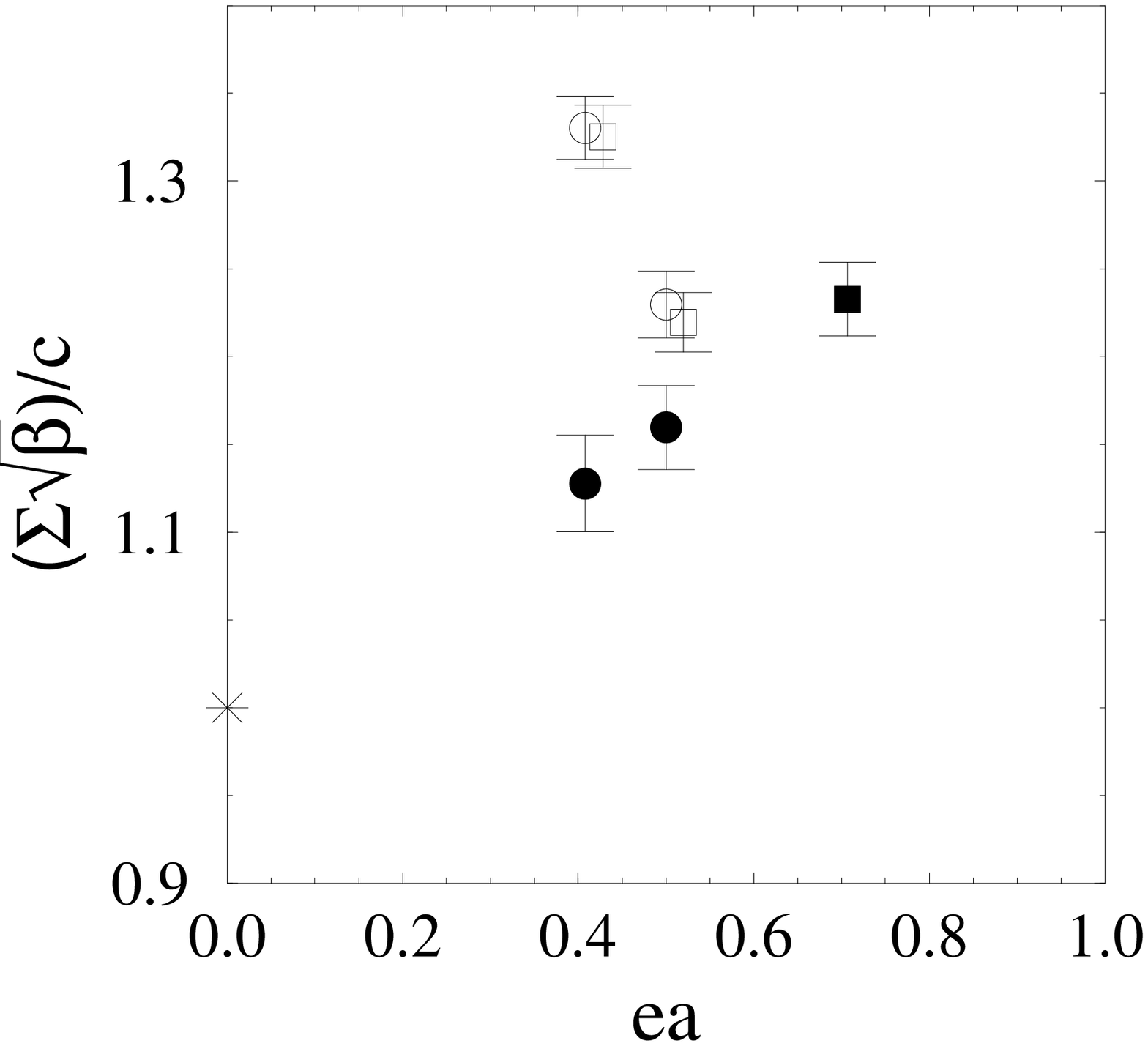,width=6 truecm, angle=-90}
\end{center}
\caption{\label{fig:cond}
The ratio of $\Sigma\,\sqrt{\beta}$ (in units of the  continuum value
$-\contcond/e=c$, cf. (\ref{SigmaPhys})),  obtained from  the
distribution of the smallest eigenvalue for the quenched (left)  and
unquenched (right) setup as a function of $e\,a=1/\sqrt{\beta}$.  Full
(open) symbols denote  results of the fit in the large (small) physical
volume region,  with unitary (symplectic) prediction;  circles denote
the FP action, boxes   Neuberger's operator.} \end{figure}

The values of $\Sigma$ found here provide an estimate of the fermion
condensate in the infinite volume limit. In Table \ref{tab:rec_cond} we
summarize these results for this quantity. These should also be
compared with the direct determination in the finite volume of Section
\ref{subs:res_cond} in Table \ref{tab:cond}. Typically, the statistical
error are smaller for the values obtained by the RMT techniques.

In Fig. \ref{fig:cond} we display these estimates for the infinite
volume condensate. The results obtained in the large volume region (using
the chUE distributions) seem to approach the continuum with a linear
dependence in the lattice spacing; those obtained in the small volume
region appear more scattered. 

\begin{table}[htb]
\centering
\begin{tabular}{llllllll}
\hline
Op. &$L$& $\beta$ & $N_f=0$ & $N_f=1$ & Ensemble & Cont.\\
\hline
{\rm \DOV} & 16 & 2 & 0.1437(39) & 0.1394(26) &chUE& 0.1131  \\
{\rm \DFP} & 16 & 4 & 0.0946(14) & 0.0983(15) &chSE& 0.0804  \\
{\rm \DOV} & 16 & 4 & 0.0927(14) & 0.0975(14) &chSE& 0.0804  \\
{\rm \DFP} & 24 & 4 & 0.0933(11) & 0.0927(19) &chUE& 0.0804  \\
{\rm \DFP} & 16 & 6 & 0.0791(12) & 0.0868(12) &chSE& 0.0653  \\
{\rm \DOV} & 16 & 6 & 0.0783(12) & 0.0865(12) &chSE& 0.0653  \\
{\rm \DFP} & 24 & 6 & 0.0696(15) & 0.0736(18) &chUE& 0.0653  \\
\hline
\end{tabular}
\caption{\label{tab:rec_cond} 
Values of the fermion condensate in the infinite volume obtained from
comparison of lattice data with chRMT.}
\end{table}

Another check \cite{BeMeSc98} of the chRMT  predictions related to  the  microscopic
spectral distribution involves the quantity \cite{LeSm92}
\be
S_2\equiv\frac{1}{\Sigma^2\,V^2}\,\gaugeexp{\sum_{\lambda>0}
\frac{1}{\lambda^2}}
\quad
\stackrel{V\to\infty}{\longrightarrow}\quad
\int dz\;\frac{\rho_s(z)}{z^2}\;.
\ee
Inserting the predictions from chRMT for $\rho_s(z)$, one obtains
\cite{Ve94}:
\be \label{LeSmilga}
S_2 = \frac{b}{8\,(\frac{b\,\nu}{2}+\frac{b}{2}+N_f-1)}\;\;,
\ee
with $b=2$ and 4 for chUE and chSE, respectively.

In Tables \ref{tab:sumr_unit} and \ref{tab:sumr_symp} we compare our
values for $S_2$ with (\ref{LeSmilga}) in the large  and the small 
(physical) volume region. We find better agreement for large volumes
and for higher topological sectors.

\begin{table}[tb]
\centering
\begin{tabular}{cllllll}
\hline
& \multicolumn{2}{c}{\small $N_f=0$} & \hspace{2mm} & \multicolumn{2}{c}{\small $N_f=1$} &  \\
\cline{2 - 3}    \cline{5 - 6}
{\small $N_f+\nu$} \hspace{2mm} & {\small $L\!\!=\!\!24$ $\beta\!\!=\!\!4$ } & {\small $L\!\!=\!\!16$ $\beta\!\!=\!\!2$ 
} &  & 
{\small $L\!\!=\!\!24$ $\beta\!\!=\!\!4$} &{\small $L\!\!=\!\!16$ $\beta\!\!=\!\!2$ 
} & {\small \rm chUE} \\
\hline
1 \hspace{2mm} & 0.369(47) & 0.30(12)   & & 0.288(29)  & 0.326(17)  & 0.25 \\
2 \hspace{2mm} & 0.157(6)  & 0.159(7)   & & 0.1401(24) & 0.1567(20) & 0.125 \\
3 \hspace{2mm} & 0.0938(24)& 0.0922(36) & & 0.0941(10) & 0.1042(12) & 0.0833 \\
4 \hspace{2mm} & 0.0666(15)& 0.0607(13) & & 0.0692(8)  & 0.0758(9)& 0.0625 \\
5 \hspace{2mm} & 0.0508(16)& 0.0525(36) & & 0.0554(8)  & 0.0563(8)& 0.05\\
6 \hspace{2mm} & 0.0407(16)& 0.0402(20) & & 0.0452(8)  & 0.0455(11)& 0.0417\\
7 \hspace{2mm} &           &            & & 0.0392(12) & 0.0380(16)& 0.0357\\
\hline
\end{tabular}
\caption{\label{tab:sumr_unit}
Our values for $S_2$ in the large-volume region
in the quenched and unquenched setup: $L=24$ and $\beta=4$ 
(FP action), $L=16$ and $\beta=2$ (Neuberger's Dirac operator); chUE
denotes the prediction (\ref{LeSmilga}) for $b=2$.} 
\end{table}
\begin{table}[tb]
\centering
\begin{tabular}{llllllll}
\hline
& \multicolumn{3}{c}{\small $N_f=0$} & & \multicolumn{3}{c}{\small $N_f=1$}  \\
\cline{2 - 4}    \cline{6 - 8}
{\small $\nu$} \hspace{1mm} & {\small \rm FP} & {\small \rm Neub.} & {\small \rm chSE} &  & 
{\small \rm FP} &{\small \rm Neub.} & {\small \rm chSE} \\
\hline
0 \hspace{2mm} & 0.994(84)  & 1.16(18)   & 0.5     & & 0.318(12)   & 0.300(10) & 0.25 \\
1 \hspace{2mm} & 0.2020(25) & 0.2003(22) & 0.1666  & & 0.1562(11)  & 0.1520(10)& 0.125 \\
2 \hspace{2mm} & 0.1150(13) & 0.1143(13) & 0.1     & & 0.1034(8)   & 0.0997(7) & 0.0833 \\
3 \hspace{2mm} & 0.0773(14) & 0.0768(14) & 0.0714  & & 0.0739(9)   & 0.0725(8) & 0.0625 \\
4 \hspace{2mm} & 0.0552(19) & 0.0551(18) & 0.0555  & & 0.0543(12)  & 0.0536(12) & 0.05 \\
5 \hspace{2mm} & 0.0468(48) & 0.0472(49) & 0.0454  & & 0.0385(25) & 0.0457(43)& 0.0417\\
\hline
\end{tabular}
\caption{\label{tab:sumr_symp}
The values $S_2$ in the small-volume region 
in the quenched and unquenched setup: $L=16$ and $\beta=4$; chSE
denotes the prediction (\ref{LeSmilga}) for $b=4$.}
\end{table}

\section{Discussion and conclusions}
\label{sec:con}

Both, the FP action and Neuberger's Dirac operator, guarantee the
restoration of the main features of chiral symmetry in the continuum
limit. We checked this through the analysis of their spectrum, in
particular the zero modes and their chirality.  The GW relation allows
one to define an index for the Dirac operator which is useful to define
a topological charge on the lattice. The additional property of the FP
action to be perfect at the classical level (which is not the case for
Neuberger's operator) makes this definition particularly  reliable,
since it is related to a renormalization group guided procedure of
interpolation of the gauge configuration.  This results, as we observe,
in an excellent agreement --  good even at low $\beta$ --   with the
geometric definition, which in this context is the most  close-fitting
to the continuum. The agreement is poor in the case of Neuberger's
operator.

At the practical level, Neuberger's approach seems to be  most suitable
when strictness of the chiral properties is required, as in the case of
the calculation of the fermion condensate by subtraction.
Parametrization uncertainties of the FP action seem to be a problem in
this case, for small or even vanishing fermion mass.  On the other
hand, some experience \cite{FaHiLa98} indicates that e.g. for
dispersion relations,  even the approximate perfectness
of the FP action is a great advantage,  while Neuberger's operator
introduces discretization effects as large as  for the Wilson action.
Of course, the present parametrization of the FP action could be
improved (e.g. by sampling according to the compact gauge action in the
determination of the FP fermion action); there may be even the chance
to reduce the effective number of parameters \cite{Bi98b}.
However, in view of the four-dimensional  environment, where only few
operators can be included in the action for practical reasons, the
parametrization effects are likely to be a serious problem, at least
for small fermion mass.

We also studied the statistical fluctuations of the spectra of the Dirac
operators $\DFP$ and $\DOV$. In all cases considered here (different values of
$\beta$ and lattice size)  $\DFP$ and $\DOV$ display very similar statistical
properties. Concerning comparison with chRMT, the results show an unexpected
complementarity between the  symplectic and the unitary universal behavior, 
the latter coming into play for larger physical volumes. However, several
aspects lead us to conclude that the symplectic  behavior is an artifact of the
small volume where RMT breaks down. The spectrum is {\em not} doubly
degenerate as would be expected  in the symplectic framework;
it just looks like one half of a chSE spectrum. The number
variance shows deviation from chSE already from very small values of $z$. The
distribution shape depends on the physical size of the volume. This is in
agreement with the expected range of applicability of RMT,
$1/(a\,m_{Schwinger})\ll L$ as discussed in Section \ref{subs:trivial}. We
conclude that RMT necessitates randomness at a lattice size $L$ much larger
than the correlation length of the system. In the (in physical units) large
volume region  chUE describes our data up to $z_{max}\simeq2$ (for the trivial
sector).  This distribution then should govern the continuum limit.

RMT provides a powerful tool to extract the infinite volume chiral condensate
from finite (and hopefully small) lattices in the case of dynamical fermions.
Our findings indicate, that some caution is in order. We argue that the
physical volume should be not {\em too} small, since chRMT interpretation may
produce misleading results in this situation. 

Concerning the condensate as determined from the chUE fits in the 
large volume domain, we observe linear cut-off effects for both actions 
(Fig. \ref{fig:cond}). If this is not an artifact of this approach, the
natural explanation would be, that one still has to improve the field
operators. This is also true for the FP action, since perfectness
applies to spectral quantities and  its extension to other observables
passes through the implementation of the RG improvement to the operator
(in this case $\bar{\psi}\psi$) as well. Also, we did not attempt to
introduce a renormalized coupling constant. 

We find that studying the spectrum of GW operators is extremely helpful
to identify chirality properties. With chRMT one learns how to
disentangle universal properties from the dynamical parameters like the
fermion condensate. Our FP action and Neuberger's operator lead to similar
results in this context. 

\vspace{.5cm}

{\bf Acknowledgment:}

We want to thank K. Splittorff for stimulating discussions on Random
Matrix Theory. We also thank T. Wettig for useful discussions and for
providing some of his unpublished notes, and J.-Z. Ma for supplying
analytical data. We are grateful to W. Bietenholz,  S.
Chandrasekharan, Ph. de Forcrand,  P. Hasenfratz and F. Niedermayer for
various discussions.  Support by Fonds zur F\"orderung der
Wissenschaftlichen Forschung in \"Osterreich, Project P11502-PHY is
gratefully acknowledged.


\newpage

\begin{thebibliography}{10}

\bibitem{NiNi81a}
H. Nielsen and M. Ninomiya,
\newblock Nucl. Phys. B 193 (1981) 173.

\bibitem{AtSi71}
M. Atiyah and I.~M. Singer,
\newblock Ann. Math. 93 (1971) 139.

\bibitem{GiWi82}
P.~H. Ginsparg and K.~G. Wilson,
\newblock Phys. Rev. D 25 (1982) 2649.

\bibitem{Ha98c}
P. Hasenfratz,
\newblock Nucl. Phys. B (Proc. Suppl.) 63A-C (1998) 53.

\bibitem{NaNe93}
R. Narayanan and H. Neuberger,
\newblock Phys. Rev. Lett. 71 (1993) 3251.

\bibitem{Ne98}
H. Neuberger,
\newblock Phys. Lett. B 417 (1998) 141.

\bibitem{Ne98a}
H. Neuberger,
\newblock Phys. Lett. B 427 (1998) 353.

\bibitem{HaLaNi98}
P. Hasenfratz, V. Laliena, and F. Niedermayer,
\newblock Phys. Lett. B 427 (1998) 125.

\bibitem{Ha98a}
P. Hasenfratz,
\newblock Nucl. Phys. B 525 (1998) 401.

\bibitem{Lu98}
M. L{\"u}scher,
\newblock Phys. Lett. B 428 (1998) 342.

\bibitem{Ch98a}
S. Chandrasekharan,
\newblock hep-lat/9805015.

\bibitem{Ho98}
I. Horvath,
\newblock Phys. Rev. Lett. 81 (1998) 4063.

\bibitem{LaPa98}
C.~B. Lang and T.~K. Pany,
\newblock Nucl. Phys. B (Proc. Suppl.) 63A-C (1998) 898;
\newblock Nucl. Phys. B 513 (1998) 645.

\bibitem{FaLaWo98}
F. Farchioni, C.~B. Lang, and M. Wohlgenannt,
\newblock Phys. Lett. B 433 (1998) 377.

\bibitem{FaHiLa98a}
F. Farchioni, I. Hip, C.~B. Lang, and M. Wohlgenannt,
\newblock hep-lat/9809049.

\bibitem{FaHiLa98}
F. Farchioni, I. Hip, and C.~B. Lang,
\newblock Phys. Lett. B 443 (1998) 214.

\bibitem{Ch98g}
S. Chandrasekharan,
\newblock hep-lat/9810007.

\bibitem{NaNeVr95}
R. Narayanan, H. Neuberger, and P. Vranas,
\newblock Phys. Lett. B 353 (1995) 507.

\bibitem{HeJaLu98}
P. Hern\'andez, K. Jansen, and M. L\"uscher,
\newblock hep-lat/9808010.

\bibitem{LeSm92}
H. Leutwyler and A. Smilga,
\newblock Phys. Rev. D 46 (1992) 5607.

\bibitem{ShVe93}
E.~V. Shuryak and J.~J.~M. Verbaarschot,
\newblock Nucl. Phys. A 560 (1993) 306.

\bibitem{GuMuWe98}
T. Guhr, A. M{\"u}ller-Groeling, and H.~A. Weidenm{\"u}ller,
\newblock Phys. Rep. 299 (1998) 189.

\bibitem{DaAkOsDoVe}
P.~H. Damgaard, \newblock Phys. Lett. B 424 (1998) 322;
G. Akemann and P.~H. Damgaard, \newblock Nucl. Phys. B 528 (1998) 411;
J.~C. Osborn, D. Toublan and J.~J.~M. Verbaarschot, \newblock hep-th/9806110;
P.~H. Damgaard, J.~C. Osborn, D. Toublan and J.~J.~M. Verbaarschot,
\newblock hep-th/9811212.

\bibitem{VeZa93}
J.~J.~M. Verbaarschot and I. Zahed,
\newblock Phys. Rev. Lett. 70 (1993) 3852.

\bibitem{Ve94}
J.~J.~M. Verbaarschot,
\newblock Phys. Rev. Lett. 72 (1994) 2531.

\bibitem{JaSp98}
K. Splittorff and A.~D. Jackson,
\newblock hep-lat/9805018.

\bibitem{FaLa98}
E. Farchioni and V. Laliena,
\newblock Phys. Rev. D 58 (1998) 054501.

\bibitem{BlBuHa96}
M. Blatter, R. Burkhalter, P. Hasenfratz, and F. Niedermayer,
\newblock Phys. Rev. D 53 (1996) 923.

\bibitem{Ni98}
F. Niedermayer,
\newblock hep-lat/9810026.

\bibitem{VanTh}
J. Kiskis,
\newblock Phys. Rev. D 15 (1977) 2329;
N.~K. Nielsen and B. Schroer,
\newblock Nucl. Phys. B 127 (1977) 493;
M.~M. Ansourian,
\newblock Phys. Lett. 70B (1977) 301.

\bibitem{Ne98d}
H. Neuberger,
\newblock Phys. Rev. D 57 (1998) 5417.

\bibitem{BaCa80}
T. Banks and A. Casher,
\newblock Nucl. Phys. 169 (1980) 103.

\bibitem{ChZe98}
T.~W. Chiu and S.~V. Zenkin, hep-lat/9806019.

\bibitem{BeMeSc98}
M.~E. Berbenni-Bitsch, S. Meyer, A. Sch\"afer, J.~J.~M. Verbaarschot, and T.
Wettig,
\newblock Phys. Rev. Lett. 80 (1998) 1146.

\bibitem{Sp98}
K. Splittorff,
\newblock hep-th/9810248.

\bibitem{MaGuWe98}
J.-Z. Ma, T. Guhr, and T. Wettig,
\newblock Eur. Phys. J. A 2 (1998) 87, 425.

\bibitem{BeMeWe98}
M.~E. Berbenni-Bitsch, S. Meyer, and T. Wettig,
\newblock Phys. Rev. D 58 (1998) 071502.

\bibitem{BeGoMe98}
M.~E. Berbenni-Bitsch, M. G\"ockeler, S. Meyer, 
A. Sch\"afer, and T. Wettig,
\newblock hep-lat/9809058.

\bibitem{DaHeKrGoHeRa98}
P.~H. Damgaard, U.~M. Heller, and A. Krasnitz,
\newblock hep-lat/9810060.
M. G\"ockeler, H. Hehl, P.~E.~L. Rakow, A. Sch\"afer, and T. Wettig,
\newblock hep-lat/9811018.

\bibitem{AkDaMaNiSeVe}
G. Akeman, P.~H. Damgaard, U. Magnea and S.~M. Nishigaki, 
\newblock Nucl. Phys. B 487 (1997) 721;
P.~H. Damgaard and S.~M. Nishigaki, \newblock Nucl. Phys. B 518 (1998) 495;
M.~K. Sener and J.~J.~M. Verbaarschot, \newblock Phys. Rev. Lett. 81 (1998) 248.
 

\bibitem{Ne98e}
H. Neuberger,
\newblock Phys. Rev. Lett. 81 (1998) 4060.

\bibitem{EdHeNa98}
R.~G. Edwards, U.~M. Heller, and R. Narayanan,
\newblock Nucl. Phys. B 535 (1998) 403.

\bibitem{Bo98}
A. Borici,
\newblock hep-lat/9810064.

\bibitem{Li98}
C. Liu,
\newblock hep-lat/9811008.

\bibitem{SaWi92}
I. Sachs and A. Wipf,
\newblock Helv. Phys. Acta 65 (1992) 653.

\bibitem{Fo93}
P.~J. Forrester,
\newblock Nucl. Phys. B 402 (1993) 709.

\bibitem{NaFo95}
T. Nagao and P.~J. Forrester,
\newblock Nucl. Phys. B 435 (1995) 401.

\bibitem{BeGoGu98}
M.~E. Berbenni-Bitsch, M. G{\"o}ckeler, T. Guhr et~al.,
\newblock Phys. Lett. B 438 (1998) 14.

\bibitem{StJaNoPaOsVe98}
J. Stern,
\newblock hep-ph/9812082.
R.~A. Janik, M.~A. Nowak, G. Papp, and I. Zahed,
\newblock Phys. Rev. Lett. 81 (1998) 264.
J.~C. Osborn and J.~J.~M. Verbaarshot,
\newblock Nucl. Phys. B 525 (1998) 738; Phys. Rev. Lett. 81 (1998) 268.

\bibitem{NaFo98}
T. Nagao and P.~J. Forrester,
\newblock Nucl. Phys. B 509 (1998) 561.

\bibitem{Bi98b}
W. Bietenholz,
\newblock Eur. Phys. J. C 6 (1999) 537.
\end{thebibliography}
\end{document}